\documentclass[aps,prd,nofootinbib,twocolumn,groupedaddress,showpacs]{revtex4}

\usepackage{amssymb,amsmath}
\usepackage[dvips]{epsfig}
\usepackage{epsf}
\usepackage{graphics}
\usepackage{ulem}

\def\mprp{\mbox{\tiny $\bot$}}
\def\mprl{\mbox{\tiny $\|$}}

\def\beq{\begin{eqnarray}}
\def\eeq{\end{eqnarray}}

\newcommand{\ii}{\mathrm{i}} 
\newcommand{\dd}{\mathrm{d}} 
\newcommand{\eee}{\mathrm{e}} 

\begin{document}

\title{Generalized two-point tree-level amplitude $jf \to j^{\, \prime} f^{\, \prime}$ 
in a magnetized medium (extended version)}

\author{A. V. Kuznetsov} \email[]{avkuzn@uniyar.ac.ru}
\author{D. A. Rumyantsev} \email[]{rda@uniyar.ac.ru}
\author{D. M. Shlenev}
\affiliation{Division of Theoretical Physics Yaroslavl State (P.G.~Demidov) 
University, Sovietskaya 14, 150000 Yaroslavl, Russian Federation}

\date{\today}

\begin{abstract}

{The tree-level two-point amplitudes for the transitions 
$jf \to j^{\, \prime} f^{\, \prime}$, where $f$ is a fermion and $j$ is a generalized current, 
in a constant uniform magnetic field of an arbitrary strength and in  
charged fermion plasma, for the $jf$ interaction vertices of the scalar, pseudoscalar, vector and  
axial-vector types have been investigated. The particular cases of a very strong magnetic field, 
and of the coherent scattering off the real fermions without change of
their states (the ``forward'' scattering) have been analysed.  
}
\end{abstract}





\maketitle

\unitlength 1mm

\section{Introduction}

Nowadays, there exists rather keen interest to astrophysical objects
with the scale of the magnetic field strength near  
the critical value of $B_e = m^2/e \simeq 4.41\times 10^{13}$~G
~\footnote{We use natural units
$c = \hbar = k_{\rm{B}} = 1$, $m_f$ is the fermion mass, and $e_f$ is the fermion
charge.}. 
This group of objects includes the radio pulsars and the so-called
magnetars, which are the neutron stars featuring
the magnetic field strengths from $10^{12}$~G (radio pulsars) to  
$4\times 10^{14}$~G 
(magnetars)~\cite{Kouveliotou:1998ze,Kouveliotou:1998fd,Gavriil:2002mc,Ibrahim:2002zw}.
The spectra analysis of these objects also provides an evidence for the
presence of electron-positron plasma in the radio pulsars and magnetars environment, 
with the minimum magnetospheric plasma density being of the order 
of the Goldreich-Julian density~\cite{GJ:1969}: 
\begin{eqnarray}
\label{eq:ngj}
n_{GJ} \simeq 
 3\cdot 10^{13}\, \mbox{cm}^{-3} 
\left (\frac{B}{100B_e} \right )\left (\frac{10\,\mbox{s}}{P} \right ) \, .
\end{eqnarray}                                         
It is well-known that strong magnetic field and/or plasma could have an essential influence on various quantum processes~\cite{Lai:2001,Harding:2006qn,KM_Book_2003,KM_Book_2013}, because the external active medium 
catalyses the processes, by changing their kinematics and inducing new interactions. Therefore,
the effects of magnetized plasma on microscopic physics should be
incorporated in the magnetosphere models of strongly magnetized neutron stars.  
In the present paper we consider the two-point processes, because such reactions 
can have  possible resonant behavior, and therefore they 
are very interesting for astrophysical applications~\cite{Lyutikov:2006}.  
                                                           
The investigation of the two-point processes in an external active medium (electromagnetic 
field and/or plasma) has a rather long history. 
The most general expression for a
two-vertex loop amplitude of the form $j \to f \bar f \to j^{\, \prime}$
in a pure constant uniform magnetic field and in a crossed
field was obtained previously in Ref.~\cite{Borovkov:1999}, where all possible combinations
of scalar, pseudoscalar, vector, and pseudovector
interactions of the generalized currents $j$ and $j^{\, \prime}$
with fermions were considered.

The typical example of a tree-level process 
with two vector vertices in the presence of magnetized plasma is the
Compton scattering  as a possible channel
of the radiation spectra formation.  This process  was studied in 
a number of papers, see e.g.~\cite{Herold:1979, Melrose:1983b, 
Harding:1986, Harding:2000, Miller:1995, Bulik:1997}), 
but the results were presented there in the form without taking 
account of the photon dispersion properties. In the recent paper~\cite{RCh09} 
this neglect was corrected. The expression for the Compton scattering 
amplitude, with the initial and final electrons being on the lowest Landau level 
was presented in Ref.~\cite{RCh09} in the explicit Lorentz invariant form. 
The other example of the Compton like process with the vector  and 
axial-vector vertices, the photon transition into the neutrino pair, $\gamma \to \nu \bar \nu$, in the presence of magnetized plasma, was studied in Ref.~\cite{Kennett:1998}. However, the 
results in that paper were presented in rather cumbersome form with an implicit covariance. 
Those results would be poorly applicable for an analysis of the other photon-fermion scattering processes 
with the production of exotic particles, such as axion, neutralino, etc.

Thus, it is interesting to consider the tree-level two-point amplitude for the transition of the type
$jf \to j^{\, \prime} f^{\prime}$ in a constant uniform magnetic field and 
charged fermion plasma, for different combinations of the vertices that were used in the
paper~\cite{Borovkov:1999}.  Particularly, we generalize the results, obtained 
in Ref.~\cite{Borovkov:1999} to the case of magnetized plasma, since such a situation looks 
the most realistic for astrophysical objects. Such a generalization was performed in part 
in Ref.~\cite{Shabad} for the case of the photon polarization operator in a magnetized electron-positron plasma.                   
The paper is organized as follows. In Sec.~\ref{Sec:2}, we calculate the scattering 
amplitudes for different spin states  
of the initial and final fermions and for generalized vertices of the scalar, pseudoscalar, vector or  
axial vector types. All the amplitudes are presented in the explicit Lorentz and gauge invariant 
forms. In Sec.~\ref{Sec:3}, we consider the particular case, when all the 
fermions occupy the ground Landau level (the strong field limit). 
A coherent scattering of neutral particles off the real fermions without change of
their states (``forward'' scattering) is analysed in Sec.~\ref{Sec:4}.
Final comments and discussion of the obtained results and possible astrophysical applications 
are given in Sec.~\ref{Sec:5}.


\section{The set of expressions for the amplitudes}
\label{Sec:2}


The generalized amplitude of the transition $jf \to j^{\, \prime} f^{\, \prime}$ 
will be analyzed 
by using the effective Lagrangian for the interaction of the current $j$ 
with fermions in the form
\begin{eqnarray}
{\cal L}(x) \, = \, \sum \limits_{k} g_k 
[\bar {\psi_f}(x) \Gamma_k \psi_f(x)] j_k(x), 
\label{eq:L}
\end{eqnarray}
\noindent
where the generalized index $k = S, P, V, A$ numbers the matrices 
$\Gamma_k$,   
$\Gamma_S = 1, \, \Gamma_P = \gamma_5, \, \Gamma_V = \gamma_{\alpha},
\, \Gamma_A = \gamma_{\alpha} \gamma_5$; 
$j_k(x)$ are the generalized currents ($j_S$, $j_P$, $j_{V\alpha}$ or $j_{A\alpha}$) 
or the photon polarization vectors, $g_k$ are the coupling constants, and
$\psi_f(x)$ are the fermion wave functions. 
The $\gamma_5$ matrix is defined as $\gamma_5 = \ii \, \gamma^0 \gamma^1 \gamma^2 \gamma^3$.
\begin{figure}
\includegraphics[width=8cm]{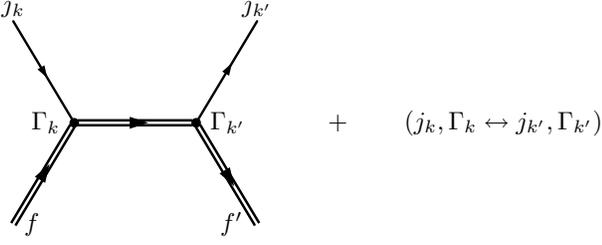}
\caption{ The Feynman diagrams for the reaction $jf \to j^{\, \prime} f^{\, \prime}$. 
Double lines 
mean that the effect of the external field on 
the initial and final states is exactly taken into account. }
\label{fig:Diagjj}
\end{figure}

The $S$-matrix element in the tree approximation is described by the Feynman diagrams 
shown in Fig.~\ref{fig:Diagjj} and has the form
\beq                          
\label{eq:S1}
S^{s^{\, \prime} s}_{k^{\, \prime} k} &=& - g_k g_{k'}\int \dd^4 X \dd^4 Y j_k (X) j_{k'} (Y)
\\
\nonumber
&\times& 
\left [\bar \Psi^{s'}_{p',\ell'}(Y) \Gamma_{k'} 
\hat S(Y,X) 
\Gamma_k \Psi^{s}_{p,\ell}(X) \right ]\, .
\eeq

\noindent Here,  $p^{\mu} = (E_{\ell}, {\bf p})$ and $p^{\, \prime \mu} =
(E^{\, \prime}_{\ell'}, {\bf p}^{\, \prime})$ 
are the four-momenta of the initial and final fermions correspondingly, $\Psi^s_{p,n}(X)$ 
are the fermion wave functions in the presence of external magnetic field, 
$X^{\mu} = (X_0, X_1, X_2, X_3)$. 

There exist several descriptions of a procedure of obtaining the fermion wave functions in the presence of external 
magnetic field by solving the Dirac equation, see
e.g.~\cite{Johnson:1949,Akhiezer:1965,Sokolov:1968,Melrose:1983a,Sokolov:1986,Bhattacharya:2004,Balantsev:2011} and 
also~\cite{KM_Book_2003,KM_Book_2013}. 
In the most cases, the solutions are presented in the form with the upper two components 
of the bispinor corresponding to the fermion states with the spin projections 1/2 and -1/2
on the magnetic field direction.
In this approach, we use a 
representation of the fermion wave functions as the 
eigenstates of the covariant operator 
$\hat{\mu}_z = m_f \Sigma_z - \ii \gamma_0 \gamma_5 [{\bf \Sigma} \times \hat{\bf P}]_z $~\cite{Sokolov:1968}. 
Here, $\hat{\bf P} =  - \ii {\bf \nabla} - e_f {\bf A}$ is the generalized momentum operator. 
We take the frame where the field is directed 
along the $z$ axis, and the Landau gauge where the four-potential is: $A^\mu = (0, 0, x B, 0)$. 
It is convenient to use the notation $\beta = |e_f| B$, and to introduce the sign of the fermion 
charge as $\eta = e_f/|e_f|$. 

Our choice of the Dirac equation solutions as the eigenfunctions of the operator $\hat{\mu}_z$ 
is caused by the following arguments. Calculations of the process widths with two or more vertices 
in an external magnetic field by the standard method, including the squaring the amplitude 
with all the Feynman diagrams and with summation or averaging over the fermion polarization states, 
contain significant computational difficulties. In this case, it is convenient to calculate partial 
contributions to the amplitude from the channels with different fermion polarization states and 
for each diagram separately, by direct multiplication of the bispinors and the Dirac matrices. 
The result, up to a total for both diagrams non-invariant phase will have an explicit Lorentz 
invariant structure. On the contrary, the amplitudes obtained with using the solutions for 
a fixed direction of the spin, do not have Lorentz invariant structure. Only the amplitude squared, 
summed over the fermion polarization states, is manifestly Lorentz-invariant.

The fermion wave functions having the form
\beq
\label{eq:psie}
\Psi^s_{p,n}(X) = \frac{e^{-\ii(E_{n} X_0 - p_y X_2 - p_z X_3)}\; U^s_{n} (\xi)}
{\sqrt{4E_{n}M_n (E_{n} + M_n)(M_n + m_f) L_y L_z}} \, ,  
\eeq
where 
\beq
\label{eq:E_n,M_n}
E_n = \sqrt{M_n^2 + p_z^2}\, , \quad  M_n = \sqrt{m_f^2 + 2 \beta n}\, ,
\eeq
are the solutions of the equation
\beq
\label{eq:mu_z_Eq}
\hat{\mu}_z \,\Psi^s_{p,n}(X) = s \, M_n \, \Psi^s_{p,n}(X) \, , \quad s = \pm 1\,.
\eeq
%
It is convenient to present the bispinors $U^s_{n} (\xi)$ in the form of 
decomposition over the solutions for negative and positive fermion charge, $U^s_{n, \eta} (\xi)$:
\beq
\label{eq:U^s}
U^s_{n} (\xi) = \frac{1-\eta}{2} \, U^{s}_{n,-} (\xi) + \frac{1+\eta}{2} \, U^{s}_{n,+} (\xi) \,,
\eeq
where
\beq
\label{eq:U--}
&&U^{-}_{n,-} (\xi) = \left ( 
\begin{array}{c}
-\ii\sqrt{2\beta n} \, p_z V_{n-1} (\xi)\\[2mm]
(E_n + M_n)(M_n + m_f) V_n (\xi)\\[2mm]
-\ii\sqrt{2\beta n} (E_n + M_n) V_{n-1} (\xi)\\[2mm]
-p_z (M_n + m_f) V_n (\xi)
\end{array}
\right )  ,   
\\ [3mm]
\label{eq:U+-}
&&U^{+}_{n,-} (\xi) = \left ( 
\begin{array}{c}
(E_n + M_n) (M_n + m_f) V_{n-1} (\xi)\\[2mm]
-\ii\sqrt{2\beta n} \, p_z V_n (\xi)\\[2mm]
p_z (M_n + m_f) V_{n-1} (\xi)\\[2mm]
\ii \sqrt{2 \beta n} (E_n + M_n) V_n (\xi)
\end{array}
\right )\! , 
\eeq

\beq
\label{eq:U-+}
&&U^{-}_{n,+} (\xi) = \left ( 
\begin{array}{c}
\ii\sqrt{2\beta n} \, p_z V_{n} (\xi)\\[2mm]
(E_n + M_n)(M_n + m_f) V_{n-1} (\xi)\\[2mm]
\ii\sqrt{2\beta n} (E_n + M_n) V_{n} (\xi)\\[2mm]
-p_z (M_n + m_f) V_{n-1} (\xi)
\end{array}
\right ) \!\! ,   
\\ [3mm]
\label{eq:U++}
&&U^{+}_{n,+} (\xi) = \left ( 
\begin{array}{c}
(E_n + M_n) (M_n + m_f) V_{n} (\xi)\\[2mm]
\ii\sqrt{2\beta n} \, p_z V_{n-1} (\xi)\\[2mm]
p_z (M_n + m_f) V_{n} (\xi)\\[2mm]
-\ii \sqrt{2 \beta n} (E_n + M_n) V_{n-1} (\xi)
\end{array}
\right ) , 
\eeq
$V_n(\xi) \, (n = 0,1,2, \dots)$ 
are the normalized harmonic oscillator functions, which are expressed in terms of 
Hermite polynomials $H_n(\xi)$ \cite{Gradshtein}:
\beq
\label{eq:V_n}
V_n (\xi) = \frac{\beta^{1/4}\eee^{-\xi^2/2}}{\sqrt{2^n n! \sqrt{\pi}}} \, H_n(\xi)\, ,
\eeq
\beq
\label{eq:xi}
\xi = \sqrt{\beta} \left (X_1 - \eta \frac{p_y}{\beta} \right ) .
\eeq
The currents $j_k$ in Eq.~(\ref{eq:S1}) can be expressed through
the amplitudes in the momentum space:
%
\beq
\label{eq:j_k}
j_k (X) = \frac{e^{-\ii(qX)}}{\sqrt{2q_0 V}} \, j_k(q) \,.
\eeq

\noindent We use the fermion propagator in the form of the sum over 
the Landau levels~\cite{Kuznetsov:2011,KM_Book_2013}:
\begin{equation}
S (X, X^{\,\prime}) = \sum\limits_{n=0}^{\infty} \; S_n (X, X^{\,\prime}) \,,
\label{eq:propagator_sum_n}
\end{equation}
\beq                          
&&S_n (X, X^{\,\prime}) = \frac{\ii}{2^n \, n!} \, \sqrt{\frac{\beta}{\pi}} \, 
\exp \left(- \, \beta \, \frac{X_1^2 + X_1^{\,\prime\,2}}{2} \right)
\nonumber\\[3mm]
&&\times 
\int\frac{\dd p_0 \, \dd p_y \, \dd p_z}{(2 \pi)^3 }
\frac{\eee^{- \, \ii \, \left( p \,(X - X^{\,\prime}) \right)_{\parallel}}}
{p_{\shortparallel}^2 - m^2 - 2 \, \beta \, n + \ii\,\varepsilon}\,
\nonumber\\[3mm]
&&\times 
\exp \left\{ - \, \frac{p_y^2}{\beta} 
- p_y \left[\,X_1 + X_1^{\,\prime} - \ii \, (X_2 - X_2^{\,\prime})\right] \right\} 
\nonumber\\[3mm]
&&\times 
\bigg\{ 
\left[ (p \gamma)_{\shortparallel} + m \right]
\left[ \varPi_- \, H_n (\xi) \, H_n (\xi^{\,\prime})  \right.
\nonumber\\[3mm]
&&+ \left. \varPi_+ \, 2 n \, H_{n-1} (\xi) \, H_{n-1} (\xi^{\,\prime}) 
\right]  
\nonumber\\[3mm] 
&& 
+ \, \ii \, 2 n \, \sqrt{\beta} \, \gamma^1 
\left[ \varPi_- \, H_{n-1} (\xi) \, H_n (\xi^{\,\prime}) 
\right.
\nonumber\\[3mm] 
&&-\left.  
\varPi_+ \, H_n (\xi) \, H_{n-1} (\xi^{\,\prime}) 
\right]
\bigg\} ,
\label{eq:propn2}
\end{eqnarray}
where $\xi$ and $\xi^{\,\prime}$ are defined similarly to Eq.~(\ref{eq:xi}).

Hereafter we use the following notations:
four-vectors with the indices $\bot$ and $\parallel$
belong
to the Euclidean \{1, 2\} subspace and the Minkowski \{0, 3\} subspace correspondingly.
Then for arbitrary 4-vectors
$A_\mu$, $B_\mu$ one has
\beq
&&A_{\mprp}^\mu = (0, A_1, A_2, 0), \quad  A_{\mprl}^\mu = (A_0, 0, 0, A_3), \nonumber \\
&&(A B)_{\mprp} = (A \Lambda B) =  A_1 B_1 + A_2 B_2 , \nonumber \\
&&(A B)_{\mprl} = (A \widetilde \Lambda B) = A_0 B_0 - A_3 B_3, \nonumber
\end{eqnarray}

\noindent where the matrices
$\Lambda_{\mu \nu} = (\varphi \varphi)_{\mu \nu}$,\,
$\widetilde \Lambda_{\mu \nu} =
(\tilde \varphi \tilde \varphi)_{\mu \nu}$ are constructed with
the dimensionless tensor of the external
magnetic field, $\varphi_{\mu \nu} =  F_{\mu \nu} /B$,
and the dual tensor,
${\tilde \varphi}_{\mu \nu} = \frac{1}{2}
\varepsilon_{\mu \nu \rho \sigma} \varphi_{\rho \sigma}$.
The matrices $\Lambda_{\mu \nu}$ and  $\widetilde \Lambda_{\mu \nu}$
are connected by the relation
$\widetilde \Lambda_{\mu \nu} - \Lambda_{\mu \nu} =
g_{\mu \nu} = \rm{diag} (1, -1, -1, -1)$,
and play the role of the metric tensors in the perpendicular ($\bot$)
and the parallel ($\parallel$) subspaces respectively.

After integration in Eq.~(\ref{eq:S1}) over $\dd^4X$ and $\dd^4Y$ 
we obtain
\beq
\label{eq:S2}                                  
&&S^{s^{\,\prime} s}_{k^{\, \prime} k} = \frac{\ii (2\pi)^3 
\delta^{(3)} (P - p^{\, \prime} - q^{\, \prime})}
{\sqrt{2q_0 V 2q^{\, \prime}_0 V 2 E_{\ell} L_y L_z 2 E^{\, \prime}_{\ell'}L_y L_z}}\, 
{\cal M}^{s^{\, \prime} s}_{k^{\, \prime} k} \, , 
\eeq
\noindent where    $\delta^3 (P - p^{\, \prime} - q^{\, \prime}) 
= \delta (P_0 - E^{\, \prime}_{\ell'}-q^{\, \prime}_0) 
\delta (P_y - p^{\, \prime}_y - q^{\, \prime}_y) 
\delta (P_z - p^{\, \prime}_z - q^{\, \prime}_z)$,  $P_\alpha = (p+q)_\alpha , 
\,\, \alpha =0,2,3$,  and the partial amplitudes 
${\cal M}^{s^{\, \prime} s}_{k^{\, \prime} k}$  can be  
presented in the following form:
\beq
&&{\cal M}^{s^{\, \prime} s}_{k^{\, \prime} k} = \frac{-
\exp{\left [-\ii\theta \right ]}}
{2\; \sqrt{M_\ell M_{\ell'} (M_\ell + m_f)(M_{\ell'} + m_f)}} 
\\
\nonumber
&&\times \bigg \{ \exp \left [\frac{\ii (q \varphi q^{\, \prime})}{2\beta} \right ]
\, \left [\frac{q_y + \ii q_x}{\sqrt{q^2_{\mprp}}} \right ]^{-\ell}
\, \left [\frac{q^{\, \prime}_y - \ii q^{\, \prime}_x}
{\sqrt{q^{\, \prime 2}_{\mprp}}} \right ]^{-\ell'} 
\\
\nonumber
&&\times \sum \limits_{n=0}^{\infty}
\; \left (\frac{(q\Lambda q^{\, \prime})-\ii (q\varphi q^{\, \prime})}
{\sqrt{q^2_{\mprp} q^{\, \prime 2}_{\mprp}}} \right )^n 
\frac{{\cal R}^{s^{\, \prime} s}_{k^{\, \prime} k}}{P^2_{\mprl} - m_f^2 - 2\beta n}  
\\
\nonumber
&&+ (-1)^{\ell + \ell'} \exp \left [-\frac{\ii (q \varphi q^{\, \prime})}{2\beta} \right ]
\, \left [\frac{q^{\, \prime}_y + \ii q^{\, \prime}_x}{\sqrt{q^{\, \prime  2}_{\mprp}}} \right ]^{-\ell}
\, \left [\frac{q_y - \ii q_x}
{\sqrt{q^{2}_{\mprp}}} \right ]^{-\ell'} 
\\
\nonumber
&&\times \sum \limits_{n=0}^{\infty}
\; \left (\frac{(q\Lambda q^{\, \prime})+\ii (q\varphi q^{\, \prime})}
{\sqrt{q^2_{\mprp} q^{\, \prime 2}_{\mprp}}} \right )^n 
\frac{{\cal R}^{s^{\, \prime} s}_{k k^{\, \prime}}}{P^{\, \prime 2}_{\mprl} - m_f^2 - 2\beta n} \bigg \}
\, ,
\label{eq:M11}
\eeq
\noindent where  $\theta = (q_x - q'_x)(p_y+p'_y)/(2\beta)$ is the general phase for both 
diagrams in Fig.~\ref{fig:Diagjj}.

The main part of the problem is to calculate the values
${\cal R}^{s^{\, \prime}s}_{k^{\, \prime} k}$ which are expressed via 
the following Lorentz covariants in the $\{0, 3\}$-subspace
\beq
\nonumber
&&{\cal K}_{1\alpha} = \sqrt{\frac{2}{(p\widetilde \Lambda p^{\, \prime}) + 
M_\ell M_{\ell'}}} 
\\
&&\times \left \{M_\ell (\widetilde \Lambda p^{\, \prime})_\alpha + 
M_{\ell'} (\widetilde \Lambda p)_\alpha  \right \}\, ,
\label{eq:K1}
\eeq
\beq
\nonumber
&&{\cal K}_{2\alpha} = \sqrt{\frac{2}{(p\widetilde \Lambda p^{\, \prime}) + 
M_\ell M_{\ell'}}} 
\\
&&\times \left \{M_\ell (\widetilde \varphi p^{\, \prime})_\alpha + 
M_{\ell'} (\widetilde \varphi p)_\alpha  \right \}\, ,
\label{eq:K2}
\eeq
\beq
{\cal K}_{3} = \sqrt{2\left [(p\widetilde \Lambda p^{\, \prime}) + 
M_\ell M_{\ell'} \right]} \, ,  
\eeq
\beq
{\cal K}_4 = 
- \sqrt{\frac{2}{(p\widetilde \Lambda p^{\, \prime}) + M_\ell M_{\ell'}}}\, 
(p\widetilde \varphi p^{\, \prime}) \, .
\label{eq:K34}
\eeq

\noindent The following integrals appear in the calculations: 
\beq
\label{eq:Ilnnl}
&&\frac{1}{\sqrt{\pi}}\int \dd Z \, \eee^{-Z^2} 
H_n \left (Z + \frac{q_y + \ii q_x}{2\sqrt{\beta}} \right )   
\\[3mm]
\nonumber
&&\times H_{\ell} \left (Z - \frac{q_y - \ii q_x}{2\sqrt{\beta}} \right ) 
\\[3mm]
\nonumber
&&= 2^{(n+\ell)/2} \sqrt{n ! \, \ell !} 
\left [\frac{q_y + \ii q_x}{\sqrt{q^2_{\mprp}}} \right ]^{n-\ell} 
\eee^{{q^2_{\mprp}}/{(4\beta)}} 
{\cal I}_{n, \ell} \left (\frac{q^{2}_{\mprp}}{2 \beta} \right ) \, , 
\eeq
\noindent where
\beq
\nonumber
&&{\cal I}_{n, \ell} (x) = (-1)^{n-\ell} {\cal I}_{\ell, n} (x) 
\\
&&= \sqrt{\frac{\ell !}{n !}} \; \eee^{-x/2} x^{(n-\ell)/2} L_\ell^{n-\ell} (x) \, ,
\label{eq:Inl}
\eeq
\noindent and $L^k_n (x)$ are the generalized Laguerre polynomials~\cite{Gradshtein}.

The results for ${\cal R}^{s^{\, \prime} s}_{k^{\, \prime} k}$ are presented below. 
Hereafter we use the following definitions: $P_\alpha = (p+q)_\alpha$, $P'_\alpha = (p-q')_\alpha$, 
${\cal I}_{n, \ell} \equiv {\cal I}_{n, \ell}
\left ({q^{2}_{\mprp}}/{(2 \beta)} \right )$ and 
${\cal I}^{\, \prime}_{n, \ell'} \equiv {\cal I}_{n, \ell'}
\left ({q^{\, \prime 2}_{\mprp}}/{(2 \beta)} \right )$. 
For definiteness, we further consider the fermion with a negative charge, $\eta = -1$. 

\begin{widetext}

%
\begin{enumerate}


\item
In the case when $j$ and $j^{\, \prime}$ are scalar currents ($k,k^{\, \prime} = S$) 
the calculation yields
\beq
\nonumber
&&{\cal R}^{++}_{SS} = g_s g_s^{\, \prime} j_s j_s^{\, \prime} 
\bigg \{ 2\beta 
\sqrt{\ell \ell^{\, \prime}} \left [({\cal K}_1 P) - m_f {\cal K}_3 \right ] 
{\cal I}^{\, \prime}_{n, \ell'} {\cal I}_{n, \ell}  + (M_{\ell} + m_f) (M_{\ell'} + m_f) 
\left [({\cal K}_1 P) + m_f {\cal K}_3 \right ] 
\\[3mm]
\label{eq:Rssupup}
&&\times {\cal I}^{\, \prime}_{n-1, \ell'-1} {\cal I}_{n-1, \ell-1} -
2\beta \sqrt{n} {\cal K}_3 \big [\sqrt{\ell} (M_{\ell'} + m_f) 
{\cal I}^{\, \prime}_{n-1, \ell'-1} {\cal I}_{n, \ell} + 
\sqrt{\ell^{\, \prime}} (M_{\ell} + m_f) 
{\cal I}^{\, \prime}_{n, \ell'} {\cal I}_{n-1, \ell-1} \big ]  
\bigg \}\, ; 
\eeq
\beq
\nonumber
&&{\cal R}^{+-}_{SS} = \ii g_s g_s^{\, \prime} j_s j_s^{\, \prime} 
\bigg \{ \sqrt{2\beta \ell^{\, \prime}}\; (M_{\ell} + m_f) 
\left [({\cal K}_2 P) - m_f {\cal K}_4 \right ] 
{\cal I}^{\, \prime}_{n, \ell'} {\cal I}_{n, \ell}  - \sqrt{2\beta \ell}\; (M_{\ell'} + m_f) 
\left [({\cal K}_2 P) + m_f {\cal K}_4 \right ] 
\\[3mm]
\label{eq:Rssupdown}
&&\times {\cal I}^{\, \prime}_{n-1, \ell'-1} {\cal I}_{n-1, \ell-1} -
\sqrt{2\beta n}\; {\cal K}_4 \big [ (M_{\ell} + m_f) (M_{\ell'} + m_f)
{\cal I}^{\, \prime}_{n-1, \ell'-1} {\cal I}_{n, \ell} - 
2\beta
\sqrt{\ell \ell^{\, \prime}}
{\cal I}^{\, \prime}_{n, \ell'} {\cal I}_{n-1, \ell-1} \big ]  
\bigg \}\, ; 
\eeq
\beq
\nonumber
&&{\cal R}^{-+}_{SS} =  - \ii g_s g_s^{\, \prime} j_s j_s^{\, \prime} 
\bigg \{ \sqrt{2\beta \ell}\; (M_{\ell'} + m_f) 
\left [({\cal K}_2 P) + m_f {\cal K}_4 \right ] 
{\cal I}^{\, \prime}_{n, \ell'} {\cal I}_{n, \ell} -
\sqrt{2\beta \ell^{\, \prime}}\; (M_{\ell} + m_f) 
\left [({\cal K}_2 P) - m_f {\cal K}_4 \right ]  
\\[3mm]
\label{eq:Rssdownup}
&&\times {\cal I}^{\, \prime}_{n-1, \ell'-1} {\cal I}_{n-1, \ell-1} -
\sqrt{2\beta n}\; {\cal K}_4 \big [2\beta
\sqrt{\ell \ell^{\, \prime}} 
{\cal I}^{\, \prime}_{n-1, \ell'-1} {\cal I}_{n, \ell} - 
(M_{\ell} + m_f) (M_{\ell'} + m_f)
{\cal I}^{\, \prime}_{n, \ell'} {\cal I}_{n-1, \ell-1} \big ]  
\bigg \}\, ; 
\eeq
\beq
\nonumber
&&{\cal R}^{--}_{SS} = g_s g_s^{\, \prime} j_s j_s^{\, \prime} 
\bigg \{ (M_{\ell} + m_f) (M_{\ell'} + m_f)
\left [({\cal K}_1 P) + m_f {\cal K}_3 \right ]  
{\cal I}^{\, \prime}_{n, \ell'} {\cal I}_{n, \ell} +
2\beta 
\sqrt{\ell \ell^{\, \prime}} \left [({\cal K}_1 P) - m_f {\cal K}_3 \right ] 
\\[3mm]
\label{eq:Rssdowndown}
&&\times {\cal I}^{\, \prime}_{n-1, \ell'-1} {\cal I}_{n-1, \ell-1} -
2\beta \sqrt{n} {\cal K}_3 \big [\sqrt{\ell^{\, \prime}} (M_{\ell} + m_f) 
{\cal I}^{\, \prime}_{n-1, \ell'-1} {\cal I}_{n, \ell} + 
\sqrt{\ell} (M_{\ell'} + m_f) 
{\cal I}^{\, \prime}_{n, \ell'} {\cal I}_{n-1, \ell-1} \big ]  
\bigg \}\, . 
\eeq

For second diagram we have the following replacement $P_{\alpha} \to P_{\alpha}^{\, \prime}$, 
${\cal I}_{m,n} \leftrightarrow {\cal I}_{m,n}^{\, \prime}$.


\item 
In the case where $j$ is scalar current and $j^{\, \prime}$ is pseudoscalar current 
($k = S, \, k^{\, \prime} = P$) we obtain

\beq
\nonumber
&&{\cal R}^{++}_{PS} = g_s g_p^{\, \prime} j_s j_p^{\, \prime} 
\bigg \{ 2\beta 
\sqrt{\ell \ell^{\, \prime}} \left [({\cal K}_2 P) + m_f {\cal K}_4 \right ] 
{\cal I}^{\, \prime}_{n, \ell'} {\cal I}_{n, \ell} -
(M_{\ell} + m_f) (M_{\ell'} + m_f) 
\left [({\cal K}_2 P) - m_f {\cal K}_4 \right ] 
\\[3mm]
\label{eq:Rspupup}
&&\times {\cal I}_{n-1, \ell'-1} {\cal I}^{\, \prime}_{n-1, \ell-1} -
2\beta \sqrt{n} {\cal K}_4 \big [\sqrt{\ell} (M_{\ell'} + m_f) 
{\cal I}^{\, \prime}_{n-1, \ell'-1} {\cal I}_{n, \ell} - 
\sqrt{\ell^{\, \prime}} (M_{\ell} + m_f) 
{\cal I}^{\, \prime}_{n, \ell'} {\cal I}_{n-1, \ell-1} \big ]  
\bigg \}\, ; 
\eeq
\beq
\nonumber
&&{\cal R}^{++}_{SP} = -g_s g_p^{\, \prime} j_s j_p^{\, \prime}
\bigg \{ 2\beta 
\sqrt{\ell \ell^{\, \prime}} \left [({\cal K}_2 P^{\, \prime}) - m_f {\cal K}_4 \right ] 
{\cal I}_{n, \ell'} {\cal I}^{\, \prime}_{n, \ell} -
(M_{\ell} + m_f) (M_{\ell'} + m_f) 
\left [({\cal K}_2 P^{\, \prime}) + m_f {\cal K}_4 \right ] 
\\[3mm]
\label{eq:Rspupup2}
&&\times {\cal I}_{n-1, \ell'-1} {\cal I}^{\, \prime}_{n-1, \ell-1} -
2\beta \sqrt{n} {\cal K}_4 \big [\sqrt{\ell} (M_{\ell'} + m_f) 
{\cal I}_{n-1, \ell'-1} {\cal I}^{\, \prime}_{n, \ell} - 
\sqrt{\ell^{\, \prime}} (M_{\ell} + m_f) 
{\cal I}_{n, \ell'} {\cal I}^{\, \prime}_{n-1, \ell-1} \big ]  
\bigg \}\, ; 
\eeq


\beq
\nonumber
&&{\cal R}^{+-}_{PS} =  \ii g_s g_p^{\, \prime} j_s j_p^{\, \prime} 
\bigg \{ \sqrt{2\beta \ell^{\, \prime}}\; (M_{\ell} + m_f) 
\left [({\cal K}_1 P) + m_f {\cal K}_3 \right ] 
{\cal I}^{\, \prime}_{n, \ell'} {\cal I}_{n, \ell} +
\sqrt{2\beta \ell}\; (M_{\ell'} + m_f) 
\left [({\cal K}_1 P) - m_f {\cal K}_3 \right ] 
\\[3mm]
\label{eq:Rspupdown}
&&\times {\cal I}^{\, \prime}_{n-1, \ell'-1} {\cal I}_{n-1, \ell-1} -
\sqrt{2\beta n}\; {\cal K}_3 \big [ (M_{\ell} + m_f) (M_{\ell'} + m_f)
{\cal I}^{\, \prime}_{n-1, \ell'-1} {\cal I}_{n, \ell} + 
2\beta
\sqrt{\ell \ell^{\, \prime}}
{\cal I}^{\, \prime}_{n, \ell'} {\cal I}_{n-1, \ell-1} \big ]  
\bigg \}\, ;
\eeq
\beq
\nonumber
&&{\cal R}^{+-}_{SP} = -\ii g_s g_p^{\, \prime} j_s j_p^{\, \prime} 
\bigg \{ \sqrt{2\beta \ell^{\, \prime}}\; (M_{\ell} + m_f) 
\left [({\cal K}_1 P^{\, \prime}) - m_f {\cal K}_3 \right ] 
{\cal I}_{n, \ell'} {\cal I}^{\, \prime}_{n, \ell} +
\sqrt{2\beta \ell}\; (M_{\ell'} + m_f) 
\left [({\cal K}_1 P^{\, \prime}) + m_f {\cal K}_3 \right ]  
\\[3mm]
\label{eq:Rspupdown2}
&&\times {\cal I}_{n-1, \ell'-1} {\cal I}^{\, \prime}_{n-1, \ell-1} -
\sqrt{2\beta n}\; {\cal K}_3 \big [ (M_{\ell} + m_f) (M_{\ell'} + m_f)
{\cal I}_{n-1, \ell'-1} {\cal I}^{\, \prime}_{n, \ell} + 
2\beta
\sqrt{\ell \ell^{\, \prime}}
{\cal I}_{n, \ell'} {\cal I}^{\, \prime}_{n-1, \ell-1} \big ]  
\bigg \}\, ;
\eeq


\beq
\nonumber
&&{\cal R}^{-+}_{PS} =  - \ii g_s g_p^{\, \prime} j_s j_p^{\, \prime} 
\bigg \{ \sqrt{2\beta \ell}\; (M_{\ell'} + m_f) 
\left [({\cal K}_1 P) - m_f {\cal K}_3 \right ] 
{\cal I}^{\, \prime}_{n, \ell'} {\cal I}_{n, \ell} +
\sqrt{2\beta \ell^{\, \prime}}\; (M_{\ell} + m_f) 
\left [({\cal K}_1 P) + m_f {\cal K}_3 \right ]  
\\[3mm]
\label{eq:Rspdownup}
&&\times {\cal I}^{\, \prime}_{n-1, \ell'-1} {\cal I}_{n-1, \ell-1} -
\sqrt{2\beta n}\; {\cal K}_3 \big [2\beta
\sqrt{\ell \ell^{\, \prime}} 
{\cal I}^{\, \prime}_{n-1, \ell'-1} {\cal I}_{n, \ell} + 
(M_{\ell} + m_f) (M_{\ell'} + m_f)
{\cal I}^{\, \prime}_{n, \ell'} {\cal I}_{n-1, \ell-1} \big ]  
\bigg \}\, ; 
\eeq
\beq
\nonumber
&&{\cal R}^{-+}_{SP} =   \ii g_s g_p^{\, \prime} j_s j_p^{\, \prime}
\bigg \{ \sqrt{2\beta \ell}\; (M_{\ell'} + m_f) 
\left [({\cal K}_1 P^{\, \prime}) + m_f {\cal K}_3 \right ] 
{\cal I}_{n, \ell'} {\cal I}^{\, \prime}_{n, \ell} +
\sqrt{2\beta \ell^{\, \prime}}\; (M_{\ell} + m_f) 
\left [({\cal K}_1 P^{\, \prime}) - m_f {\cal K}_3 \right ]  
\\[3mm]
\label{eq:Rspdownup2}
&&\times {\cal I}_{n-1, \ell'-1} {\cal I}^{\, \prime}_{n-1, \ell-1} -
\sqrt{2\beta n}\; {\cal K}_3 \big [2\beta
\sqrt{\ell \ell^{\, \prime}} 
{\cal I}_{n-1, \ell'-1} {\cal I}^{\, \prime}_{n, \ell} + 
(M_{\ell} + m_f) (M_{\ell'} + m_f)
{\cal I}_{n, \ell'} {\cal I}^{\, \prime}_{n-1, \ell-1} \big ]  
\bigg \}\, ; 
\eeq


\beq
\nonumber
&&{\cal R}^{--}_{PS} =  g_s g_p^{\, \prime} j_s j_p^{\, \prime} 
\bigg \{ (M_{\ell} + m_f) (M_{\ell'} + m_f)
\left [({\cal K}_2 P) - m_f {\cal K}_4 \right ]  
{\cal I}^{\, \prime}_{n, \ell'} {\cal I}_{n, \ell} -
2\beta 
\sqrt{\ell \ell^{\, \prime}} \left [({\cal K}_2 P) + m_f {\cal K}_4 \right ]  
\\[3mm]
\label{eq:Rspdowndown}
&&\times {\cal I}^{\, \prime}_{n-1, \ell'-1} {\cal I}_{n-1, \ell-1} -
2\beta \sqrt{n} {\cal K}_4 \big [\sqrt{\ell^{\, \prime}} (M_{\ell} + m_f) 
{\cal I}^{\, \prime}_{n-1, \ell'-1} {\cal I}_{n, \ell} - 
\sqrt{\ell} (M_{\ell'} + m_f) 
{\cal I}^{\, \prime}_{n, \ell'} {\cal I}_{n-1, \ell-1} \big ]  
\bigg \}\, ; 
\eeq
\beq
\nonumber
&&{\cal R}^{--}_{SP} = - g_s g_p^{\, \prime} j_s j_p^{\, \prime} 
\bigg \{ (M_{\ell} + m_f) (M_{\ell'} + m_f)
\left [({\cal K}_2 P^{\, \prime}) + m_f {\cal K}_4 \right ]  
{\cal I}_{n, \ell'} {\cal I}^{\, \prime}_{n, \ell} -
2\beta 
\sqrt{\ell \ell^{\, \prime}} \left [({\cal K}_2 P^{\, \prime}) - m_f {\cal K}_4 \right ]  
\\[3mm]
\label{eq:Rspdowndown2}
&&\times {\cal I}_{n-1, \ell'-1} {\cal I}^{\, \prime}_{n-1, \ell-1} -
2\beta \sqrt{n} {\cal K}_4 \big [\sqrt{\ell^{\, \prime}} (M_{\ell} + m_f) 
{\cal I}_{n-1, \ell'-1} {\cal I}^{\, \prime}_{n, \ell} - 
\sqrt{\ell} (M_{\ell'} + m_f) 
{\cal I}_{n, \ell'} {\cal I}^{\, \prime}_{n-1, \ell-1} \big ]  
\bigg \}\, . 
\eeq




\item 
In the case where $j$ is scalar current and $j^{\, \prime}$ is a vector current 
($k = S, \, k^{\, \prime} = V$) we obtain


%
\beq
\label{eq:Rsvupup}
&&{\cal R}^{++}_{VS} = g_s g_v^{\, \prime} j_s \bigg \{ - 2\beta 
\sqrt{\ell \ell^{\, \prime}} \left [(P \tilde \Lambda j^{\, \prime}){\cal K}_3 + 
(P \tilde \varphi j^{\, \prime}) {\cal K}_4 
- m_f ({\cal K}_1 j^{\, \prime}) \right ] 
{\cal I}^{\, \prime}_{n, \ell'} {\cal I}_{n, \ell}   
\\
\nonumber
&&+(M_{\ell} + m_f) (M_{\ell'} + m_f) 
\left [(P \tilde \Lambda j^{\, \prime}){\cal K}_3 + 
(P \tilde \varphi j^{\, \prime}) {\cal K}_4 
+ m_f ({\cal K}_1 j^{\, \prime}) \right ] {\cal I}^{\, \prime}_{n-1, \ell'-1} {\cal I}_{n-1, \ell-1} 
\\[3mm]
\nonumber
&& -2\beta \sqrt{n} ({\cal K}_1 j^{\, \prime}) \left [\sqrt{\ell} (M_{\ell'} + m_f) 
{\cal I}^{\, \prime}_{n-1, \ell'-1} {\cal I}_{n, \ell} - 
\sqrt{\ell^{\, \prime}} (M_{\ell} + m_f) 
{\cal I}^{\, \prime}_{n, \ell'} {\cal I}_{n-1, \ell-1} \right ]  
\\[3mm]
\nonumber 
&&+\sqrt{\frac{2 \beta}{q_{\mprp}^{\, \prime 2}}}\; (M_{\ell'} + m_f)  
\left [(q^{\, \prime} \Lambda j^{\, \prime}) + 
\ii (q^{\, \prime} \varphi j^{\, \prime}) \right ] 
\left [\sqrt{\ell} \; [({\cal K}_1 P) - m_f {\cal K}_3] 
{\cal I}^{\, \prime}_{n, \ell'-1} {\cal I}_{n, \ell} - 
\sqrt{n} (M_{\ell} + m_f) {\cal K}_3 {\cal I}^{\, \prime}_{n, \ell'-1} 
{\cal I}_{n-1, \ell-1} \right ] 
\\[3mm]
\nonumber
&&-\sqrt{\frac{2 \beta \ell^{\, \prime}}
{q_{\mprp}^{\, \prime 2}}}\;  
\left [(q^{\, \prime} \Lambda j^{\, \prime}) - 
\ii (q^{\, \prime} \varphi j^{\, \prime}) \right ] 
\left [(M_{\ell} + m_f) [({\cal K}_1 P) + m_f {\cal K}_3] 
{\cal I}^{\, \prime}_{n-1, \ell'} {\cal I}_{n-1, \ell-1}  - 
2 \beta \sqrt{\ell n} \, {\cal K}_3 {\cal I}^{\, \prime}_{n-1, \ell'} 
{\cal I}_{n, \ell} \right ]
\bigg \}\, ; 
\eeq
\beq
\label{eq:Rsvupup2}
&&{\cal R}^{++}_{SV} = g_s g_v^{\, \prime} j_s \bigg \{ - 2\beta 
\sqrt{\ell \ell^{\, \prime}} \left [(P^{\, \prime} \tilde \Lambda j^{\, \prime}){\cal K}_3 - 
(P^{\, \prime} \tilde \varphi j^{\, \prime}) {\cal K}_4 
- m_f ({\cal K}_1 j^{\, \prime}) \right ] 
{\cal I}_{n, \ell'} {\cal I}^{\, \prime}_{n, \ell}   
\\
\nonumber
&&+(M_{\ell} + m_f) (M_{\ell'} + m_f) 
\left [(P^{\, \prime} \tilde \Lambda j^{\, \prime}){\cal K}_3 - 
(P^{\, \prime} \tilde \varphi j^{\, \prime}) {\cal K}_4 
+ m_f ({\cal K}_1 j^{\, \prime}) \right ] {\cal I}_{n-1, \ell'-1} {\cal I}^{\, \prime}_{n-1, \ell-1} 
\\[3mm]
\nonumber  
&& -2\beta \sqrt{n} ({\cal K}_1 j^{\, \prime}) \left [\sqrt{\ell} (M_{\ell'} + m_f) 
{\cal I}_{n-1, \ell'-1} {\cal I}^{\, \prime}_{n, \ell} - 
\sqrt{\ell^{\, \prime}} (M_{\ell} + m_f) 
{\cal I}_{n, \ell'} {\cal I}^{\, \prime}_{n-1, \ell-1} \right ]  
\\[3mm]
\nonumber 
&&+ \sqrt{\frac{2 \beta \ell}{q^{\, \prime 2}_{\mprp}}}\;  
\left [(q^{\, \prime} \Lambda j^{\, \prime}) + 
\ii (q^{\, \prime} \varphi j^{\, \prime}) \right ] 
\left [(M_{\ell'} + m_f) \; [({\cal K}_1 P^{\, \prime}) + m_f {\cal K}_3] 
{\cal I}_{n-1, \ell'-1} {\cal I}^{\, \prime}_{n-1, \ell} - 
2 \beta \sqrt{\ell^{\, \prime} n}\, {\cal K}_3 {\cal I}_{n, \ell'} 
{\cal I}^{\, \prime}_{n-1, \ell} \right ] 
\\[3mm]
\nonumber
&&- \sqrt{\frac{2 \beta }
{q^{\, \prime 2}_{\mprp}}}\;  (M_{\ell} + m_f)
\left [(q^{\, \prime} \Lambda j^{\, \prime}) - 
\ii (q^{\, \prime} \varphi j^{\, \prime}) \right ] 
\left [\sqrt{\ell^{\, \prime}}\; [({\cal K}_1 P^{\, \prime}) - m_f {\cal K}_3] 
{\cal I}_{n, \ell'} {\cal I}^{\, \prime}_{n, \ell-1}  - 
\sqrt{n} (M_{\ell'} + m_f) {\cal K}_3 {\cal I}_{n-1, \ell'-1} 
{\cal I}^{\, \prime}_{n, \ell-1} \right ]
\bigg \}\, ; 
\eeq

%
\beq
\label{eq:Rsvupdown}
&&{\cal R}^{+-}_{VS} = -\ii g_s g_v^{\, \prime} j_s \bigg \{  
\sqrt{2 \beta \ell^{\, \prime}} (M_{\ell} + m_f) \left [(P \tilde \Lambda j^{\, \prime}) {\cal K}_4 + 
(P \tilde \varphi j^{\, \prime}) {\cal K}_3 
- m_f ({\cal K}_2 j^{\, \prime}) \right ] {\cal I}^{\, \prime}_{n, \ell'} {\cal I}_{n, \ell} 
\\
\nonumber
&& + \sqrt{2 \beta \ell} (M_{\ell'} + m_f) 
\left [(P \tilde \Lambda j^{\, \prime}) {\cal K}_4 + 
(P \tilde \varphi j^{\, \prime}) {\cal K}_3 
+ m_f ({\cal K}_2 j^{\, \prime}) \right ] 
{\cal I}^{\, \prime}_{n-1, \ell'-1} {\cal I}_{n-1, \ell-1} 
\\[3mm]
\nonumber
&&+ \sqrt{2 \beta n}\; ({\cal K}_2 j^{\, \prime}) 
\left [(M_{\ell} + m_f) (M_{\ell'} + m_f) 
{\cal I}^{\, \prime}_{n-1, \ell'-1} {\cal I}_{n, \ell} + 
2 \beta \sqrt{\ell \ell^{\, \prime}}  
{\cal I}^{\, \prime}_{n, \ell'} {\cal I}_{n-1, \ell-1} \right ] 
- \frac{(q^{\, \prime} \Lambda j^{\, \prime}) + 
\ii (q^{\, \prime} \varphi j^{\, \prime})}{\sqrt{q_{\mprp}^{\, \prime 2}}}
\\[3mm]
\nonumber 
&& \times  (M_{\ell'} + m_f) \;  
\left [(M_{\ell} + m_f)  [({\cal K}_2 P) - m_f {\cal K}_4] 
{\cal I}^{\, \prime}_{n, \ell'-1} {\cal I}_{n, \ell} + 
2\beta \sqrt{n \ell}  \;{\cal K}_4 {\cal I}^{\, \prime}_{n, \ell'-1} 
{\cal I}_{n-1, \ell-1} \right ]  
\\ [3mm]
\nonumber
&& 
-2 \beta \; \sqrt{\ell^{\, \prime}} 
\; \frac{(q^{\, \prime} \Lambda j^{\, \prime}) - 
\ii (q^{\, \prime} \varphi j^{\, \prime})}{\sqrt{q_{\mprp}^{\, \prime 2}}} 
\left [\sqrt{\ell} \; [({\cal K}_2 P) + m_f {\cal K}_4]  
{\cal I}^{\, \prime}_{n-1, \ell'} {\cal I}_{n-1, \ell-1} + 
 \sqrt{ n}\; (M_{\ell} + m_f) \; {\cal K}_4 \; {\cal I}^{\, \prime}_{n-1, \ell'} 
{\cal I}_{n, \ell} \right ]
\bigg \}\, ; 
\eeq
%

%
\beq
\label{eq:Rsvupdown2}
&&{\cal R}^{+-}_{SV} = \ii g_s g_v^{\, \prime} j_s \bigg \{  
\sqrt{2 \beta \ell^{\, \prime}} (M_{\ell} + m_f) \left [ 
(P^{\, \prime} \tilde \varphi j^{\, \prime}) {\cal K}_3 
-(P^{\, \prime} \tilde \Lambda j^{\, \prime}) {\cal K}_4 +
 m_f ({\cal K}_2 j^{\, \prime}) \right ] {\cal I}_{n, \ell'} {\cal I}^{\, \prime}_{n, \ell} 
\\
\nonumber
&& + \sqrt{2 \beta \ell} (M_{\ell'} + m_f) 
\left [ (P^{\, \prime} \tilde \varphi j^{\, \prime}) {\cal K}_3 -
(P^{\, \prime} \tilde \Lambda j^{\, \prime}) {\cal K}_4 -
m_f ({\cal K}_2 j^{\, \prime}) \right ] 
{\cal I}_{n-1, \ell'-1} {\cal I}^{\, \prime}_{n-1, \ell-1} 
\\[3mm]
\nonumber
&&+ \sqrt{2 \beta n}\; ({\cal K}_2 j^{\, \prime}) 
\left [(M_{\ell} + m_f) (M_{\ell'} + m_f) 
{\cal I}_{n-1, \ell'-1} {\cal I}^{\, \prime}_{n, \ell} + 
2 \beta \sqrt{\ell \ell^{\, \prime}}  
{\cal I}_{n, \ell'} {\cal I}^{\, \prime}_{n-1, \ell-1} \right ] 
+ \frac{(q^{\, \prime} \Lambda j^{\, \prime}) + 
\ii (q^{\, \prime} \varphi j^{\, \prime})}{\sqrt{q^{\, \prime 2}_{\mprp}}}
\\[3mm]
\nonumber 
&& \times (M_{\ell} + m_f) \;   
\left [(M_{\ell'} + m_f)  [({\cal K}_2 P^{\, \prime}) + m_f {\cal K}_4] 
{\cal I}_{n-1, \ell'-1} {\cal I}^{\, \prime}_{n-1, \ell} - 
2\beta \sqrt{n \ell^{\, \prime}}  \;{\cal K}_4 {\cal I}_{n, \ell'} 
{\cal I}^{\, \prime}_{n-1, \ell} \right ]  
\\ [3mm]
\nonumber
&& 
+2 \beta \; \sqrt{\ell} 
\; \frac{(q^{\, \prime} \Lambda j^{\, \prime}) + 
\ii (q^{\, \prime} \varphi j^{\, \prime})}{\sqrt{q^{\, \prime 2}_{\mprp}}} 
\left [\sqrt{\ell^{\, \prime}} \; [({\cal K}_2 P^{\, \prime}) - m_f {\cal K}_4]  
{\cal I}_{n, \ell'} {\cal I}^{\, \prime}_{n, \ell-1} - 
 \sqrt{ n}\; (M_{\ell'} + m_f) \; {\cal K}_4 \; {\cal I}_{n, \ell'-1} 
{\cal I}^{\, \prime}_{n-1, \ell-1} \right ]
\bigg \}\, ; 
\eeq
%

%
\beq
\label{eq:Rsvdownup}
&&{\cal R}^{-+}_{VS} = - \ii g_s g_v^{\, \prime} j_s
\bigg \{ \sqrt{2 \beta \ell} (M_{\ell'} + m_f)
\left [(P \tilde \Lambda j^{\, \prime}) {\cal K}_4 +
(P \tilde \varphi j^{\, \prime}) {\cal K}_3
+ m_f ({\cal K}_2 j^{\, \prime}) \right ] {\cal I}^{\, \prime}_{n, \ell'} {\cal I}_{n, \ell}
\\
\nonumber
&&+ \sqrt{2 \beta \ell^{\, \prime}} (M_{\ell} + m_f) 
\left [(P \tilde \Lambda j^{\, \prime}) {\cal K}_4 +
(P \tilde \varphi j^{\, \prime})  {\cal K}_3
- m_f ({\cal K}_2 j^{\, \prime}) \right ] 
{\cal I}^{\, \prime}_{n-1, \ell'-1} {\cal I}_{n-1, \ell-1} 
\\[3mm]
\nonumber
&& + \sqrt{2 \beta n}\; ({\cal K}_2 j^{\, \prime})
\left [2 \beta \sqrt{\ell \ell^{\, \prime}}
{\cal I}^{\, \prime}_{n-1, \ell'-1} {\cal I}_{n, \ell} +
(M_{\ell} + m_f) (M_{\ell'} + m_f)
{\cal I}^{\, \prime}_{n, \ell'} {\cal I}_{n-1, \ell-1} \right ]  
\\[3mm]
\nonumber
&& - 2\beta \; \sqrt{\ell^{\, \prime}} \; 
\frac{(q^{\, \prime} \Lambda j^{\, \prime}) +
\ii (q^{\, \prime} \varphi j^{\, \prime})}{\sqrt{q_{\mprp}^{\, \prime 2}}}
\left [\sqrt{\ell} \; [({\cal K}_2 P) + m_f {\cal K}_4] 
{\cal I}^{\, \prime}_{n, \ell'-1} {\cal I}_{n, \ell} +
\sqrt{ n} (M_{\ell} + m_f) \;{\cal K}_4 {\cal I}^{\, \prime}_{n, \ell'-1}
{\cal I}_{n-1, \ell-1} \right ] 
\\[3mm]
\nonumber
&&-
\frac{(q^{\, \prime} \Lambda j^{\, \prime}) -
\ii (q^{\, \prime} \varphi j^{\, \prime})}{\sqrt{q_{\mprp}^{\, \prime 2}}}
(M_{\ell'} + m_f) \; \Big [(M_{\ell} + m_f) [({\cal K}_2 P) - m_f {\cal K}_4] 
{\cal I}^{\, \prime}_{n-1, \ell'} {\cal I}_{n-1, \ell-1} 
- 2\beta \sqrt{n \ell}\;  {\cal K}_4 \; {\cal I}^{\, \prime}_{n-1, \ell'}
{\cal I}_{n, \ell} \Big ]
\bigg \}\, ;
\eeq

%
\beq
\label{eq:Rsvdownup2}
&&{\cal R}^{-+}_{SV} =  \ii g_s g_v^{\, \prime} j_s
\bigg \{ \sqrt{2 \beta \ell} (M_{\ell'} + m_f)
\left [(P^{\, \prime} \tilde \varphi j^{\, \prime}) {\cal K}_3
-(P^{\, \prime} \tilde \Lambda j^{\, \prime}) {\cal K}_4 
- m_f ({\cal K}_2 j^{\, \prime}) \right ] {\cal I}_{n, \ell'} {\cal I}^{\, \prime}_{n, \ell}
\\
\nonumber
&&+ \sqrt{2 \beta \ell^{\, \prime}} (M_{\ell} + m_f) 
\left [(P^{\, \prime} \tilde \varphi j^{\, \prime})  {\cal K}_3
-(P^{\, \prime} \tilde \Lambda j^{\, \prime}) {\cal K}_4 
+ m_f ({\cal K}_2 j^{\, \prime}) \right ] 
{\cal I}_{n-1, \ell'-1} {\cal I}^{\, \prime}_{n-1, \ell-1} 
\\[3mm]
\nonumber
&& + \sqrt{2 \beta n}\; ({\cal K}_2 j^{\, \prime})
\left [2 \beta \sqrt{\ell \ell^{\, \prime}}
{\cal I}_{n-1, \ell'-1} {\cal I}^{\, \prime}_{n, \ell} +
(M_{\ell} + m_f) (M_{\ell'} + m_f)
{\cal I}_{n, \ell'} {\cal I}^{\, \prime}_{n-1, \ell-1} \right ]  
\\[3mm]
\nonumber
&& + 2\beta \; \sqrt{\ell} \; 
\frac{(q^{\, \prime} \Lambda j^{\, \prime}) +
\ii (q^{\, \prime} \varphi j^{\, \prime})}{\sqrt{q^{\, \prime 2}_{\mprp}}}
\left [\sqrt{\ell^{\, \prime}} \; [({\cal K}_2 P^{\, \prime}) - m_f {\cal K}_4] 
{\cal I}_{n-1, \ell'-1} {\cal I}^{\, \prime}_{n-1, \ell} -
\sqrt{ n} (M_{\ell'} + m_f) \;{\cal K}_4 {\cal I}_{n-1, \ell'}
{\cal I}^{\, \prime}_{n, \ell} \right ] 
\\[3mm]
\nonumber
&&+
\frac{(q^{\, \prime} \Lambda j^{\, \prime}) -
\ii (q^{\, \prime} \varphi j^{\, \prime})}{\sqrt{q^{\, \prime 2}_{\mprp}}}
(M_{\ell} + m_f) \; \Big [(M_{\ell'} + m_f) [({\cal K}_2 P^{\, \prime}) + m_f {\cal K}_4] 
{\cal I}_{n, \ell'} {\cal I}^{\, \prime}_{n, \ell-1} 
- 2\beta \sqrt{n \ell^{\, \prime}}\;  {\cal K}_4 \; {\cal I}_{n, \ell'-1}
{\cal I}^{\, \prime}_{n-1, \ell-1} \Big ]
\bigg \}\, ;
\eeq


%
\beq
\label{eq:Rsvdowndown}
&&{\cal R}^{--}_{VS} = g_s g_v^{\, \prime} j_s 
\bigg \{ (M_{\ell} + m_f) (M_{\ell'} + m_f)  \left [(P \tilde \Lambda j^{\, \prime}) {\cal K}_3 + 
(P \tilde \varphi j^{\, \prime})  {\cal K}_4 
+ m_f ({\cal K}_1 j^{\, \prime}) \right ] {\cal I}^{\, \prime}_{n, \ell'} {\cal I}_{n, \ell}  
\\
\nonumber
&& - 2\beta 
\sqrt{\ell \ell^{\, \prime}} 
\left [(P \tilde \Lambda j^{\, \prime}) {\cal K}_3 + 
(P \tilde \varphi j^{\, \prime})  {\cal K}_4 
- m_f ({\cal K}_1 j^{\, \prime}) \right ] 
{\cal I}^{\, \prime}_{n-1, \ell'-1} {\cal I}_{n-1, \ell-1} 
\\[3mm]
\nonumber
&&+ 2\beta \sqrt{n} ({\cal K}_1 j^{\, \prime}) \left [\sqrt{\ell^{\, \prime}} (M_{\ell} + m_f) 
{\cal I}^{\, \prime}_{n-1, \ell'-1} {\cal I}_{n, \ell} - 
\sqrt{\ell} (M_{\ell'} + m_f) 
{\cal I}^{\, \prime}_{n, \ell'} {\cal I}_{n-1, \ell-1} \right ]   
\\[3mm]
\nonumber 
&& -
\sqrt{\frac{2 \beta \ell^{\, \prime}}{q_{\mprp}^{\, \prime 2}}}\; 
\left [(q^{\, \prime} \Lambda j^{\, \prime}) + 
\ii (q^{\, \prime} \varphi j^{\, \prime}) \right ] 
\left [(M_{\ell} + m_f) [({\cal K}_1 P) + m_f {\cal K}_3] 
{\cal I}^{\, \prime}_{n, \ell'-1} {\cal I}_{n, \ell} - 
2 \beta \sqrt{\ell n} {\cal K}_3 {\cal I}^{\, \prime}_{n, \ell'-1} 
{\cal I}_{n-1, \ell-1} \right ]  
\\[3mm]
\nonumber
&&+\sqrt{\frac{2 \beta}
{q_{\mprp}^{\, \prime 2}}}\; (M_{\ell'} + m_f)  
\left [(q^{\, \prime} \Lambda j^{\, \prime}) - 
\ii (q^{\, \prime} \varphi j^{\, \prime}) \right ] 
\left [ \sqrt{\ell}\;  [({\cal K}_1 P) - m_f {\cal K}_3] 
{\cal I}^{\, \prime}_{n-1, \ell'} {\cal I}_{n-1, \ell-1} - 
 \sqrt{n}\; (M_{\ell} + m_f) {\cal K}_3 {\cal I}^{\, \prime}_{n-1, \ell'} 
{\cal I}_{n, \ell} \right ]
\bigg \}\, . 
\eeq


%
\beq
\label{eq:Rsvdowndown2}
&&{\cal R}^{--}_{SV} = g_s g_v^{\, \prime} j_s 
\bigg \{ (M_{\ell} + m_f) (M_{\ell'} + m_f)  
\left [(P^{\, \prime} \tilde \Lambda j^{\, \prime}) {\cal K}_3 - 
(P^{\, \prime} \tilde \varphi j^{\, \prime})  {\cal K}_4 
+ m_f ({\cal K}_1 j^{\, \prime}) \right ] {\cal I}_{n, \ell'} {\cal I}^{\, \prime}_{n, \ell}  
\\
\nonumber
&& - 2\beta 
\sqrt{\ell \ell^{\, \prime}} 
\left [(P^{\, \prime} \tilde \Lambda j^{\, \prime}) {\cal K}_3 - 
(P^{\, \prime} \tilde \varphi j^{\, \prime})  {\cal K}_4 
- m_f ({\cal K}_1 j^{\, \prime}) \right ] 
{\cal I}_{n-1, \ell'-1} {\cal I}^{\, \prime}_{n-1, \ell-1} 
\\[3mm]
\nonumber
&&+ 2\beta \sqrt{n} ({\cal K}_1 j^{\, \prime}) \left [\sqrt{\ell^{\, \prime}} (M_{\ell} + m_f) 
{\cal I}_{n-1, \ell'-1} {\cal I}^{\, \prime}_{n, \ell} - 
\sqrt{\ell} (M_{\ell'} + m_f) 
{\cal I}_{n, \ell'} {\cal I}^{\, \prime}_{n-1, \ell-1} \right ]   
\\[3mm]
\nonumber 
&& -
\sqrt{\frac{2 \beta}{q^{\, \prime 2}_{\mprp}}}\;  (M_{\ell} + m_f)
\left [(q^{\, \prime} \Lambda j^{\, \prime}) + 
\ii (q^{\, \prime} \varphi j^{\, \prime}) \right ] 
\left [\sqrt{\ell^{\, \prime}}\; [({\cal K}_1 P^{\, \prime}) - m_f {\cal K}_3] 
{\cal I}_{n-1, \ell'-1} {\cal I}^{\, \prime}_{n-1, \ell} - 
\sqrt{n}\; (M_{\ell'} + m_f) {\cal K}_3 {\cal I}_{n, \ell'} 
{\cal I}^{\, \prime}_{n-1, \ell} \right ]  
\\[3mm]
\nonumber
&&+\sqrt{\frac{2 \beta \ell}
{q^{\, \prime 2}_{\mprp}}}\;   
\left [(q^{\, \prime} \Lambda j^{\, \prime}) - 
\ii (q^{\, \prime} \varphi j^{\, \prime}) \right ] 
\left [  (M_{\ell'} + m_f)\;  [({\cal K}_1 P^{\, \prime}) + m_f {\cal K}_3] 
{\cal I}_{n, \ell'} {\cal I}^{\, \prime}_{n, \ell-1} - 
2 \beta \sqrt{\ell^{\, \prime} n} {\cal K}_3 {\cal I}_{n-1, \ell'-1} 
{\cal I}^{\, \prime}_{n, \ell-1}  \right ]
\bigg \}\, . 
\eeq



\newpage
\item
In the case where $j$ is scalar current and $j^{\, \prime}$ is a pseudovector current 
($k = S, \, k^{\, \prime} = A$) we obtain


%
\beq
\label{eq:Rsaupup}
&&{\cal R}^{++}_{AS} = -g_s g_a^{\, \prime} j_s  \bigg \{ 2\beta 
\sqrt{\ell \ell^{\, \prime}} \left [(P \tilde \Lambda j^{\, \prime}) {\cal K}_4 + 
(P \tilde \varphi j^{\, \prime}) {\cal K}_3 
+ m_f ({\cal K}_2 j^{\, \prime}) \right ] 
{\cal I}^{\, \prime}_{n, \ell'} {\cal I}_{n, \ell}  
\\
\nonumber
&&+ (M_{\ell} + m_f) (M_{\ell'} + m_f) 
\left [(P \tilde \Lambda j^{\, \prime}) {\cal K}_4 + 
(P \tilde \varphi j^{\, \prime})  {\cal K}_3 
- m_f ( {\cal K}_2 j^{\, \prime}) \right ] 
{\cal I}^{\, \prime}_{n-1, \ell'-1} {\cal I}_{n-1, \ell-1}
\\[3mm]
\nonumber
&& +
2\beta \sqrt{n} ({\cal K}_2 j^{\, \prime}) \left [\sqrt{\ell} (M_{\ell'} + m_f) 
{\cal I}^{\, \prime}_{n-1, \ell'-1} {\cal I}_{n, \ell} + 
\sqrt{\ell^{\, \prime}} (M_{\ell} + m_f) 
{\cal I}^{\, \prime}_{n, \ell'} {\cal I}_{n-1, \ell-1} \right ]  
\\[3mm]
\nonumber 
&&- \sqrt{\frac{2 \beta}{q_{\mprp}^{\, \prime 2}}}\; (M_{\ell'} + m_f)  
\left [(q^{\, \prime} \Lambda j^{\, \prime}) + 
\ii (q^{\, \prime} \varphi j^{\, \prime}) \right ] 
\left [\sqrt{\ell}\; [({\cal K}_2 P) + m_f {\cal K}_4] 
{\cal I}^{\, \prime}_{n, \ell'-1} {\cal I}_{n, \ell} + 
\sqrt{n} (M_{\ell} + m_f) {\cal K}_4 {\cal I}^{\, \prime}_{n, \ell'-1} 
{\cal I}_{n-1, \ell-1} \right ]  
\\[3mm]
\nonumber
&&- \sqrt{\frac{2 \beta \ell^{\, \prime}}
{q_{\mprp}^{\, \prime 2}}}\;  
\left [(q^{\, \prime} \Lambda j^{\, \prime}) - 
\ii (q^{\, \prime} \varphi j^{\, \prime}) \right ] 
\left [(M_{\ell} + m_f) [({\cal K}_2 P) - m_f {\cal K}_4] 
{\cal I}^{\, \prime}_{n-1, \ell'} {\cal I}_{n-1, \ell-1} + 
2 \beta \sqrt{\ell n}\; {\cal K}_4 {\cal I}^{\, \prime}_{n-1, \ell'} 
{\cal I}_{n, \ell} \right ]
\bigg \}\, ; 
\eeq
%

%
\beq
\label{eq:Rsaupup2}
&&{\cal R}^{++}_{SA} = - g_s g_a^{\, \prime} j_s  \bigg \{ 2\beta 
\sqrt{\ell \ell^{\, \prime}} \left [ (P^{\, \prime} \tilde \varphi j^{\, \prime}) {\cal K}_3 
-(P^{\, \prime} \tilde \Lambda j^{\, \prime}) {\cal K}_4 
+ m_f ({\cal K}_2 j^{\, \prime}) \right ] 
{\cal I}_{n, \ell'} {\cal I}^{\, \prime}_{n, \ell}  
\\
\nonumber
&&+ (M_{\ell} + m_f) (M_{\ell'} + m_f) 
\left [ (P^{\, \prime} \tilde \varphi j^{\, \prime})  {\cal K}_3 
-(P^{\, \prime} \tilde \Lambda j^{\, \prime}) {\cal K}_4 
- m_f ( {\cal K}_2 j^{\, \prime}) \right ] 
{\cal I}_{n-1, \ell'-1} {\cal I}^{\, \prime}_{n-1, \ell-1}
\\[3mm]
\nonumber
&& +
2\beta \sqrt{n} ({\cal K}_2 j^{\, \prime}) \left [\sqrt{\ell} (M_{\ell'} + m_f) 
{\cal I}_{n-1, \ell'-1} {\cal I}^{\, \prime}_{n, \ell} + 
\sqrt{\ell^{\, \prime}} (M_{\ell} + m_f) 
{\cal I}_{n, \ell'} {\cal I}^{\, \prime}_{n-1, \ell-1} \right ]  
\\[3mm]
\nonumber 
&&+ \sqrt{\frac{2 \beta \ell}{q^{\, \prime 2}_{\mprp}}}\;  
\left [(q^{\, \prime} \Lambda j^{\, \prime}) + 
\ii (q^{\, \prime} \varphi j^{\, \prime}) \right ] 
\left [(M_{\ell'} + m_f) \; [({\cal K}_2 P^{\, \prime}) + m_f {\cal K}_4] 
{\cal I}_{n-1, \ell'-1} {\cal I}^{\, \prime}_{n-1, \ell} - 
2 \beta \sqrt{\ell^{\, \prime} n}\; {\cal K}_4 {\cal I}_{n, \ell'} 
{\cal I}^{\, \prime}_{n-1, \ell} \right ]  
\\[3mm]
\nonumber
&&+ \sqrt{\frac{2 \beta}
{q^{\, \prime 2}_{\mprp}}}\; (M_{\ell} + m_f)  
\left [(q^{\, \prime} \Lambda j^{\, \prime}) - 
\ii (q^{\, \prime} \varphi j^{\, \prime}) \right ] 
\left [\sqrt{\ell^{\, \prime}}\; [({\cal K}_2 P^{\, \prime}) - m_f {\cal K}_4] 
{\cal I}_{n, \ell'} {\cal I}^{\, \prime}_{n, \ell-1} - 
\sqrt{n} (M_{\ell'} + m_f) {\cal K}_4 {\cal I}_{n-1, \ell'-1} 
{\cal I}^{\, \prime}_{n, \ell-1} \right ]
\bigg \}\, ; 
\eeq
%


%
\beq
\label{eq:Rsaupdown}
&&{\cal R}^{+-}_{AS} = - \ii g_s g_a^{\, \prime} j_s 
\bigg \{ \sqrt{2 \beta \ell^{\, \prime}} (M_{\ell} + m_f)
\left [(P \tilde \Lambda j^{\, \prime}) {\cal K}_3 + 
(P \tilde \varphi j^{\, \prime}) {\cal K}_4 
+ m_f ({\cal K}_1 j^{\, \prime}) \right ] 
{\cal I}^{\, \prime}_{n, \ell'}  
{\cal I}_{n, \ell}  
\\
\nonumber
&&- \sqrt{2 \beta \ell} (M_{\ell'} + m_f) 
\left [(P \tilde \Lambda j^{\, \prime}) {\cal K}_3 + 
(P \tilde \varphi j^{\, \prime})  {\cal K}_4 
- m_f ({\cal K}_1 j^{\, \prime}) \right ] 
{\cal I}^{\, \prime}_{n-1, \ell'-1}  
{\cal I}_{n-1, \ell-1} 
\\[3mm]
\nonumber
&&+ \sqrt{2 \beta n}\; ({\cal K}_1 j^{\, \prime}) 
\left [(M_{\ell} + m_f) (M_{\ell'} + m_f) 
{\cal I}^{\, \prime}_{n-1, \ell'-1} 
{\cal I}_{n, \ell}  - 
2 \beta \sqrt{\ell \ell^{\, \prime}}  
{\cal I}^{\, \prime}_{n, \ell'}  
{\cal I}_{n-1, \ell-1} \right ]  
\\[3mm]
\nonumber 
&& - \frac{(q^{\, \prime} \Lambda j^{\, \prime}) + 
\ii (q^{\, \prime} \varphi j^{\, \prime})}{\sqrt{q_{\mprp}^{\, \prime 2}}}  
\; (M_{\ell'} + m_f) \left [(M_{\ell} + m_f)  [({\cal K}_1 P) + m_f {\cal K}_3] 
{\cal I}^{\, \prime}_{n, \ell'-1}  
{\cal I}_{n, \ell} - 
2\beta \sqrt{n \ell}  \;{\cal K}_3 {\cal I}^{\, \prime}_{n, \ell'-1}  
{\cal I}_{n-1, \ell-1} \right ] 
\\[3mm]
\nonumber
&&+  2 \beta \sqrt{\ell^{\, \prime}}\;
\frac{(q^{\, \prime} \Lambda j^{\, \prime}) - 
\ii (q^{\, \prime} \varphi j^{\, \prime})}{\sqrt{q_{\mprp}^{\, \prime 2}}} \;
\left [\sqrt{\ell} \; [({\cal K}_1 P) - m_f {\cal K}_3] 
 {\cal I}^{\, \prime}_{n-1, \ell'}  
{\cal I}_{n-1, \ell-1} - 
\sqrt{n}\; (M_{\ell} + m_f) \; {\cal K}_3 \; {\cal I}^{\, \prime}_{n-1, \ell'}  
{\cal I}_{n, \ell} \right ]
\bigg \}\, ; 
\eeq
%


%
\beq
\label{eq:Rsaupdown2}
&&{\cal R}^{+-}_{SA} =  \ii g_s g_a^{\, \prime} j_s 
\bigg \{ \sqrt{2 \beta \ell^{\, \prime}} (M_{\ell} + m_f)
\left [(P^{\, \prime} \tilde \Lambda j^{\, \prime}) {\cal K}_3 - 
(P^{\, \prime} \tilde \varphi j^{\, \prime}) {\cal K}_4 
+ m_f ({\cal K}_1 j^{\, \prime}) \right ] 
{\cal I}_{n, \ell'}  
{\cal I}^{\, \prime}_{n, \ell}  
\\
\nonumber
&&- \sqrt{2 \beta \ell} (M_{\ell'} + m_f) 
\left [(P^{\, \prime} \tilde \Lambda j^{\, \prime}) {\cal K}_3 - 
(P^{\, \prime} \tilde \varphi j^{\, \prime})  {\cal K}_4 
- m_f ({\cal K}_1 j^{\, \prime}) \right ] 
{\cal I}_{n-1, \ell'-1}  
{\cal I}^{\, \prime}_{n-1, \ell-1} 
\\[3mm]
\nonumber
&&- \sqrt{2 \beta n}\; ({\cal K}_1 j^{\, \prime}) 
\left [(M_{\ell} + m_f) (M_{\ell'} + m_f) 
{\cal I}_{n-1, \ell'-1} 
{\cal I}^{\, \prime}_{n, \ell}  - 
2 \beta \sqrt{\ell \ell^{\, \prime}}  
{\cal I}_{n, \ell'}  
{\cal I}^{\, \prime}_{n-1, \ell-1} \right ]  
\\[3mm]
\nonumber 
&& - \frac{(q^{\, \prime} \Lambda j^{\, \prime}) + 
\ii (q^{\, \prime} \varphi j^{\, \prime})}{\sqrt{q^{\, \prime 2}_{\mprp}}} \; (M_{\ell} + m_f) 
\left [(M_{\ell'} + m_f) [({\cal K}_1 P^{\, \prime}) + m_f {\cal K}_3] 
{\cal I}_{n-1, \ell'-1}  
{\cal I}^{\, \prime}_{n-1, \ell^{\, \prime}} - 
2\beta \sqrt{n \ell^{\, \prime}}  \;{\cal K}_3 {\cal I}_{n-1, \ell'}  
{\cal I}^{\, \prime}_{n, \ell} \right ] 
\\[3mm]
\nonumber
&&+  2 \beta \sqrt{\ell} \; 
\frac{(q^{\, \prime} \Lambda j^{\, \prime}) + 
\ii (q^{\, \prime} \varphi j^{\, \prime})}{\sqrt{q^{\, \prime 2}_{\mprp}}}\;  
\left [\sqrt{ \ell^{\, \prime}} \; [({\cal K}_1 P^{\, \prime}) - m_f {\cal K}_3] 
 {\cal I}_{n, \ell'}  
{\cal I}^{\, \prime}_{n, \ell-1} - 
\sqrt{ n}\; (M_{\ell'} + m_f) \; {\cal K}_3 \; {\cal I}_{n, \ell'-1}  
{\cal I}^{\, \prime}_{n-1, \ell-1} \right ]
\bigg \}\, ; 
\eeq
%


\beq
\label{eq:Rsadownup}
&&{\cal R}^{-+}_{AS} = -\ii g_s g_a^{\, \prime} j_s
\bigg \{ \sqrt{2 \beta \ell} (M_{\ell'} + m_f) 
\left [(P \tilde \Lambda j^{\, \prime}) {\cal K}_3 +
(P \tilde \varphi j^{\, \prime}) {\cal K}_4
- m_f ({\cal K}_1 j^{\, \prime}) \right ] 
{\cal I}^{\, \prime}_{n, \ell'} {\cal I}_{n, \ell} 
\\
\nonumber
&&- \sqrt{2 \beta \ell^{\, \prime}} (M_{\ell} + m_f) 
\left [(P \tilde \Lambda j^{\, \prime}) {\cal K}_3 +
(P \tilde \varphi j^{\, \prime}) {\cal K}_4
+ m_f ({\cal K}_1 j^{\, \prime}) \right ] 
{\cal I}^{\, \prime}_{n-1, \ell'-1} {\cal I}_{n-1, \ell-1} 
\\[3mm]
\nonumber
&&+ \sqrt{2 \beta n}\; ({\cal K}_1 j^{\, \prime})
\left [2 \beta \sqrt{\ell \ell^{\, \prime}}
{\cal I}^{\, \prime}_{n-1, \ell'-1} {\cal I}_{n, \ell} -
(M_{\ell} + m_f) (M_{\ell'} + m_f)
{\cal I}^{\, \prime}_{n, \ell'} {\cal I}_{n-1, \ell-1} \right ] 
\\[3mm]
\nonumber
&&-  2\beta \sqrt{\ell^{\, \prime}} \; 
\frac{(q^{\, \prime} \Lambda j^{\, \prime}) +
\ii (q^{\, \prime} \varphi j^{\, \prime})}{\sqrt{q_{\mprp}^{\, \prime 2}}} \;
\left [\sqrt{ \ell}\; [({\cal K}_1 P) - m_f {\cal K}_3] 
{\cal I}^{\, \prime}_{n, \ell'-1} {\cal I}_{n, \ell} -
\sqrt{ n} (M_{\ell} + m_f) \;{\cal K}_3 {\cal I}^{\, \prime}_{n, \ell'-1}
{\cal I}_{n-1, \ell-1} \right ] 
\\[3mm]
\nonumber
&&+ 
\frac{(q^{\, \prime} \Lambda j^{\, \prime}) -
\ii (q^{\, \prime} \varphi j^{\, \prime})}{\sqrt{q_{\mprp}^{\, \prime 2}}} \; (M_{\ell'} + m_f)
\left [(M_{\ell} + m_f)  [({\cal K}_1 P) + m_f {\cal K}_3]  
{\cal I}^{\, \prime}_{n-1, \ell'} {\cal I}_{n-1, \ell-1} +
2\beta \sqrt{n \ell}\;  {\cal K}_3 \; {\cal I}^{\, \prime}_{n-1, \ell'}
{\cal I}_{n, \ell} \right ]
\bigg \}\, ;
\eeq
%


\beq
\label{eq:Rsadownup2}
&&{\cal R}^{-+}_{SA} =  \ii g_s g_a^{\, \prime} j_s
\bigg \{ \sqrt{2 \beta \ell} (M_{\ell'} + m_f) 
\left [(P^{\, \prime} \tilde \Lambda j^{\, \prime}) {\cal K}_3 -
(P^{\, \prime} \tilde \varphi j^{\, \prime}) {\cal K}_4
- m_f ({\cal K}_1 j^{\, \prime}) \right ] 
{\cal I}_{n, \ell'} {\cal I}^{\, \prime}_{n, \ell} 
\\
\nonumber
&&- \sqrt{2 \beta \ell^{\, \prime}} (M_{\ell} + m_f) 
\left [(P^{\, \prime} \tilde \Lambda j^{\, \prime}) {\cal K}_3 -
(P^{\, \prime} \tilde \varphi j^{\, \prime}) {\cal K}_4
+ m_f ({\cal K}_1 j^{\, \prime}) \right ] 
{\cal I}_{n-1, \ell'-1} {\cal I}^{\, \prime}_{n-1, \ell-1} 
\\[3mm]
\nonumber
&&- \sqrt{2 \beta n}\; ({\cal K}_1 j^{\, \prime})
\left [2 \beta \sqrt{\ell \ell^{\, \prime}}
{\cal I}_{n-1, \ell'-1} {\cal I}^{\, \prime}_{n, \ell} -
(M_{\ell} + m_f) (M_{\ell'} + m_f)
{\cal I}_{n, \ell'} {\cal I}^{\, \prime}_{n-1, \ell-1} \right ] 
\\[3mm]
\nonumber
&&-  2\beta \sqrt{\ell} \; 
\frac{(q^{\, \prime} \Lambda j^{\, \prime}) +
\ii (q^{\, \prime} \varphi j^{\, \prime})}{\sqrt{q^{\, \prime 2}_{\mprp}}} \;
\left [\sqrt{ \ell^{\, \prime}}\; [({\cal K}_1 P^{\, \prime}) - m_f {\cal K}_3] 
{\cal I}_{n-1, \ell'-1} {\cal I}^{\, \prime}_{n-1, \ell} -
\sqrt{ n} (M_{\ell} + m_f) \;{\cal K}_3 {\cal I}_{n-1, \ell'}
{\cal I}^{\, \prime}_{n, \ell} \right ] 
\\[3mm]
\nonumber
&&+ 
\frac{(q^{\, \prime} \Lambda j^{\, \prime}) +
\ii (q^{\, \prime} \varphi j^{\, \prime})}{\sqrt{q^{\, \prime 2}_{\mprp}}} \; (M_{\ell} + m_f)
\left [(M_{\ell'} + m_f)  [({\cal K}_1 P^{\, \prime}) + m_f {\cal K}_3]  
{\cal I}_{n, \ell'} {\cal I}^{\, \prime}_{n, \ell-1} +
2\beta \sqrt{n \ell^{\, \prime}}\;  {\cal K}_3 \; {\cal I}_{n, \ell'-1}
{\cal I}^{\, \prime}_{n-1, \ell-1} \right ]
\bigg \}\, ;
\eeq
%

\beq
\label{eq:Rsadowndown}
&&{\cal R}^{--}_{AS} = - g_s g_a^{\, \prime} j_s 
\bigg \{- (M_{\ell} + m_f) (M_{\ell'} + m_f) \left [(P \tilde \Lambda j^{\, \prime}) {\cal K}_4 + 
(P \tilde \varphi j^{\, \prime})  {\cal K}_3 
- m_f ({\cal K}_2 j^{\, \prime}) \right ] 
{\cal I}^{\, \prime}_{n, \ell'} {\cal I}_{n, \ell}  
\\
\nonumber
&&- 2\beta \sqrt{\ell \ell^{\, \prime}} 
\left [(P \tilde \Lambda j^{\, \prime}) {\cal K}_4 + 
(P \tilde \varphi j^{\, \prime})  {\cal K}_3 
+ m_f ({\cal K}_2 j^{\, \prime}) \right ] 
{\cal I}^{\, \prime}_{n-1, \ell'-1} {\cal I}_{n-1, \ell-1} 
\\[3mm]
\nonumber
&&- 2\beta \sqrt{n} ( {\cal K}_2 j^{\, \prime}) \left [\sqrt{\ell^{\, \prime}} (M_{\ell} + m_f) 
{\cal I}^{\, \prime}_{n-1, \ell'-1} {\cal I}_{n, \ell} + 
\sqrt{\ell} (M_{\ell'} + m_f) 
{\cal I}^{\, \prime}_{n, \ell'} {\cal I}_{n-1, \ell-1} \right ] 
\\[3mm]
\nonumber 
&&+
\sqrt{\frac{2 \beta \ell^{\, \prime}}{q_{\mprp}^{\, \prime 2}}} \;    
\left [(q^{\, \prime} \Lambda j^{\, \prime}) + 
\ii (q^{\, \prime} \varphi j^{\, \prime}) \right ] 
\left [(M_{\ell} + m_f) [({\cal K}_2 P) - m_f {\cal K}_4] 
{\cal I}^{\, \prime}_{n, \ell'-1} {\cal I}_{n, \ell} + 
2 \beta \sqrt{\ell n}\; {\cal K}_4 {\cal I}^{\, \prime}_{n-1, \ell'-1}  
{\cal I}_{n, \ell-1}\right ]  
\\[3mm]
\nonumber
&&+ \sqrt{\frac{2 \beta}
{q_{\mprp}^{\, \prime 2}}}\; (M_{\ell'} + m_f)  
\left [(q^{\, \prime} \Lambda j^{\, \prime}) - 
\ii (q^{\, \prime} \varphi j^{\, \prime}) \right ] 
\left [ \sqrt{\ell}\;  [({\cal K}_2 P) + m_f {\cal K}_4] 
{\cal I}^{\, \prime}_{n-1, \ell'} {\cal I}_{n-1, \ell-1} + 
 \sqrt{n}\; (M_{\ell} + m_f) {\cal K}_4 {\cal I}^{\, \prime}_{n, \ell'}  
{\cal I}_{n-1, \ell} \right ]
\bigg \}\, ; 
\eeq

\beq
\label{eq:Rsadowndown2}
&&{\cal R}^{--}_{SA} = - g_s g_a^{\, \prime} j_s 
\bigg \{- (M_{\ell} + m_f) (M_{\ell'} + m_f) 
\left [(P^{\, \prime} \tilde \varphi j^{\, \prime})  {\cal K}_3 
- (P^{\, \prime} \tilde \Lambda j^{\, \prime}) {\cal K}_4  
- m_f ({\cal K}_2 j^{\, \prime}) \right ] 
{\cal I}_{n, \ell'} {\cal I}^{\, \prime}_{n, \ell}  
\\
\nonumber
&&- 2\beta \sqrt{\ell \ell^{\, \prime}} 
\left [ (P^{\, \prime} \tilde \varphi j^{\, \prime})  {\cal K}_3  
- (P^{\, \prime} \tilde \Lambda j^{\, \prime}) {\cal K}_4 
+ m_f ({\cal K}_2 j^{\, \prime}) \right ] 
{\cal I}_{n-1, \ell'-1} {\cal I}^{\, \prime}_{n-1, \ell-1} 
\\[3mm]
\nonumber
&&- 2\beta \sqrt{n} ( {\cal K}_2 j^{\, \prime}) \left [\sqrt{\ell^{\, \prime}} (M_{\ell} + m_f) 
{\cal I}_{n-1, \ell'-1} {\cal I}^{\, \prime}_{n, \ell} + 
\sqrt{\ell} (M_{\ell'} + m_f) 
{\cal I}_{n, \ell'} {\cal I}^{\, \prime}_{n-1, \ell-1} \right ] 
\\[3mm]
\nonumber 
&&-
\sqrt{\frac{2 \beta }{q^{\, \prime 2}_{\mprp}}} \; (M_{\ell} + m_f)   
\left [(q^{\, \prime} \Lambda j^{\, \prime}) + 
\ii (q^{\, \prime} \varphi j^{\, \prime}) \right ] 
\left [ \sqrt{\ell^{\, \prime}} \; [({\cal K}_2 P^{\, \prime}) + m_f {\cal K}_4] 
{\cal I}_{n-1, \ell'-1} {\cal I}^{\, \prime}_{n-1, \ell} - 
 \sqrt{n}\; (M_{\ell'} + m_f) {\cal K}_4 {\cal I}_{n, \ell'}  
{\cal I}^{\, \prime}_{n-1, \ell} \right ]  
\\[3mm]
\nonumber
&&- \sqrt{\frac{2 \beta \ell}
{q^{\, \prime 2}_{\mprp}}}\;   
\left [(q^{\, \prime} \Lambda j^{\, \prime}) - 
\ii (q^{\, \prime} \varphi j^{\, \prime}) \right ] 
\left [(M_{\ell'} + m_f) \;  [({\cal K}_2 P^{\, \prime}) - m_f {\cal K}_4] 
{\cal I}_{n, \ell'} {\cal I}^{\, \prime}_{n, \ell-1} - 
2 \beta \sqrt{\ell^{\, \prime} n}\; {\cal K}_4 {\cal I}_{n-1, \ell'-1}  
{\cal I}^{\, \prime}_{n, \ell-1} \right ]
\bigg \}\, ; 
\eeq



\item 
In the case where $j$ and $j^{\, \prime}$ are pseudoscalar currents 
($k =  k^{\, \prime} = P$) we obtain
%
%
\beq
\nonumber
&&{\cal R}^{++}_{PP} = - g_p g_p^{\, \prime} j_p j_p^{\, \prime} \bigg \{ 2\beta 
\sqrt{\ell \ell^{\, \prime}} \left [({\cal K}_1 P) + m_f {\cal K}_3 \right ] 
{\cal I}^{\, \prime}_{n, \ell'} {\cal I}_{n, \ell} +
(M_{\ell} + m_f) (M_{\ell'} + m_f) 
\left [({\cal K}_1 P) - m_f {\cal K}_3 \right ] 
\times 
\\[3mm]
\label{eq:Rppupup}
&&{\cal I}^{\, \prime}_{n-1, \ell'-1} {\cal I}_{n-1, \ell-1} -
2\beta \sqrt{n} {\cal K}_3 \big [\sqrt{\ell} (M_{\ell'} + m_f) 
{\cal I}^{\, \prime}_{n-1, \ell'-1} {\cal I}_{n, \ell} + 
\sqrt{\ell^{\, \prime}} (M_{\ell} + m_f) 
{\cal I}^{\, \prime}_{n, \ell'} {\cal I}_{n-1, \ell-1} \big ]  
\bigg \}\, ; 
\eeq
\beq
\nonumber
&&{\cal R}^{+-}_{PP} = - \ii g_p g_p^{\, \prime} j_p j_p^{\, \prime} 
\bigg \{ \sqrt{2\beta \ell^{\, \prime}}\; (M_{\ell} + m_f) 
\left [({\cal K}_2 P) + m_f {\cal K}_4 \right ] 
{\cal I}^{\, \prime}_{n, \ell'} {\cal I}_{n, \ell} -
\sqrt{2\beta \ell}\; (M_{\ell'} + m_f) 
\left [({\cal K}_2 P) - m_f {\cal K}_4 \right ] \times 
\\[3mm]
\label{eq:Rppupdown}
&&{\cal I}^{\, \prime}_{n-1, \ell'-1} {\cal I}_{n-1, \ell-1} -
\sqrt{2\beta n}\; {\cal K}_4 \big [ (M_{\ell} + m_f) (M_{\ell'} + m_f)
{\cal I}^{\, \prime}_{n-1, \ell'-1} {\cal I}_{n, \ell} - 
2\beta
\sqrt{\ell \ell^{\, \prime}}
{\cal I}^{\, \prime}_{n, \ell'} {\cal I}_{n-1, \ell-1} \big ]  
\bigg \}\, ; 
\eeq
\beq
\nonumber
&&{\cal R}^{-+}_{PP} =   \ii g_p g_p^{\, \prime} j_p j_p^{\, \prime} 
\bigg \{ \sqrt{2\beta \ell}\; (M_{\ell'} + m_f) 
\left [({\cal K}_2 P) - m_f {\cal K}_4 \right ] 
{\cal I}^{\, \prime}_{n, \ell'} {\cal I}_{n, \ell} -
\sqrt{2\beta \ell^{\, \prime}}\; (M_{\ell} + m_f) 
\left [({\cal K}_2 P) + m_f {\cal K}_4 \right ] \times 
\\[3mm]
\label{eq:Rppdownup}
&&{\cal I}^{\, \prime}_{n-1, \ell'-1} {\cal I}_{n-1, \ell-1} -
\sqrt{2\beta n}\; {\cal K}_4 \big [2\beta
\sqrt{\ell \ell^{\, \prime}} 
{\cal I}^{\, \prime}_{n-1, \ell'-1} {\cal I}_{n, \ell} - 
(M_{\ell} + m_f) (M_{\ell'} + m_f)
{\cal I}^{\, \prime}_{n, \ell'} {\cal I}_{n-1, \ell-1} \big ]  
\bigg \}\, ; 
\eeq
\beq
\nonumber
&&{\cal R}^{--}_{PP} = - g_p g_p^{\, \prime} j_p j_p^{\, \prime} 
\bigg \{ (M_{\ell} + m_f) (M_{\ell'} + m_f)
\left [({\cal K}_1 P) - m_f {\cal K}_3 \right ]  
{\cal I}^{\, \prime}_{n, \ell'} {\cal I}_{n, \ell} +
2\beta 
\sqrt{\ell \ell^{\, \prime}} \left [({\cal K}_1 P) + m_f {\cal K}_3 \right ] \times 
\\[3mm]
\label{eq:Rppdowndown}
&&{\cal I}^{\, \prime}_{n-1, \ell'-1} {\cal I}_{n-1, \ell-1} -
2\beta \sqrt{n} {\cal K}_3 \big [\sqrt{\ell^{\, \prime}} (M_{\ell} + m_f) 
{\cal I}^{\, \prime}_{n-1, \ell'-1} {\cal I}_{n, \ell} + 
\sqrt{\ell} (M_{\ell'} + m_f) 
{\cal I}^{\, \prime}_{n, \ell'} {\cal I}_{n-1, \ell-1} \big ]  
\bigg \}\, . 
\eeq

For second diagram we have the following replacement $P_{\alpha} \to P_{\alpha}^{\, \prime}$, 
${\cal I}_{m,n} \leftrightarrow {\cal I}_{m,n}^{\, \prime}$.


\item
In the case where $j$ is pseudoscalar current and $j^{\, \prime}$ is a vector current 
($k = P, \, k^{\, \prime} = V$) we obtain


%
\beq
\label{eq:Rpvupup}
&&{\cal R}^{++}_{VP} = g_p g_v^{\, \prime} j_p  \bigg \{ 2\beta 
\sqrt{\ell \ell^{\, \prime}} \left [(P \tilde \Lambda j^{\, \prime}) {\cal K}_4 + 
(P \tilde \varphi j^{\, \prime}) {\cal K}_3 
- m_f ({\cal K}_2 j^{\, \prime}) \right ] 
{\cal I}^{\, \prime}_{n, \ell'} {\cal I}_{n, \ell}  
\\
\nonumber
&&+ (M_{\ell} + m_f) (M_{\ell'} + m_f) 
\left [(P \tilde \Lambda j^{\, \prime}) {\cal K}_4 + 
(P \tilde \varphi j^{\, \prime})  {\cal K}_3 
+ m_f ( {\cal K}_2 j^{\, \prime}) \right ] 
{\cal I}^{\, \prime}_{n-1, \ell'-1} {\cal I}_{n-1, \ell-1}
\\[3mm]
\nonumber
&& +
2\beta \sqrt{n} ({\cal K}_2 j^{\, \prime}) \left [\sqrt{\ell} (M_{\ell'} + m_f) 
{\cal I}^{\, \prime}_{n-1, \ell'-1} {\cal I}_{n, \ell} + 
\sqrt{\ell^{\, \prime}} (M_{\ell} + m_f) 
{\cal I}^{\, \prime}_{n, \ell'} {\cal I}_{n-1, \ell-1} \right ]  
\\[3mm]
\nonumber 
&&- \sqrt{\frac{2 \beta}{q_{\mprp}^{\, \prime 2}}}\; (M_{\ell'} + m_f)  
\left [(q^{\, \prime} \Lambda j^{\, \prime}) + 
\ii (q^{\, \prime} \varphi j^{\, \prime}) \right ] 
\left [\sqrt{\ell}\; [({\cal K}_2 P) - m_f {\cal K}_4] 
{\cal I}^{\, \prime}_{n, \ell'-1} {\cal I}_{n, \ell} + 
\sqrt{n} (M_{\ell} + m_f) {\cal K}_4 {\cal I}^{\, \prime}_{n, \ell'-1} 
{\cal I}_{n-1, \ell-1} \right ]  
\\[3mm]
\nonumber
&&- \sqrt{\frac{2 \beta \ell^{\, \prime}}
{q_{\mprp}^{\, \prime 2}}}\;  
\left [(q^{\, \prime} \Lambda j^{\, \prime}) - 
\ii (q^{\, \prime} \varphi j^{\, \prime}) \right ] 
\left [(M_{\ell} + m_f) [({\cal K}_2 P) + m_f {\cal K}_4] 
{\cal I}^{\, \prime}_{n-1, \ell'} {\cal I}_{n-1, \ell-1} + 
2 \beta \sqrt{\ell n}\; {\cal K}_4 {\cal I}^{\, \prime}_{n-1, \ell'} 
{\cal I}_{n, \ell} \right ]
\bigg \}\, ; 
\eeq
%

%
\beq
\label{eq:Rpvupup2}
&&{\cal R}^{++}_{PV} = -g_p g_v^{\, \prime} j_p  \bigg \{ 2\beta 
\sqrt{\ell \ell^{\, \prime}} \left [ (P^{\, \prime} \tilde \varphi j^{\, \prime}) {\cal K}_3 
-(P^{\, \prime} \tilde \Lambda j^{\, \prime}) {\cal K}_4 
- m_f ({\cal K}_2 j^{\, \prime}) \right ] 
{\cal I}_{n, \ell'} {\cal I}^{\, \prime}_{n, \ell}  
\\
\nonumber
&&+ (M_{\ell} + m_f) (M_{\ell'} + m_f) 
\left [ (P^{\, \prime} \tilde \varphi j^{\, \prime})  {\cal K}_3 
-(P^{\, \prime} \tilde \Lambda j^{\, \prime}) {\cal K}_4 
+ m_f ( {\cal K}_2 j^{\, \prime}) \right ] 
{\cal I}_{n-1, \ell'-1} {\cal I}^{\, \prime}_{n-1, \ell-1}
\\[3mm]
\nonumber
&& +
2\beta \sqrt{n} ({\cal K}_2 j^{\, \prime}) \left [\sqrt{\ell} (M_{\ell'} + m_f) 
{\cal I}_{n-1, \ell'-1} {\cal I}^{\, \prime}_{n, \ell} + 
\sqrt{\ell^{\, \prime}} (M_{\ell} + m_f) 
{\cal I}_{n, \ell'} {\cal I}^{\, \prime}_{n-1, \ell-1} \right ]  
\\[3mm]
\nonumber 
&&+ \sqrt{\frac{2 \beta \ell}{q^{\, \prime 2}_{\mprp}}}\;  
\left [(q^{\, \prime} \Lambda j^{\, \prime}) + 
\ii (q^{\, \prime} \varphi j^{\, \prime}) \right ] 
\left [(M_{\ell'} + m_f) \; [({\cal K}_2 P^{\, \prime}) - m_f {\cal K}_4] 
{\cal I}_{n-1, \ell'-1} {\cal I}^{\, \prime}_{n-1, \ell} - 
2 \beta \sqrt{\ell^{\, \prime} n}\; {\cal K}_4 {\cal I}_{n, \ell'} 
{\cal I}^{\, \prime}_{n-1, \ell} \right ]  
\\[3mm]
\nonumber
&&+ \sqrt{\frac{2 \beta}
{q^{\, \prime 2}_{\mprp}}}\; (M_{\ell} + m_f)  
\left [(q^{\, \prime} \Lambda j^{\, \prime}) - 
\ii (q^{\, \prime} \varphi j^{\, \prime}) \right ] 
\left [\sqrt{\ell^{\, \prime}}\; [({\cal K}_2 P^{\, \prime}) + m_f {\cal K}_4] 
{\cal I}_{n, \ell'} {\cal I}^{\, \prime}_{n, \ell-1} - 
\sqrt{n} (M_{\ell'} + m_f) {\cal K}_4 {\cal I}_{n-1, \ell'-1} 
{\cal I}^{\, \prime}_{n, \ell-1} \right ]
\bigg \}\, ; 
\eeq
%


%
\beq
\label{eq:Rpvupdown}
&&{\cal R}^{+-}_{VP} = \ii g_p g_v^{\, \prime} j_p 
\bigg \{ \sqrt{2 \beta \ell^{\, \prime}} (M_{\ell} + m_f)
\left [(P \tilde \Lambda j^{\, \prime}) {\cal K}_3 + 
(P \tilde \varphi j^{\, \prime}) {\cal K}_4 
- m_f ({\cal K}_1 j^{\, \prime}) \right ] 
{\cal I}^{\, \prime}_{n, \ell'}  
{\cal I}_{n, \ell}  
\\
\nonumber
&&- \sqrt{2 \beta \ell} (M_{\ell'} + m_f) 
\left [(P \tilde \Lambda j^{\, \prime}) {\cal K}_3 + 
(P \tilde \varphi j^{\, \prime})  {\cal K}_4 
+ m_f ({\cal K}_1 j^{\, \prime}) \right ] 
{\cal I}^{\, \prime}_{n-1, \ell'-1}  
{\cal I}_{n-1, \ell-1} 
\\[3mm]
\nonumber
&&+ \sqrt{2 \beta n}\; ({\cal K}_1 j^{\, \prime}) 
\left [(M_{\ell} + m_f) (M_{\ell'} + m_f) 
{\cal I}^{\, \prime}_{n-1, \ell'-1} 
{\cal I}_{n, \ell}  - 
2 \beta \sqrt{\ell \ell^{\, \prime}}  
{\cal I}^{\, \prime}_{n, \ell'}  
{\cal I}_{n-1, \ell-1} \right ]  
\\[3mm]
\nonumber 
&& - \frac{(q^{\, \prime} \Lambda j^{\, \prime}) + 
\ii (q^{\, \prime} \varphi j^{\, \prime})}{\sqrt{q_{\mprp}^{\, \prime 2}}}  
\; (M_{\ell'} + m_f) \left [(M_{\ell} + m_f)  [({\cal K}_1 P) - m_f {\cal K}_3] 
{\cal I}^{\, \prime}_{n, \ell'-1}  
{\cal I}_{n, \ell} - 
2\beta \sqrt{n \ell}  \;{\cal K}_3 {\cal I}^{\, \prime}_{n, \ell'-1}  
{\cal I}_{n-1, \ell-1} \right ] 
\\[3mm]
\nonumber
&&+  2 \beta \sqrt{\ell^{\, \prime}}\;
\frac{(q^{\, \prime} \Lambda j^{\, \prime}) - 
\ii (q^{\, \prime} \varphi j^{\, \prime})}{\sqrt{q_{\mprp}^{\, \prime 2}}} \;
\left [\sqrt{\ell} \; [({\cal K}_1 P) + m_f {\cal K}_3] 
 {\cal I}^{\, \prime}_{n-1, \ell'}  
{\cal I}_{n-1, \ell-1} - 
\sqrt{n}\; (M_{\ell} + m_f) \; {\cal K}_3 \; {\cal I}^{\, \prime}_{n-1, \ell'}  
{\cal I}_{n, \ell} \right ]
\bigg \}\, ; 
\eeq
%


%
\beq
\label{eq:Rpvupdown2}
&&{\cal R}^{+-}_{PV} =  \ii g_p g_v^{\, \prime} j_p 
\bigg \{ \sqrt{2 \beta \ell^{\, \prime}} (M_{\ell} + m_f)
\left [(P^{\, \prime} \tilde \Lambda j^{\, \prime}) {\cal K}_3 - 
(P^{\, \prime} \tilde \varphi j^{\, \prime}) {\cal K}_4 
- m_f ({\cal K}_1 j^{\, \prime}) \right ] 
{\cal I}_{n, \ell'}  
{\cal I}^{\, \prime}_{n, \ell}  
\\
\nonumber
&&- \sqrt{2 \beta \ell} (M_{\ell'} + m_f) 
\left [(P^{\, \prime} \tilde \Lambda j^{\, \prime}) {\cal K}_3 - 
(P^{\, \prime} \tilde \varphi j^{\, \prime})  {\cal K}_4 
+ m_f ({\cal K}_1 j^{\, \prime}) \right ] 
{\cal I}_{n-1, \ell'-1}  
{\cal I}^{\, \prime}_{n-1, \ell-1} 
\\[3mm]
\nonumber
&&- \sqrt{2 \beta n}\; ({\cal K}_1 j^{\, \prime}) 
\left [(M_{\ell} + m_f) (M_{\ell'} + m_f) 
{\cal I}_{n-1, \ell'-1} 
{\cal I}^{\, \prime}_{n, \ell}  - 
2 \beta \sqrt{\ell \ell^{\, \prime}}  
{\cal I}_{n, \ell'}  
{\cal I}^{\, \prime}_{n-1, \ell-1} \right ]  
\\[3mm]
\nonumber 
&& - \frac{(q^{\, \prime} \Lambda j^{\, \prime}) + 
\ii (q^{\, \prime} \varphi j^{\, \prime})}{\sqrt{q^{\, \prime 2}_{\mprp}}} \; (M_{\ell} + m_f) 
\left [(M_{\ell'} + m_f) [({\cal K}_1 P^{\, \prime}) - m_f {\cal K}_3] 
{\cal I}_{n-1, \ell'-1}  
{\cal I}^{\, \prime}_{n-1, \ell^{\, \prime}} - 
2\beta \sqrt{n \ell^{\, \prime}}  \;{\cal K}_3 {\cal I}_{n-1, \ell'}  
{\cal I}^{\, \prime}_{n, \ell} \right ] 
\\[3mm]
\nonumber
&&+  2 \beta \sqrt{\ell} \; 
\frac{(q^{\, \prime} \Lambda j^{\, \prime}) + 
\ii (q^{\, \prime} \varphi j^{\, \prime})}{\sqrt{q^{\, \prime 2}_{\mprp}}}\;  
\left [\sqrt{ \ell^{\, \prime}} \; [({\cal K}_1 P^{\, \prime}) + m_f {\cal K}_3] 
 {\cal I}_{n, \ell'}  
{\cal I}^{\, \prime}_{n, \ell-1} - 
\sqrt{ n}\; (M_{\ell'} + m_f) \; {\cal K}_3 \; {\cal I}_{n, \ell'-1}  
{\cal I}^{\, \prime}_{n-1, \ell-1} \right ]
\bigg \}\, ; 
\eeq
%


\beq
\label{eq:Rpvdownup}
&&{\cal R}^{-+}_{VP} = \ii g_p g_v^{\, \prime} j_p
\bigg \{ \sqrt{2 \beta \ell} (M_{\ell'} + m_f) 
\left [(P \tilde \Lambda j^{\, \prime}) {\cal K}_3 +
(P \tilde \varphi j^{\, \prime}) {\cal K}_4
+ m_f ({\cal K}_1 j^{\, \prime}) \right ] 
{\cal I}^{\, \prime}_{n, \ell'} {\cal I}_{n, \ell} 
\\
\nonumber
&&- \sqrt{2 \beta \ell^{\, \prime}} (M_{\ell} + m_f) 
\left [(P \tilde \Lambda j^{\, \prime}) {\cal K}_3 +
(P \tilde \varphi j^{\, \prime}) {\cal K}_4
- m_f ({\cal K}_1 j^{\, \prime}) \right ] 
{\cal I}^{\, \prime}_{n-1, \ell'-1} {\cal I}_{n-1, \ell-1} 
\\[3mm]
\nonumber
&&+ \sqrt{2 \beta n}\; ({\cal K}_1 j^{\, \prime})
\left [2 \beta \sqrt{\ell \ell^{\, \prime}}
{\cal I}^{\, \prime}_{n-1, \ell'-1} {\cal I}_{n, \ell} -
(M_{\ell} + m_f) (M_{\ell'} + m_f)
{\cal I}^{\, \prime}_{n, \ell'} {\cal I}_{n-1, \ell-1} \right ] 
\\[3mm]
\nonumber
&&-  2\beta \sqrt{\ell^{\, \prime}} \; 
\frac{(q^{\, \prime} \Lambda j^{\, \prime}) +
\ii (q^{\, \prime} \varphi j^{\, \prime})}{\sqrt{q_{\mprp}^{\, \prime 2}}} \;
\left [\sqrt{ \ell}\; [({\cal K}_1 P) + m_f {\cal K}_3] 
{\cal I}^{\, \prime}_{n, \ell'-1} {\cal I}_{n, \ell} -
\sqrt{ n} (M_{\ell} + m_f) \;{\cal K}_3 {\cal I}^{\, \prime}_{n, \ell'-1}
{\cal I}_{n-1, \ell-1} \right ] 
\\[3mm]
\nonumber
&&+ 
\frac{(q^{\, \prime} \Lambda j^{\, \prime}) -
\ii (q^{\, \prime} \varphi j^{\, \prime})}{\sqrt{q_{\mprp}^{\, \prime 2}}} \; (M_{\ell'} + m_f)
\left [(M_{\ell} + m_f)  [({\cal K}_1 P) - m_f {\cal K}_3]  
{\cal I}^{\, \prime}_{n-1, \ell'} {\cal I}_{n-1, \ell-1} +
2\beta \sqrt{n \ell}\;  {\cal K}_3 \; {\cal I}^{\, \prime}_{n-1, \ell'}
{\cal I}_{n, \ell} \right ]
\bigg \}\, ;
\eeq
%


\beq
\label{eq:Rpvdownup2}
&&{\cal R}^{-+}_{PV} =  \ii g_p g_v^{\, \prime} j_p
\bigg \{ \sqrt{2 \beta \ell} (M_{\ell'} + m_f) 
\left [(P^{\, \prime} \tilde \Lambda j^{\, \prime}) {\cal K}_3 -
(P^{\, \prime} \tilde \varphi j^{\, \prime}) {\cal K}_4
+ m_f ({\cal K}_1 j^{\, \prime}) \right ] 
{\cal I}_{n, \ell'} {\cal I}^{\, \prime}_{n, \ell} 
\\
\nonumber
&&- \sqrt{2 \beta \ell^{\, \prime}} (M_{\ell} + m_f) 
\left [(P^{\, \prime} \tilde \Lambda j^{\, \prime}) {\cal K}_3 -
(P^{\, \prime} \tilde \varphi j^{\, \prime}) {\cal K}_4
- m_f ({\cal K}_1 j^{\, \prime}) \right ] 
{\cal I}_{n-1, \ell'-1} {\cal I}^{\, \prime}_{n-1, \ell-1} 
\\[3mm]
\nonumber
&&- \sqrt{2 \beta n}\; ({\cal K}_1 j^{\, \prime})
\left [2 \beta \sqrt{\ell \ell^{\, \prime}}
{\cal I}_{n-1, \ell'-1} {\cal I}^{\, \prime}_{n, \ell} -
(M_{\ell} + m_f) (M_{\ell'} + m_f)
{\cal I}_{n, \ell'} {\cal I}^{\, \prime}_{n-1, \ell-1} \right ] 
\\[3mm]
\nonumber
&&-  2\beta \sqrt{\ell} \; 
\frac{(q^{\, \prime} \Lambda j^{\, \prime}) +
\ii (q^{\, \prime} \varphi j^{\, \prime})}{\sqrt{q^{\, \prime 2}_{\mprp}}} \;
\left [\sqrt{ \ell^{\, \prime}}\; [({\cal K}_1 P^{\, \prime}) + m_f {\cal K}_3] 
{\cal I}_{n-1, \ell'-1} {\cal I}^{\, \prime}_{n-1, \ell} -
\sqrt{ n} (M_{\ell} + m_f) \;{\cal K}_3 {\cal I}_{n-1, \ell'}
{\cal I}^{\, \prime}_{n, \ell} \right ] 
\\[3mm]
\nonumber
&&+ 
\frac{(q^{\, \prime} \Lambda j^{\, \prime}) +
\ii (q^{\, \prime} \varphi j^{\, \prime})}{\sqrt{q^{\, \prime 2}_{\mprp}}} \; (M_{\ell} + m_f)
\left [(M_{\ell'} + m_f)  [({\cal K}_1 P^{\, \prime}) - m_f {\cal K}_3]  
{\cal I}_{n, \ell'} {\cal I}^{\, \prime}_{n, \ell-1} +
2\beta \sqrt{n \ell^{\, \prime}}\;  {\cal K}_3 \; {\cal I}_{n, \ell'-1}
{\cal I}^{\, \prime}_{n-1, \ell-1} \right ]
\bigg \}\, ;
\eeq
%

\beq
\label{eq:Rpvdowndown}
&&{\cal R}^{--}_{VP} = g_p g_v^{\, \prime} j_p
\bigg \{ -(M_{\ell} + m_f) (M_{\ell'} + m_f) \left [(P \tilde \Lambda j^{\, \prime}) {\cal K}_4 + 
(P \tilde \varphi j^{\, \prime})  {\cal K}_3 
+ m_f ({\cal K}_2 j^{\, \prime}) \right ] 
{\cal I}^{\, \prime}_{n, \ell'} {\cal I}_{n, \ell}  
\\
\nonumber
&&- 2\beta \sqrt{\ell \ell^{\, \prime}} 
\left [(P \tilde \Lambda j^{\, \prime}) {\cal K}_4 + 
(P \tilde \varphi j^{\, \prime})  {\cal K}_3 
- m_f ({\cal K}_2 j^{\, \prime}) \right ] 
{\cal I}^{\, \prime}_{n-1, \ell'-1} {\cal I}_{n-1, \ell-1} 
\\[3mm]
\nonumber
&&- 2\beta \sqrt{n} ( {\cal K}_2 j^{\, \prime}) \left [\sqrt{\ell^{\, \prime}} (M_{\ell} + m_f) 
{\cal I}^{\, \prime}_{n-1, \ell'-1} {\cal I}_{n, \ell} + 
\sqrt{\ell} (M_{\ell'} + m_f) 
{\cal I}^{\, \prime}_{n, \ell'} {\cal I}_{n-1, \ell-1} \right ] 
\\[3mm]
\nonumber 
&&+
\sqrt{\frac{2 \beta \ell^{\, \prime}}{q_{\mprp}^{\, \prime 2}}} \;    
\left [(q^{\, \prime} \Lambda j^{\, \prime}) + 
\ii (q^{\, \prime} \varphi j^{\, \prime}) \right ] 
\left [(M_{\ell} + m_f) [({\cal K}_2 P) + m_f {\cal K}_4] 
{\cal I}^{\, \prime}_{n, \ell'-1} {\cal I}_{n, \ell} + 
2 \beta \sqrt{\ell n}\; {\cal K}_4 {\cal I}^{\, \prime}_{n-1, \ell'-1}  
{\cal I}_{n, \ell-1}\right ]  
\\[3mm]
\nonumber
&&+ \sqrt{\frac{2 \beta}
{q_{\mprp}^{\, \prime 2}}}\; (M_{\ell'} + m_f)  
\left [(q^{\, \prime} \Lambda j^{\, \prime}) - 
\ii (q^{\, \prime} \varphi j^{\, \prime}) \right ] 
\left [ \sqrt{\ell}\;  [({\cal K}_2 P) - m_f {\cal K}_4] 
{\cal I}^{\, \prime}_{n-1, \ell'} {\cal I}_{n-1, \ell-1} + 
 \sqrt{n}\; (M_{\ell} + m_f) {\cal K}_4 {\cal I}^{\, \prime}_{n, \ell'}  
{\cal I}_{n-1, \ell} \right ]
\bigg \}\, ; 
\eeq

\beq
\label{eq:Rpvdowndown2}
&&{\cal R}^{--}_{PV} = -g_p g_v^{\, \prime} j_p 
\bigg \{- (M_{\ell} + m_f) (M_{\ell'} + m_f) 
\left [(P^{\, \prime} \tilde \varphi j^{\, \prime})  {\cal K}_3 
- (P^{\, \prime} \tilde \Lambda j^{\, \prime}) {\cal K}_4  
+ m_f ({\cal K}_2 j^{\, \prime}) \right ] 
{\cal I}_{n, \ell'} {\cal I}^{\, \prime}_{n, \ell}  
\\
\nonumber
&&- 2\beta \sqrt{\ell \ell^{\, \prime}} 
\left [ (P^{\, \prime} \tilde \varphi j^{\, \prime})  {\cal K}_3  
- (P^{\, \prime} \tilde \Lambda j^{\, \prime}) {\cal K}_4 
- m_f ({\cal K}_2 j^{\, \prime}) \right ] 
{\cal I}_{n-1, \ell'-1} {\cal I}^{\, \prime}_{n-1, \ell-1} 
\\[3mm]
\nonumber
&&- 2\beta \sqrt{n} ( {\cal K}_2 j^{\, \prime}) \left [\sqrt{\ell^{\, \prime}} (M_{\ell} + m_f) 
{\cal I}_{n-1, \ell'-1} {\cal I}^{\, \prime}_{n, \ell} + 
\sqrt{\ell} (M_{\ell'} + m_f) 
{\cal I}_{n, \ell'} {\cal I}^{\, \prime}_{n-1, \ell-1} \right ] 
\\[3mm]
\nonumber 
&&-
\sqrt{\frac{2 \beta }{q^{\, \prime 2}_{\mprp}}} \; (M_{\ell} + m_f)   
\left [(q^{\, \prime} \Lambda j^{\, \prime}) + 
\ii (q^{\, \prime} \varphi j^{\, \prime}) \right ] 
\left [ \sqrt{\ell^{\, \prime}} \; [({\cal K}_2 P^{\, \prime}) - m_f {\cal K}_4] 
{\cal I}_{n-1, \ell'-1} {\cal I}^{\, \prime}_{n-1, \ell} - 
 \sqrt{n}\; (M_{\ell'} + m_f) {\cal K}_4 {\cal I}_{n, \ell'}  
{\cal I}^{\, \prime}_{n-1, \ell} \right ]  
\\[3mm]
\nonumber
&&- \sqrt{\frac{2 \beta \ell}
{q^{\, \prime 2}_{\mprp}}}\;   
\left [(q^{\, \prime} \Lambda j^{\, \prime}) - 
\ii (q^{\, \prime} \varphi j^{\, \prime}) \right ] 
\left [(M_{\ell'} + m_f) \;  [({\cal K}_2 P^{\, \prime}) + m_f {\cal K}_4] 
{\cal I}_{n, \ell'} {\cal I}^{\, \prime}_{n, \ell-1} - 
2 \beta \sqrt{\ell^{\, \prime} n}\; {\cal K}_4 {\cal I}_{n-1, \ell'-1}  
{\cal I}^{\, \prime}_{n, \ell-1} \right ]
\bigg \}\, ; 
\eeq




\item
In the case where $j$ is pseudoscalar current and $j^{\, \prime}$ is a pseudovector current 
($k = P, \, k^{\, \prime} = A$) we obtain



%
\beq
\label{eq:Rpaupup}
&&{\cal R}^{++}_{AP} = -g_p g_a^{\, \prime} j_p \bigg \{ - 2\beta 
\sqrt{\ell \ell^{\, \prime}} \left [(P \tilde \Lambda j^{\, \prime}){\cal K}_3 + 
(P \tilde \varphi j^{\, \prime}) {\cal K}_4 
+ m_f ({\cal K}_1 j^{\, \prime}) \right ] 
{\cal I}^{\, \prime}_{n, \ell'} {\cal I}_{n, \ell}   
\\
\nonumber
&&+(M_{\ell} + m_f) (M_{\ell'} + m_f) 
\left [(P \tilde \Lambda j^{\, \prime}){\cal K}_3 + 
(P \tilde \varphi j^{\, \prime}) {\cal K}_4 
- m_f ({\cal K}_1 j^{\, \prime}) \right ] {\cal I}^{\, \prime}_{n-1, \ell'-1} {\cal I}_{n-1, \ell-1} 
\\[3mm]
\nonumber
&& -2\beta \sqrt{n} ({\cal K}_1 j^{\, \prime}) \left [\sqrt{\ell} (M_{\ell'} + m_f) 
{\cal I}^{\, \prime}_{n-1, \ell'-1} {\cal I}_{n, \ell} - 
\sqrt{\ell^{\, \prime}} (M_{\ell} + m_f) 
{\cal I}^{\, \prime}_{n, \ell'} {\cal I}_{n-1, \ell-1} \right ]  
\\[3mm]
\nonumber 
&&+\sqrt{\frac{2 \beta}{q_{\mprp}^{\, \prime 2}}}\; (M_{\ell'} + m_f)  
\left [(q^{\, \prime} \Lambda j^{\, \prime}) + 
\ii (q^{\, \prime} \varphi j^{\, \prime}) \right ] 
\left [\sqrt{\ell} \; [({\cal K}_1 P) + m_f {\cal K}_3] 
{\cal I}^{\, \prime}_{n, \ell'-1} {\cal I}_{n, \ell} - 
\sqrt{n} (M_{\ell} + m_f) {\cal K}_3 {\cal I}^{\, \prime}_{n, \ell'-1} 
{\cal I}_{n-1, \ell-1} \right ] 
\\[3mm]
\nonumber
&&-\sqrt{\frac{2 \beta \ell^{\, \prime}}
{q_{\mprp}^{\, \prime 2}}}\;  
\left [(q^{\, \prime} \Lambda j^{\, \prime}) - 
\ii (q^{\, \prime} \varphi j^{\, \prime}) \right ] 
\left [(M_{\ell} + m_f) [({\cal K}_1 P) - m_f {\cal K}_3] 
{\cal I}^{\, \prime}_{n-1, \ell'} {\cal I}_{n-1, \ell-1}  - 
2 \beta \sqrt{\ell n} \, {\cal K}_3 {\cal I}^{\, \prime}_{n-1, \ell'} 
{\cal I}_{n, \ell} \right ]
\bigg \}\, ; 
\eeq
\beq
\label{eq:Rpaupup2}
&&{\cal R}^{++}_{PA} = g_p g_a^{\, \prime} j_p \bigg \{ - 2\beta 
\sqrt{\ell \ell^{\, \prime}} \left [(P^{\, \prime} \tilde \Lambda j^{\, \prime}){\cal K}_3 - 
(P^{\, \prime} \tilde \varphi j^{\, \prime}) {\cal K}_4 
+ m_f ({\cal K}_1 j^{\, \prime}) \right ] 
{\cal I}_{n, \ell'} {\cal I}^{\, \prime}_{n, \ell}   
\\
\nonumber
&&+(M_{\ell} + m_f) (M_{\ell'} + m_f) 
\left [(P^{\, \prime} \tilde \Lambda j^{\, \prime}){\cal K}_3 - 
(P^{\, \prime} \tilde \varphi j^{\, \prime}) {\cal K}_4 
- m_f ({\cal K}_1 j^{\, \prime}) \right ] {\cal I}_{n-1, \ell'-1} {\cal I}^{\, \prime}_{n-1, \ell-1} 
\\[3mm]
\nonumber  
&& -2\beta \sqrt{n} ({\cal K}_1 j^{\, \prime}) \left [\sqrt{\ell} (M_{\ell'} + m_f) 
{\cal I}_{n-1, \ell'-1} {\cal I}^{\, \prime}_{n, \ell} - 
\sqrt{\ell^{\, \prime}} (M_{\ell} + m_f) 
{\cal I}_{n, \ell'} {\cal I}^{\, \prime}_{n-1, \ell-1} \right ]  
\\[3mm]
\nonumber 
&&+ \sqrt{\frac{2 \beta \ell}{q^{\, \prime 2}_{\mprp}}}\;  
\left [(q^{\, \prime} \Lambda j^{\, \prime}) + 
\ii (q^{\, \prime} \varphi j^{\, \prime}) \right ] 
\left [(M_{\ell'} + m_f) \; [({\cal K}_1 P^{\, \prime}) - m_f {\cal K}_3] 
{\cal I}_{n-1, \ell'-1} {\cal I}^{\, \prime}_{n-1, \ell} - 
2 \beta \sqrt{\ell^{\, \prime} n}\, {\cal K}_3 {\cal I}_{n, \ell'} 
{\cal I}^{\, \prime}_{n-1, \ell} \right ] 
\\[3mm]
\nonumber
&&- \sqrt{\frac{2 \beta }
{q^{\, \prime 2}_{\mprp}}}\;  (M_{\ell} + m_f)
\left [(q^{\, \prime} \Lambda j^{\, \prime}) - 
\ii (q^{\, \prime} \varphi j^{\, \prime}) \right ] 
\left [\sqrt{\ell^{\, \prime}}\; [({\cal K}_1 P^{\, \prime}) + m_f {\cal K}_3] 
{\cal I}_{n, \ell'} {\cal I}^{\, \prime}_{n, \ell-1}  - 
\sqrt{n} (M_{\ell'} + m_f) {\cal K}_3 {\cal I}_{n-1, \ell'-1} 
{\cal I}^{\, \prime}_{n, \ell-1} \right ]
\bigg \}\, ; 
\eeq

%
\beq
\label{eq:Rpaupdown}
&&{\cal R}^{+-}_{AP} = \ii g_p g_a^{\, \prime} j_p \bigg \{  
\sqrt{2 \beta \ell^{\, \prime}} (M_{\ell} + m_f) \left [(P \tilde \Lambda j^{\, \prime}) {\cal K}_4 + 
(P \tilde \varphi j^{\, \prime}) {\cal K}_3 
+ m_f ({\cal K}_2 j^{\, \prime}) \right ] {\cal I}^{\, \prime}_{n, \ell'} {\cal I}_{n, \ell} 
\\
\nonumber
&& + \sqrt{2 \beta \ell} (M_{\ell'} + m_f) 
\left [(P \tilde \Lambda j^{\, \prime}) {\cal K}_4 + 
(P \tilde \varphi j^{\, \prime}) {\cal K}_3 
- m_f ({\cal K}_2 j^{\, \prime}) \right ] 
{\cal I}^{\, \prime}_{n-1, \ell'-1} {\cal I}_{n-1, \ell-1} 
\\[3mm]
\nonumber
&&+ \sqrt{2 \beta n}\; ({\cal K}_2 j^{\, \prime}) 
\left [(M_{\ell} + m_f) (M_{\ell'} + m_f) 
{\cal I}^{\, \prime}_{n-1, \ell'-1} {\cal I}_{n, \ell} + 
2 \beta \sqrt{\ell \ell^{\, \prime}}  
{\cal I}^{\, \prime}_{n, \ell'} {\cal I}_{n-1, \ell-1} \right ] 
- \frac{(q^{\, \prime} \Lambda j^{\, \prime}) + 
\ii (q^{\, \prime} \varphi j^{\, \prime})}{\sqrt{q_{\mprp}^{\, \prime 2}}}
\\[3mm]
\nonumber 
&& \times (M_{\ell'} + m_f)  
\left [(M_{\ell} + m_f)  [({\cal K}_2 P) + m_f {\cal K}_4] 
{\cal I}^{\, \prime}_{n, \ell'-1} {\cal I}_{n, \ell} + 
2\beta \sqrt{n \ell}  \;{\cal K}_4 {\cal I}^{\, \prime}_{n, \ell'-1} 
{\cal I}_{n-1, \ell-1} \right ]  
\\ [3mm]
\nonumber
&& 
-2 \beta \; \sqrt{\ell^{\, \prime}} 
\; \frac{(q^{\, \prime} \Lambda j^{\, \prime}) - 
\ii (q^{\, \prime} \varphi j^{\, \prime})}{\sqrt{q_{\mprp}^{\, \prime 2}}} 
\left [\sqrt{\ell} \; [({\cal K}_2 P) - m_f {\cal K}_4]  
{\cal I}^{\, \prime}_{n-1, \ell'} {\cal I}_{n-1, \ell-1} + 
 \sqrt{ n}\; (M_{\ell} + m_f) \; {\cal K}_4 \; {\cal I}^{\, \prime}_{n-1, \ell'} 
{\cal I}_{n, \ell} \right ]
\bigg \}\, ; 
\eeq
%

%
\beq
\label{eq:Rapupdown2}
&&{\cal R}^{+-}_{PA} = \ii g_p g_a^{\, \prime} j_p \bigg \{  
\sqrt{2 \beta \ell^{\, \prime}} (M_{\ell} + m_f) \left [ 
(P^{\, \prime} \tilde \varphi j^{\, \prime}) {\cal K}_3 
-(P^{\, \prime} \tilde \Lambda j^{\, \prime}) {\cal K}_4 -
 m_f ({\cal K}_2 j^{\, \prime}) \right ] {\cal I}_{n, \ell'} {\cal I}^{\, \prime}_{n, \ell} 
\\
\nonumber
&& + \sqrt{2 \beta \ell} (M_{\ell'} + m_f) 
\left [ (P^{\, \prime} \tilde \varphi j^{\, \prime}) {\cal K}_3 -
(P^{\, \prime} \tilde \Lambda j^{\, \prime}) {\cal K}_4 +
m_f ({\cal K}_2 j^{\, \prime}) \right ] 
{\cal I}_{n-1, \ell'-1} {\cal I}^{\, \prime}_{n-1, \ell-1} 
\\[3mm]
\nonumber
&&+ \sqrt{2 \beta n}\; ({\cal K}_2 j^{\, \prime}) 
\left [(M_{\ell} + m_f) (M_{\ell'} + m_f) 
{\cal I}_{n-1, \ell'-1} {\cal I}^{\, \prime}_{n, \ell} + 
2 \beta \sqrt{\ell \ell^{\, \prime}}  
{\cal I}_{n, \ell'} {\cal I}^{\, \prime}_{n-1, \ell-1} \right ] 
+ \frac{(q^{\, \prime} \Lambda j^{\, \prime}) + 
\ii (q^{\, \prime} \varphi j^{\, \prime})}{\sqrt{q^{\, \prime 2}_{\mprp}}}
\\[3mm]
\nonumber 
&& \times (M_{\ell} + m_f) \;   
\left [(M_{\ell'} + m_f)  [({\cal K}_2 P^{\, \prime}) - m_f {\cal K}_4] 
{\cal I}_{n-1, \ell'-1} {\cal I}^{\, \prime}_{n-1, \ell} - 
2\beta \sqrt{n \ell^{\, \prime}}  \;{\cal K}_4 {\cal I}_{n, \ell'} 
{\cal I}^{\, \prime}_{n-1, \ell} \right ]  
\\ [3mm]
\nonumber
&& 
+2 \beta \; \sqrt{\ell} 
\; \frac{(q^{\, \prime} \Lambda j^{\, \prime}) + 
\ii (q^{\, \prime} \varphi j^{\, \prime})}{\sqrt{q^{\, \prime 2}_{\mprp}}} 
\left [\sqrt{\ell^{\, \prime}} \; [({\cal K}_2 P^{\, \prime}) + m_f {\cal K}_4]  
{\cal I}_{n, \ell'} {\cal I}^{\, \prime}_{n, \ell-1} - 
 \sqrt{ n}\; (M_{\ell'} + m_f) \; {\cal K}_4 \; {\cal I}_{n, \ell'-1} 
{\cal I}^{\, \prime}_{n-1, \ell-1} \right ]
\bigg \}\, ; 
\eeq
%

%
\beq
\label{eq:Rpadownup}
&&{\cal R}^{-+}_{AP} =  \ii g_p g_a^{\, \prime} j_p
\bigg \{ \sqrt{2 \beta \ell} (M_{\ell'} + m_f)
\left [(P \tilde \Lambda j^{\, \prime}) {\cal K}_4 +
(P \tilde \varphi j^{\, \prime}) {\cal K}_3
- m_f ({\cal K}_2 j^{\, \prime}) \right ] {\cal I}^{\, \prime}_{n, \ell'} {\cal I}_{n, \ell}
\\
\nonumber
&&+ \sqrt{2 \beta \ell^{\, \prime}} (M_{\ell} + m_f) 
\left [(P \tilde \Lambda j^{\, \prime}) {\cal K}_4 +
(P \tilde \varphi j^{\, \prime})  {\cal K}_3
+ m_f ({\cal K}_2 j^{\, \prime}) \right ] 
{\cal I}^{\, \prime}_{n-1, \ell'-1} {\cal I}_{n-1, \ell-1} 
\\[3mm]
\nonumber
&& + \sqrt{2 \beta n}\; ({\cal K}_2 j^{\, \prime})
\left [2 \beta \sqrt{\ell \ell^{\, \prime}}
{\cal I}^{\, \prime}_{n-1, \ell'-1} {\cal I}_{n, \ell} +
(M_{\ell} + m_f) (M_{\ell'} + m_f)
{\cal I}^{\, \prime}_{n, \ell'} {\cal I}_{n-1, \ell-1} \right ]  
\\[3mm]
\nonumber
&& - 2\beta \; \sqrt{\ell^{\, \prime}} \; 
\frac{(q^{\, \prime} \Lambda j^{\, \prime}) +
\ii (q^{\, \prime} \varphi j^{\, \prime})}{\sqrt{q_{\mprp}^{\, \prime 2}}}
\left [\sqrt{\ell} \; [({\cal K}_2 P) - m_f {\cal K}_4] 
{\cal I}^{\, \prime}_{n, \ell'-1} {\cal I}_{n, \ell} +
\sqrt{ n} (M_{\ell} + m_f) \;{\cal K}_4 {\cal I}^{\, \prime}_{n, \ell'-1}
{\cal I}_{n-1, \ell-1} \right ] 
\\[3mm]
\nonumber
&&-
\frac{(q^{\, \prime} \Lambda j^{\, \prime}) -
\ii (q^{\, \prime} \varphi j^{\, \prime})}{\sqrt{q_{\mprp}^{\, \prime 2}}}
(M_{\ell'} + m_f) \; \Big [(M_{\ell} + m_f) [({\cal K}_2 P) + m_f {\cal K}_4] 
{\cal I}^{\, \prime}_{n-1, \ell'} {\cal I}_{n-1, \ell-1} 
- 2\beta \sqrt{n \ell}\;  {\cal K}_4 \; {\cal I}^{\, \prime}_{n-1, \ell'}
{\cal I}_{n, \ell} \Big ]
\bigg \}\, ;
\eeq

%
\beq
\label{eq:Rpadownup2}
&&{\cal R}^{-+}_{PA} =   \ii g_p g_a^{\, \prime} j_p
\bigg \{ \sqrt{2 \beta \ell} (M_{\ell'} + m_f)
\left [(P^{\, \prime} \tilde \varphi j^{\, \prime}) {\cal K}_3
-(P^{\, \prime} \tilde \Lambda j^{\, \prime}) {\cal K}_4 
+ m_f ({\cal K}_2 j^{\, \prime}) \right ] {\cal I}_{n, \ell'} {\cal I}^{\, \prime}_{n, \ell}
\\
\nonumber
&&+ \sqrt{2 \beta \ell^{\, \prime}} (M_{\ell} + m_f) 
\left [(P^{\, \prime} \tilde \varphi j^{\, \prime})  {\cal K}_3
-(P^{\, \prime} \tilde \Lambda j^{\, \prime}) {\cal K}_4 
- m_f ({\cal K}_2 j^{\, \prime}) \right ] 
{\cal I}_{n-1, \ell'-1} {\cal I}^{\, \prime}_{n-1, \ell-1} 
\\[3mm]
\nonumber
&& + \sqrt{2 \beta n}\; ({\cal K}_2 j^{\, \prime})
\left [2 \beta \sqrt{\ell \ell^{\, \prime}}
{\cal I}_{n-1, \ell'-1} {\cal I}^{\, \prime}_{n, \ell} +
(M_{\ell} + m_f) (M_{\ell'} + m_f)
{\cal I}_{n, \ell'} {\cal I}^{\, \prime}_{n-1, \ell-1} \right ]  
\\[3mm]
\nonumber
&& + 2\beta \; \sqrt{\ell} \; 
\frac{(q^{\, \prime} \Lambda j^{\, \prime}) +
\ii (q^{\, \prime} \varphi j^{\, \prime})}{\sqrt{q^{\, \prime 2}_{\mprp}}}
\left [\sqrt{\ell^{\, \prime}} \; [({\cal K}_2 P^{\, \prime}) + m_f {\cal K}_4] 
{\cal I}_{n-1, \ell'-1} {\cal I}^{\, \prime}_{n-1, \ell} -
\sqrt{ n} (M_{\ell'} + m_f) \;{\cal K}_4 {\cal I}_{n-1, \ell'}
{\cal I}^{\, \prime}_{n, \ell} \right ] 
\\[3mm]
\nonumber
&&+
\frac{(q^{\, \prime} \Lambda j^{\, \prime}) -
\ii (q^{\, \prime} \varphi j^{\, \prime})}{\sqrt{q^{\, \prime 2}_{\mprp}}}
(M_{\ell} + m_f) \; \Big [(M_{\ell'} + m_f) [({\cal K}_2 P^{\, \prime}) - m_f {\cal K}_4] 
{\cal I}_{n, \ell'} {\cal I}^{\, \prime}_{n, \ell-1} 
- 2\beta \sqrt{n \ell^{\, \prime}}\;  {\cal K}_4 \; {\cal I}_{n, \ell'-1}
{\cal I}^{\, \prime}_{n-1, \ell-1} \Big ]
\bigg \}\, ;
\eeq


%
\beq
\label{eq:Rpadowndown}
&&{\cal R}^{--}_{AP} = -g_p g_a^{\, \prime} j_p 
\bigg \{ (M_{\ell} + m_f) (M_{\ell'} + m_f)  \left [(P \tilde \Lambda j^{\, \prime}) {\cal K}_3 + 
(P \tilde \varphi j^{\, \prime})  {\cal K}_4 
- m_f ({\cal K}_1 j^{\, \prime}) \right ] {\cal I}^{\, \prime}_{n, \ell'} {\cal I}_{n, \ell}  
\\
\nonumber
&& - 2\beta 
\sqrt{\ell \ell^{\, \prime}} 
\left [(P \tilde \Lambda j^{\, \prime}) {\cal K}_3 + 
(P \tilde \varphi j^{\, \prime})  {\cal K}_4 
+ m_f ({\cal K}_1 j^{\, \prime}) \right ] 
{\cal I}^{\, \prime}_{n-1, \ell'-1} {\cal I}_{n-1, \ell-1} 
\\[3mm]
\nonumber
&&+ 2\beta \sqrt{n} ({\cal K}_1 j^{\, \prime}) \left [\sqrt{\ell^{\, \prime}} (M_{\ell} + m_f) 
{\cal I}^{\, \prime}_{n-1, \ell'-1} {\cal I}_{n, \ell} - 
\sqrt{\ell} (M_{\ell'} + m_f) 
{\cal I}^{\, \prime}_{n, \ell'} {\cal I}_{n-1, \ell-1} \right ]   
\\[3mm]
\nonumber 
&& -
\sqrt{\frac{2 \beta \ell^{\, \prime}}{q_{\mprp}^{\, \prime 2}}}\; 
\left [(q^{\, \prime} \Lambda j^{\, \prime}) + 
\ii (q^{\, \prime} \varphi j^{\, \prime}) \right ] 
\left [(M_{\ell} + m_f) [({\cal K}_1 P) - m_f {\cal K}_3] 
{\cal I}^{\, \prime}_{n, \ell'-1} {\cal I}_{n, \ell} - 
2 \beta \sqrt{\ell n} {\cal K}_3 {\cal I}^{\, \prime}_{n, \ell'-1} 
{\cal I}_{n-1, \ell-1} \right ]  
\\[3mm]
\nonumber
&&+\sqrt{\frac{2 \beta}
{q_{\mprp}^{\, \prime 2}}}\; (M_{\ell'} + m_f)  
\left [(q^{\, \prime} \Lambda j^{\, \prime}) - 
\ii (q^{\, \prime} \varphi j^{\, \prime}) \right ] 
\left [ \sqrt{\ell}\;  [({\cal K}_1 P) + m_f {\cal K}_3] 
{\cal I}^{\, \prime}_{n-1, \ell'} {\cal I}_{n-1, \ell-1} - 
 \sqrt{n}\; (M_{\ell} + m_f) {\cal K}_3 {\cal I}^{\, \prime}_{n-1, \ell'} 
{\cal I}_{n, \ell} \right ]
\bigg \}\, . 
\eeq


%
\beq
\label{eq:Rpadowndown2}
&&{\cal R}^{--}_{PA} = g_p g_a^{\, \prime} j_p 
\bigg \{ (M_{\ell} + m_f) (M_{\ell'} + m_f)  
\left [(P^{\, \prime} \tilde \Lambda j^{\, \prime}) {\cal K}_3 - 
(P^{\, \prime} \tilde \varphi j^{\, \prime})  {\cal K}_4 
- m_f ({\cal K}_1 j^{\, \prime}) \right ] {\cal I}_{n, \ell'} {\cal I}^{\, \prime}_{n, \ell}  
\\
\nonumber
&& - 2\beta 
\sqrt{\ell \ell^{\, \prime}} 
\left [(P^{\, \prime} \tilde \Lambda j^{\, \prime}) {\cal K}_3 - 
(P^{\, \prime} \tilde \varphi j^{\, \prime})  {\cal K}_4 
+ m_f ({\cal K}_1 j^{\, \prime}) \right ] 
{\cal I}_{n-1, \ell'-1} {\cal I}^{\, \prime}_{n-1, \ell-1} 
\\[3mm]
\nonumber
&&+ 2\beta \sqrt{n} ({\cal K}_1 j^{\, \prime}) \left [\sqrt{\ell^{\, \prime}} (M_{\ell} + m_f) 
{\cal I}_{n-1, \ell'-1} {\cal I}^{\, \prime}_{n, \ell} - 
\sqrt{\ell} (M_{\ell'} + m_f) 
{\cal I}_{n, \ell'} {\cal I}^{\, \prime}_{n-1, \ell-1} \right ]   
\\[3mm]
\nonumber 
&& -
\sqrt{\frac{2 \beta}{q^{\, \prime 2}_{\mprp}}}\;  (M_{\ell} + m_f)
\left [(q^{\, \prime} \Lambda j^{\, \prime}) + 
\ii (q^{\, \prime} \varphi j^{\, \prime}) \right ] 
\left [\sqrt{\ell^{\, \prime}}\; [({\cal K}_1 P^{\, \prime}) + m_f {\cal K}_3] 
{\cal I}_{n-1, \ell'-1} {\cal I}^{\, \prime}_{n-1, \ell} - 
\sqrt{n}\; (M_{\ell'} + m_f) {\cal K}_3 {\cal I}_{n, \ell'} 
{\cal I}^{\, \prime}_{n-1, \ell} \right ]  
\\[3mm]
\nonumber
&&+\sqrt{\frac{2 \beta \ell}
{q^{\, \prime 2}_{\mprp}}}\;   
\left [(q^{\, \prime} \Lambda j^{\, \prime}) - 
\ii (q^{\, \prime} \varphi j^{\, \prime}) \right ] 
\left [  (M_{\ell'} + m_f)\;  [({\cal K}_1 P^{\, \prime}) - m_f {\cal K}_3] 
{\cal I}_{n, \ell'} {\cal I}^{\, \prime}_{n, \ell-1} - 
2 \beta \sqrt{\ell^{\, \prime} n} {\cal K}_3 {\cal I}_{n-1, \ell'-1} 
{\cal I}^{\, \prime}_{n, \ell-1}  \right ]
\bigg \}\, . 
\eeq

\newpage

\item
Both vertices are vectors ($k =  k^{\, \prime} = V$):
\beq
\label{eq:Rvvupup}
&&{\cal R}^{++}_{VV} =  g_v g_v^{\, \prime} \bigg \{2\beta \sqrt{\ell \ell^{\, \prime}}
\left [(P \tilde \Lambda j^{\, \prime}) ({\cal K}_1 j) +
(P \tilde \Lambda j) ({\cal K}_1 j^{\, \prime}) - 
(j \tilde \Lambda j^{\, \prime}) ({\cal K}_1 P) - m_f 
[(j \tilde \Lambda j^{\, \prime}){\cal K}_3 +  
(j \tilde \varphi j^{\, \prime}) {\cal K}_4 ]\right ] 
{\cal I}^{\, \prime}_{n, \ell'} {\cal I}_{n, \ell} 
\\[3mm]
\nonumber
&&
+ 
(M_{\ell} + m_f) (M_{\ell'} + m_f) \left [(P \tilde \Lambda j^{\, \prime}) ({\cal K}_1 j) +
(P \tilde \Lambda j) ({\cal K}_1 j^{\, \prime}) - 
(j \tilde \Lambda j^{\, \prime}) ({\cal K}_1 P) + m_f
[(j \tilde \Lambda j^{\, \prime}){\cal K}_3 +
(j \tilde \varphi j^{\, \prime}) {\cal K}_4 ] \right ] 
\\[3mm]
\nonumber
&&
\times {\cal I}^{\, \prime}_{n-1, \ell'-1} {\cal I}_{n-1, \ell-1} +
2\beta \sqrt{n} \, [(j \tilde \Lambda j^{\, \prime}){\cal K}_3 +
(j \tilde \varphi j^{\, \prime}) {\cal K}_4  ] \big [\sqrt{\ell} (M_{\ell'} + m_f)
{\cal I}^{\, \prime}_{n-1, \ell'-1} {\cal I}_{n, \ell} +
\sqrt{\ell^{\, \prime}} (M_{\ell} + m_f)
{\cal I}^{\, \prime}_{n, \ell'} {\cal I}_{n-1, \ell-1} \big ]
\\[3mm]
\nonumber
&&   -
\sqrt{\frac{2 \beta \ell}{q_{\mprp}^{\, \prime 2}}} \, 
(M_{\ell'} + m_f) [(q^{\, \prime} \Lambda j^{\, \prime}) + 
\ii (q^{\, \prime} \varphi j^{\, \prime})][(P \tilde \Lambda j) {\cal K}_3 - 
 (P \tilde \varphi j) {\cal K}_4 - m_f ({\cal K}_1 j)] 
{\cal I}^{\, \prime}_{n, \ell'-1} {\cal I}_{n, \ell}
\\[3mm]
\nonumber
&& -
\sqrt{\frac{2 \beta \ell^{\, \prime}}{q_{\mprp}^2}} \, 
(M_{\ell} + m_f) [(q \Lambda j) - 
\ii (q \varphi j)] [(P \tilde \Lambda j^{\, \prime}) {\cal K}_3 + 
 (P\tilde \varphi j^{\, \prime}) {\cal K}_4 - 
m_f ({\cal K}_1 j^{\, \prime})] 
{\cal I}^{\, \prime}_{n, \ell'} {\cal I}_{n, \ell-1} 
\\[3mm]
\nonumber
&& - 
\sqrt{\frac{2 \beta \ell^{\, \prime}}{q_{\mprp}^2}} \, 
(M_{\ell} + m_f) [(q \Lambda j^{\, \prime}) - 
\ii (q \varphi j^{\, \prime})] [(P \tilde \Lambda j) {\cal K}_3 - 
 (P \tilde \varphi j) {\cal K}_4 + 
m_f ({\cal K}_1 j)] 
{\cal I}^{\, \prime}_{n-1, \ell'} {\cal I}_{n-1, \ell-1}  
\\[3mm]
\nonumber
&&-\sqrt{\frac{2 \beta \ell}{q_{\mprp}^{\, \prime 2}}} \, 
(M_{\ell'} + m_f) [(q^{\, \prime} \Lambda j) + 
\ii (q^{\, \prime} \varphi j)]\; 
[(P \tilde \Lambda j^{\, \prime})  {\cal K}_3 + 
 (P \tilde \varphi j^{\, \prime})  {\cal K}_4 + 
m_f ({\cal K}_1 j^{\, \prime})] 
{\cal I}^{\, \prime}_{n-1, \ell'-1} {\cal I}_{n-1, \ell} 
\\[3mm]
\nonumber
&&+(M_{\ell} + m_f)(M_{\ell'} + m_f) \; [(j \Lambda j^{\, \prime}) + \ii 
(j \varphi j^{\, \prime})] \; [({\cal K}_1 P) - m_f {\cal K}_3] 
\frac{(q \Lambda q^{\, \prime}) - \ii (q \varphi q^{\, \prime})}
{\sqrt{q_{\mprp}^2  q_{\mprp}^{\, \prime 2}}} 
\; {\cal I}^{\, \prime}_{n, \ell'-1} {\cal I}_{n, \ell-1} 
\\
\nonumber
&& +
2 \beta \sqrt{\ell \ell^{\, \prime}} 
 [(j \Lambda j^{\, \prime}) - \ii 
(j \varphi j^{\, \prime})] \; [({\cal K}_1 P) + m_f {\cal K}_3] 
\frac{(q \Lambda q^{\, \prime}) + \ii (q \varphi q^{\, \prime})}
{\sqrt{q_{\mprp}^2  q_{\mprp}^{\, \prime 2}}} 
\; {\cal I}^{\, \prime}_{n-1, \ell'} {\cal I}_{n-1, \ell} 
 \\[3mm]
\nonumber
&&-
\sqrt{2\beta n}\; ({\cal K}_1 j) \; \big [ 2\beta
\sqrt{\ell \ell^{\, \prime}} 
\frac{(q^{\, \prime} \Lambda j^{\, \prime}) - 
\ii (q^{\, \prime} \varphi j^{\, \prime})}
{\sqrt{q_{\mprp}^{\, \prime 2}}}
{\cal I}^{\, \prime}_{n-1, \ell'} {\cal I}_{n, \ell}  + 
(M_{\ell} + m_f) (M_{\ell'} + m_f) \; \frac{(q^{\, \prime} \Lambda j^{\, \prime}) + 
\ii (q^{\, \prime} \varphi j^{\, \prime})}
{\sqrt{q_{\mprp}^{\, \prime 2}}} \;
{\cal I}^{\, \prime}_{n, \ell'-1} {\cal I}_{n-1, \ell-1} \big ]   
\\[3mm]
\nonumber
&&-\sqrt{2\beta n}\; ({\cal K}_1 j^{\, \prime}) \; 
\big [(M_{\ell} + m_f) (M_{\ell'} + m_f)   
\; \frac{(q \Lambda j) - 
\ii (q \varphi j)}{\sqrt{q_{\mprp}^2}}
{\cal I}^{\, \prime}_{n-1, \ell'-1}  {\cal I}_{n, \ell-1} + 
2\beta \sqrt{\ell \ell^{\, \prime}}  \frac{(q \Lambda j) + 
\ii (q \varphi j)}
{\sqrt{q_{\mprp}^2}} \;
{\cal I}^{\, \prime}_{n, \ell'} {\cal I}_{n-1, \ell} \big ] 
\\[3mm]
\nonumber
&& + 2 \beta \sqrt{n} \; {\cal K}_3 \; \big [\sqrt{\ell^{\, \prime}}\; (M_{\ell} + m_f)  
\; \frac{(q \Lambda j) - 
\ii (q \varphi j)}{\sqrt{q_{\mprp}^2}} \; \frac{(q^{\, \prime} \Lambda j^{\, \prime}) - 
\ii (q^{\, \prime} \varphi j^{\, \prime})}{\sqrt{q_{\mprp}^{\, \prime 2}}} 
{\cal I}^{\, \prime}_{n-1, \ell'} {\cal I}_{n, \ell-1}  
\\[3mm]
\nonumber
&&+\sqrt{\ell}\; (M_{\ell'} + m_f) \; \frac{(q \Lambda j) +
\ii (q \varphi j)}{\sqrt{q_{\mprp}^2}} \; \frac{(q^{\, \prime} \Lambda j^{\, \prime}) +
\ii (q^{\, \prime} \varphi j^{\, \prime})}{\sqrt{q_{\mprp}^{\, \prime 2}}} 
{\cal I}^{\, \prime}_{n, \ell'-1} {\cal I}_{n-1, \ell} \big ]
\bigg \} \, ;
\eeq

\newpage
\beq
\label{eq:Rvvupdown}
&&{\cal R}^{+-}_{VV} =  \ii g_v g_v^{\, \prime} \bigg \{
 \sqrt{2 \beta \ell^{\, \prime}} \, (M_{\ell} + m_f)
\left [(P \tilde \Lambda j^{\, \prime}) ({\cal K}_2 j) +
(P \tilde \Lambda j) ({\cal K}_2 j^{\, \prime}) - 
(j \tilde \Lambda j^{\, \prime}) ({\cal K}_2 P)  
\right.
\\[3mm]
\nonumber
&&\left. 
- m_f 
[(j \tilde \Lambda j^{\, \prime}){\cal K}_4 +  
(j \tilde \varphi j^{\, \prime}) {\cal K}_3 ]\right ] 
{\cal I}^{\, \prime}_{n, \ell'} {\cal I}_{n, \ell}  
- 
 \sqrt{2 \beta \ell} \, (M_{\ell'} + m_f) 
\left [(P \tilde \Lambda j^{\, \prime}) ({\cal K}_2 j) +
(P \tilde \Lambda j) ({\cal K}_2 j^{\, \prime})  \right. 
\\[3mm]
\nonumber
&&\left. - 
(j \tilde \Lambda j^{\, \prime}) ({\cal K}_2 P) + m_f
[(j \tilde \Lambda j^{\, \prime}){\cal K}_4 +
(j \tilde \varphi j^{\, \prime}) {\cal K}_3 ]\right ] 
{\cal I}^{\, \prime}_{n-1, \ell'-1} {\cal I}_{n-1, \ell-1} 
\\[3mm]
\nonumber
&&+
  \sqrt{2\beta n}\, [(j \tilde \Lambda j^{\, \prime}){\cal K}_4 +
(j \tilde \varphi j^{\, \prime}) {\cal K}_3 ]  
 \big [ (M_{\ell} + m_f) (M_{\ell'} + m_f)
{\cal I}^{\, \prime}_{n-1, \ell'-1} {\cal I}_{n, \ell} -
2 \beta \sqrt{\ell \ell^{\, \prime}}
{\cal I}^{\, \prime}_{n, \ell'} {\cal I}_{n-1, \ell-1} \big ]  
\\[3mm]
\nonumber
&&-
(M_{\ell} + m_f) (M_{\ell'} + m_f) \frac{(q^{\, \prime} \Lambda j^{\, \prime}) + 
\ii (q^{\, \prime} \varphi j^{\, \prime})}{\sqrt{q_{\mprp}^{\, \prime 2}}} 
[(P \tilde \Lambda j)  {\cal K}_4 - 
 (P \tilde \varphi j)  {\cal K}_3 - m_f ({\cal K}_2 j)] 
{\cal I}^{\, \prime}_{n, \ell'-1} {\cal I}_{n, \ell}  
\\[3mm]
\nonumber
&& +
\frac{ 2 \beta \sqrt{\ell \ell^{\, \prime}}}{\sqrt{q_{\mprp}^2}} \, [(q \Lambda j) - 
\ii (q \varphi j)]\; [(P \tilde \Lambda j^{\, \prime}) {\cal K}_4 + 
(P \tilde \varphi j^{\, \prime})  {\cal K}_3 - 
m_f ({\cal K}_2 j^{\, \prime})] 
{\cal I}^{\, \prime}_{n, \ell'} {\cal I}_{n, \ell-1} 
\\[3mm]
\nonumber
&& + 
\frac{ 2 \beta \sqrt{\ell \ell^{\, \prime}}}{\sqrt{q_{\mprp}^2}} \, 
 [(q \Lambda j^{\, \prime}) - 
\ii (q \varphi j^{\, \prime})]\; [(P \tilde \Lambda j)  {\cal K}_4 - 
(P \tilde \varphi j)  {\cal K}_3 + 
m_f ({\cal K}_2 j)] 
{\cal I}^{\, \prime}_{n-1, \ell'} {\cal I}_{n-1, \ell-1} 
\\[3mm]
\nonumber
&&- (M_{\ell} + m_f) (M_{\ell'} + m_f) 
\frac{(q^{\, \prime} \Lambda j) + 
\ii (q^{\, \prime} \varphi j)}{\sqrt{q_{\mprp}^{\, \prime 2}}} \; 
[(P \tilde \Lambda j^{\, \prime})  {\cal K}_4 + 
 (P \tilde \varphi j^{\, \prime}) {\cal K}_3 + m_f ({\cal K}_2 j^{\, \prime})] 
{\cal I}^{\, \prime}_{n-1, \ell'-1} {\cal I}_{n-1, \ell} 
\\[3mm]
\nonumber
&& -
 \sqrt{2 \beta \ell} \; (M_{\ell'} + m_f) [(j \Lambda j^{\, \prime}) + \ii 
(j \varphi j^{\, \prime})] \; [({\cal K}_2 P) - m_f {\cal K}_4] 
\frac{(q \Lambda q^{\, \prime}) - \ii (q \varphi q^{\, \prime})}
{\sqrt{q_{\mprp}^2  q_{\mprp}^{\, \prime 2}}} 
{\cal I}^{\, \prime}_{n, \ell'-1} {\cal I}_{n, \ell-1}  
\\[3mm]
\nonumber
&& +
 \sqrt{2 \beta \ell^{\, \prime}} \; (M_{\ell} + m_f) [(j \Lambda j^{\, \prime}) - \ii 
(j \varphi j^{\, \prime})] \; [({\cal K}_2 P) + 
m_f {\cal K}_4] \frac{(q \Lambda q^{\, \prime}) + \ii (q \varphi q^{\, \prime})}
{\sqrt{q_{\mprp}^2  q_{\mprp}^{\, \prime 2}}} 
\; {\cal I}^{\, \prime}_{n-1, \ell'} {\cal I}_{n-1, \ell} 
\\[3mm]
\nonumber
&& -2\beta \sqrt{n}\; ({\cal K}_2 j) \; \big [\sqrt{\ell^{\, \prime}} \, (M_{\ell} + m_f)  
\; \frac{(q^{\, \prime} \Lambda j^{\, \prime}) - 
\ii (q^{\, \prime} \varphi j^{\, \prime})}
{\sqrt{q_{\mprp}^{\, \prime 2}}} {\cal I}^{\, \prime}_{n-1, \ell'} {\cal I}_{n, \ell} 
- \sqrt{\ell} \, (M_{\ell'} + m_f) \; \frac{(q^{\, \prime} \Lambda j^{\, \prime}) + 
\ii (q^{\, \prime} \varphi j^{\, \prime})}
{\sqrt{q_{\mprp}^{\, \prime 2}}} {\cal I}^{\, \prime}_{n, \ell'-1} {\cal I}_{n-1, \ell-1} \big ]
\\[3mm]
\nonumber
&& +  
 2\beta \sqrt{n}\; ({\cal K}_2 j^{\, \prime}) \; 
\big [\sqrt{\ell^{\, \prime}} \, (M_{\ell} + m_f)  
\; \frac{(q \Lambda j) - 
\ii (q \varphi j)}{\sqrt{q_{\mprp}^2}}
{\cal I}^{\, \prime}_{n-1, \ell'-1} {\cal I}_{n, \ell-1}  - 
\sqrt{\ell} \, (M_{\ell'} + m_f)  \frac{(q \Lambda j) + 
\ii (q \varphi j)}
{\sqrt{q_{\mprp}^2}} \;
{\cal I}^{\, \prime}_{n, \ell'} {\cal I}_{n-1, \ell} \big ]
\\[3mm]
\nonumber
&& -
 \sqrt{2 \beta n} \; {\cal K}_4 \; \big [2 \beta \sqrt{\ell \ell^{\, \prime}}  
\; \frac{(q \Lambda j) - 
\ii (q \varphi j)}{\sqrt{q_{\mprp}^2}} \; \frac{(q^{\, \prime} \Lambda j^{\, \prime}) - 
\ii (q^{\, \prime} \varphi j^{\, \prime})}{\sqrt{q_{\mprp}^{\, \prime 2}}} 
{\cal I}^{\, \prime}_{n-1, \ell'} {\cal I}_{n, \ell-1}
\\[3mm]
\nonumber
&& - 
(M_{\ell} + m_f) (M_{\ell'} + m_f) \; \frac{(q \Lambda j) +
\ii (q \varphi j)}{\sqrt{q_{\mprp}^2}} \; \frac{(q^{\, \prime} \Lambda j^{\, \prime}) +
\ii (q^{\, \prime} \varphi j^{\, \prime})}{\sqrt{q_{\mprp}^{\, \prime 2}}} 
{\cal I}^{\, \prime}_{n, \ell'-1} {\cal I}_{n-1, \ell} \big ]
\bigg \} \, ;
\eeq

\newpage
%
\beq
\label{eq:Rvvdownup}
&&{\cal R}^{-+}_{VV} =  \ii g_v g_v^{\, \prime} \bigg \{
-  \sqrt{2 \beta \ell} \, (M_{\ell} + m_f^{\, \prime})
\left [(P \tilde \Lambda j^{\, \prime}) ({\cal K}_2 j) +
(P \tilde \Lambda j) ({\cal K}_2 j^{\, \prime}) - 
(j \tilde \Lambda j^{\, \prime}) ({\cal K}_2 P)  
\right.
\\[3mm]
\nonumber
&&\left. + m_f 
[(j \tilde \Lambda j^{\, \prime}){\cal K}_4 +  
(j \tilde \varphi j^{\, \prime}) {\cal K}_3 ] \right ] 
{\cal I}^{\, \prime}_{n, \ell'} {\cal I}_{n, \ell} + 
 \sqrt{2 \beta \ell^{\, \prime}} \, (M_{\ell} + m_f) 
\left [(P \tilde \Lambda j^{\, \prime}) ({\cal K}_2 j) +
(P \tilde \Lambda j) ({\cal K}_2 j^{\, \prime}) \right.
\\[3mm]
\nonumber
&&\left. -
(j \tilde \Lambda j^{\, \prime}) ({\cal K}_2 P) - m_f
[(j \tilde \Lambda j^{\, \prime}){\cal K}_4 +
(j \tilde \varphi j^{\, \prime}) {\cal K}_3 ]\right ] 
{\cal I}^{\, \prime}_{n-1, \ell'-1} {\cal I}_{n-1, \ell-1} 
\\[3mm]
\nonumber
&& - \sqrt{2\beta n}\, [(j \tilde \Lambda j^{\, \prime}){\cal K}_4 +
(j \tilde \varphi j^{\, \prime}) {\cal K}_3 ]  
\big [2 \beta \sqrt{\ell \ell^{\, \prime}} 
{\cal I}^{\, \prime}_{n-1, \ell'-1} {\cal I}_{n, \ell} -
(M_{\ell} + m_f) (M_{\ell'} + m_f)
{\cal I}^{\, \prime}_{n, \ell'} {\cal I}_{n-1, \ell-1} \big ] 
\\[3mm]
\nonumber
&&+
\frac{ 2 \beta \sqrt{\ell \ell^{\, \prime}}}{\sqrt{q_{\mprp}^{\, \prime 2}}} \, 
 [(q^{\, \prime} \Lambda j^{\, \prime}) + 
\ii (q^{\, \prime} \varphi j^{\, \prime})] 
[(P \tilde \Lambda j) {\cal K}_4 - 
 (P \tilde \varphi j) {\cal K}_3 + m_f ({\cal K}_2 j)] 
{\cal I}^{\, \prime}_{n, \ell'-1} {\cal I}_{n, \ell} 
\\[3mm]
\nonumber
&&-
(M_{\ell} + m_f) (M_{\ell'} + m_f) \frac{(q \Lambda j) - 
\ii (q \varphi j)}{\sqrt{q_{\mprp}^2}}  
[(P \tilde \Lambda j^{\, \prime})  {\cal K}_4 + 
(P \tilde \varphi j^{\, \prime})  {\cal K}_3 + 
m_f ({\cal K}_2 j^{\, \prime})] 
{\cal I}^{\, \prime}_{n, \ell'} {\cal I}_{n, \ell-1} 
\\[3mm]
\nonumber
&& - 
(M_{\ell} + m_f) (M_{\ell'} + m_f) 
\frac{(q \Lambda j^{\, \prime}) - 
\ii (q \varphi j^{\, \prime})}{\sqrt{q_{\mprp}^2}}\; 
[(P \tilde \Lambda j)  {\cal K}_4 - 
 (P \tilde \varphi j)  {\cal K}_3 - 
m_f ({\cal K}_2 j)] 
{\cal I}^{\, \prime}_{n-1, \ell'} {\cal I}_{n-1, \ell-1} 
\\[3mm]
\nonumber
&&+\frac{ 2 \beta \sqrt{\ell \ell^{\, \prime}}}{\sqrt{q_{\mprp}^{\, \prime 2}}} \,  
[(q^{\, \prime} \Lambda j) + 
\ii (q^{\, \prime} \varphi j)]\; 
[(P \tilde \Lambda j^{\, \prime}) {\cal K}_4 + 
(P \tilde \varphi j^{\, \prime})  {\cal K}_3 - 
m_f ({\cal K}_2 j^{\, \prime})] 
{\cal I}^{\, \prime}_{n-1, \ell'-1} {\cal I}_{n-1, \ell} 
\\[3mm]
\nonumber
&&+ \sqrt{2 \beta \ell^{\, \prime}} \; (M_{\ell} + m_f) [(j \Lambda j^{\, \prime}) + \ii 
(j \varphi j^{\, \prime})] \; [({\cal K}_2 P) + m_f {\cal K}_4] 
\frac{(q \Lambda q^{\, \prime}) - \ii (q \varphi q^{\, \prime})}
{\sqrt{q_{\mprp}^2  q_{\mprp}^{\, \prime 2}}} 
\; {\cal I}^{\, \prime}_{n, \ell'-1} {\cal I}_{n, \ell-1}
\\[3mm]
\nonumber
&&- \sqrt{2 \beta \ell} \; (M_{\ell'} + m_f) [(j \Lambda j^{\, \prime}) - \ii 
(j \varphi j^{\, \prime})] \; [({\cal K}_2 P) - m_f {\cal K}_4] 
\frac{(q \Lambda q^{\, \prime}) + \ii (q \varphi q^{\, \prime})}
{\sqrt{q_{\mprp}^2  q_{\mprp}^{\, \prime 2}}} 
\; {\cal I}^{\, \prime}_{n-1, \ell'} {\cal I}_{n-1, \ell} 
\\[3mm]
\nonumber
&&+ 2\beta \sqrt{n}\; ({\cal K}_2 j) \; \big [\sqrt{\ell} \, (M_{\ell'} + m_f)  
\; \frac{(q^{\, \prime} \Lambda j^{\, \prime}) - 
\ii (q^{\, \prime} \varphi j^{\, \prime})}
{\sqrt{q_{\mprp}^{\, \prime 2}}} {\cal I}^{\, \prime}_{n-1, \ell'} {\cal I}_{n, \ell} - 
\sqrt{\ell^{\, \prime}} \, (M_{\ell} + m_f) \; \frac{(q^{\, \prime} \Lambda j^{\, \prime}) + 
\ii (q^{\, \prime} \varphi j^{\, \prime})}
{\sqrt{q_{\mprp}^{\, \prime 2}}} {\cal I}^{\, \prime}_{n, \ell'-1} {\cal I}_{n-1, \ell-1} \big ]
\\[3mm]
\nonumber
&& -  
 2\beta \sqrt{n}\; ({\cal K}_2 j^{\, \prime}) \; 
\big [\sqrt{\ell} \, (M_{\ell'} + m_f)   
\frac{(q \Lambda j) - 
\ii (q \varphi j)}{\sqrt{q_{\mprp}^2}}
{\cal I}^{\, \prime}_{n-1, \ell'-1} {\cal I}_{n, \ell-1} - 
\sqrt{\ell^{\, \prime}} \, (M_{\ell} + m_f)  \frac{(q \Lambda j) + 
\ii (q \varphi j)}
{\sqrt{q_{\mprp}^2}} \;
{\cal I}^{\, \prime}_{n, \ell'} {\cal I}_{n-1, \ell} \big ] 
\\[3mm]
\nonumber
&&+ \sqrt{2 \beta n} \; {\cal K}_4 \; \big [(M_{\ell} + m_f) (M_{\ell'} + m_f)  
\; \frac{(q \Lambda j) - 
\ii (q \varphi j)}{\sqrt{q_{\mprp}^2}} \; \frac{(q^{\, \prime} \Lambda j^{\, \prime}) - 
\ii (q^{\, \prime} \varphi j^{\, \prime})}{\sqrt{q_{\mprp}^{\, \prime 2}}} 
{\cal I}^{\, \prime}_{n-1, \ell'} {\cal I}_{n, \ell-1} 
\\
\nonumber
&&  - 
2 \beta \sqrt{\ell \ell^{\, \prime}} \; \frac{(q \Lambda j) +
\ii (q \varphi j)}{\sqrt{q_{\mprp}^2}} \; \frac{(q^{\, \prime} \Lambda j^{\, \prime}) +
\ii (q^{\, \prime} \varphi j^{\, \prime})}{\sqrt{q_{\mprp}^{\, \prime 2}}}   
{\cal I}^{\, \prime}_{n, \ell'-1} {\cal I}_{n-1, \ell} \big ]
\bigg \} \, ;
\eeq

\newpage
%
\beq
\label{eq:Rvvdowndown}
&&{\cal R}^{--}_{VV} =  g_v g_v^{\, \prime} \bigg \{(M_{\ell} + m_f) (M_{\ell'} + m_f)
\left [(P \tilde \Lambda j^{\, \prime}) ({\cal K}_1 j) +
(P \tilde \Lambda j) ({\cal K}_1 j^{\, \prime}) - 
(j \tilde \Lambda j^{\, \prime}) ({\cal K}_1 P) \right.
\\[3mm]
\nonumber
&&\left. +  m_f 
[(j \tilde \Lambda j^{\, \prime}){\cal K}_3 +  
(j \tilde \varphi j^{\, \prime}) {\cal K}_4 ]\right ] 
{\cal I}^{\, \prime}_{n, \ell'} {\cal I}_{n, \ell}
+ 
2\beta \sqrt{\ell \ell^{\, \prime}} 
\left [(P \tilde \Lambda j^{\, \prime}) ({\cal K}_1 j) +
(P \tilde \Lambda j) ({\cal K}_1 j^{\, \prime})  \right.
\\[3mm]
\nonumber
&&\left. -
(j \tilde \Lambda j^{\, \prime}) ({\cal K}_1 P) - m_f
[(j \tilde \Lambda j^{\, \prime}){\cal K}_3 +
(j \tilde \varphi j^{\, \prime}) {\cal K}_4 ]\right ] 
{\cal I}^{\, \prime}_{n-1, \ell'-1} {\cal I}_{n-1, \ell-1} 
\\[3mm]
\nonumber
&&+
2\beta \sqrt{n} [(j \tilde \Lambda j^{\, \prime}){\cal K}_3 +
(j \tilde \varphi j^{\, \prime}) {\cal K}_4 ]  
\big [\sqrt{\ell^{\, \prime}} (M_{\ell} + m_f)
{\cal I}^{\, \prime}_{n-1, \ell'-1} {\cal I}_{n, \ell} +
\sqrt{\ell} (M_{\ell'} + m_f)
{\cal I}^{\, \prime}_{n, \ell'} {\cal I}_{n-1, \ell-1} \big ] 
\\[3mm]
\nonumber
&&-
\sqrt{\frac{2 \beta \ell^{\, \prime}}{q_{\mprp}^{\, \prime 2}}} \, 
(M_{\ell} + m_f) [(q^{\, \prime} \Lambda j^{\, \prime}) + 
\ii (q^{\, \prime} \varphi j^{\, \prime})]  
[(P \tilde \Lambda j) {\cal K}_3 - 
 (P \tilde \varphi j)  {\cal K}_4 + m_f ({\cal K}_1 j)] 
{\cal I}^{\, \prime}_{n, \ell'-1} {\cal I}_{n, \ell} 
\\[3mm]
\nonumber
&&-
\sqrt{\frac{2 \beta \ell}{q_{\mprp}^2}} \, 
(M_{\ell'} + m_f) [(q \Lambda j) - 
\ii (q \varphi j)]\;   
[(P \tilde \Lambda j^{\, \prime})  {\cal K}_3 + 
(P \tilde \varphi j^{\, \prime})  {\cal K}_4 + 
m_f ({\cal K}_1 j^{\, \prime})] 
{\cal I}^{\, \prime}_{n, \ell'} {\cal I}_{n, \ell-1} 
\\[3mm]
\nonumber
&&- 
\sqrt{\frac{2 \beta \ell}{q_{\mprp}^2}} \, 
(M_{\ell'} + m_f) [(q \Lambda j^{\, \prime}) - 
\ii (q \varphi j^{\, \prime})]  
[(P \tilde \Lambda j)  {\cal K}_3 - 
 (P \tilde \varphi j) {\cal K}_4 - m_f ({\cal K}_1 j)] 
{\cal I}^{\, \prime}_{n-1, \ell'} {\cal I}_{n-1, \ell-1} 
\\[3mm]
\nonumber
&& -
\sqrt{\frac{2 \beta \ell^{\, \prime}}{q_{\mprp}^{\, \prime 2}}} \, 
(M_{\ell} + m_f) [(q^{\, \prime} \Lambda j) + 
\ii (q^{\, \prime} \varphi j)] 
[(P \tilde \Lambda j^{\, \prime})  {\cal K}_3 + 
(P \tilde \varphi j^{\, \prime})  {\cal K}_4 - 
m_f ({\cal K}_1 j^{\, \prime})] 
{\cal I}^{\, \prime}_{n-1, \ell'-1} {\cal I}_{n-1, \ell} 
\\[3mm]
\nonumber
&&+
2 \beta \sqrt{\ell \ell^{\, \prime}} \; [(j \Lambda j^{\, \prime}) + \ii 
(j \varphi j^{\, \prime})]  
[({\cal K}_1 P) + m_f {\cal K}_3] 
\frac{(q \Lambda q^{\, \prime}) - \ii (q \varphi q^{\, \prime})}
{\sqrt{q_{\mprp}^2  q_{\mprp}^{\, \prime 2}}} 
\; {\cal I}^{\, \prime}_{n, \ell'-1} {\cal I}_{n, \ell-1}    
\\[3mm]
\nonumber
&&+
(M_{\ell} + m_f)(M_{\ell'} + m_f) \; [(j \Lambda j^{\, \prime}) - \ii 
(j \varphi j^{\, \prime})] \; [({\cal K}_1 P) - m_f {\cal K}_3] 
\frac{(q \Lambda q^{\, \prime}) + \ii (q \varphi q^{\, \prime})}
{\sqrt{q_{\mprp}^2  q_{\mprp}^{\, \prime 2}}} 
\; {\cal I}^{\, \prime}_{n-1, \ell'} {\cal I}_{n-1, \ell}  
\\[3mm]
\nonumber
&&-
\sqrt{2\beta n}\; ({\cal K}_1 j) \; \big [(M_{\ell} + m_f) (M_{\ell'} + m_f)  
\; \frac{(q^{\, \prime} \Lambda j^{\, \prime}) - 
\ii (q^{\, \prime} \varphi j^{\, \prime})}
{\sqrt{q_{\mprp}^{\, \prime 2}}}
{\cal I}^{\, \prime}_{n-1, \ell'} {\cal I}_{n, \ell} +
2\beta
\sqrt{\ell \ell^{\, \prime}} \; \frac{(q^{\, \prime} \Lambda j^{\, \prime}) + 
\ii (q^{\, \prime} \varphi j^{\, \prime})}
{\sqrt{q_{\mprp}^{\, \prime 2}}} \;
{\cal I}^{\, \prime}_{n, \ell'-1} {\cal I}_{n-1, \ell-1} \big ] 
\\[3mm]
\nonumber
&&-  
\sqrt{2\beta n}\; ({\cal K}_1 j^{\, \prime}) \; 
\big [ 2\beta \sqrt{\ell \ell^{\, \prime}}  
\; \frac{(q \Lambda j) - 
\ii (q \varphi j)}{\sqrt{q_{\mprp}^2}}
{\cal I}^{\, \prime}_{n-1, \ell'-1} {\cal I}_{n, \ell-1} + 
(M_{\ell} + m_f) (M_{\ell'} + m_f)  \frac{(q \Lambda j) + 
\ii (q \varphi j)}
{\sqrt{q_{\mprp}^2}} \;
{\cal I}^{\, \prime}_{n, \ell'} {\cal I}_{n-1, \ell} \big ] 
\\[3mm]
\nonumber
&&+
 2 \beta \sqrt{n} \; {\cal K}_3 \; \big [\sqrt{\ell}\; (M_{\ell'} + m_f)  
\; \frac{(q \Lambda j) - 
\ii (q \varphi j)}{\sqrt{q_{\mprp}^2}}  
\frac{(q^{\, \prime} \Lambda j^{\, \prime}) - 
\ii (q^{\, \prime} \varphi j^{\, \prime})}{\sqrt{q_{\mprp}^{\, \prime 2}}} 
{\cal I}^{\, \prime}_{n-1, \ell'} {\cal I}_{n, \ell-1}  
\\[3mm]
\nonumber
&& + 
\sqrt{\ell^{\, \prime}}\; (M_{\ell} + m_f) \; \frac{(q \Lambda j) +
\ii (q \varphi j)}{\sqrt{q_{\mprp}^2}} \; \frac{(q^{\, \prime} \Lambda j^{\, \prime}) +
\ii (q^{\, \prime} \varphi j^{\, \prime})}{\sqrt{q_{\mprp}^{\, \prime 2}}} 
{\cal I}^{\, \prime}_{n, \ell'-1} {\cal I}_{n-1, \ell} \big ]
\bigg \} \, .
\eeq

For second diagram we have the following replacement $P_{\alpha} \to P_{\alpha}^{\, \prime}$, 
$q_{\alpha} \leftrightarrow -q_{\alpha}^{\, \prime}$, 
$j_{\alpha} \leftrightarrow j_{\alpha}^{\, \prime}$ 
${\cal I}_{m,n} \leftrightarrow {\cal I}_{m,n}^{\, \prime}$.

\newpage

\item
In the case where $j$ is a vector current and $j^{\, \prime}$ is a pseudovector current 
($k = V, \, k^{\, \prime} = A$) we obtain
%
%
\beq
\label{eq:Rvaupup}
&&{\cal R}^{++}_{AV} =  g_v g_a^{\, \prime} \bigg \{2\beta \sqrt{\ell \ell^{\, \prime}}
\left [(P \tilde \Lambda j^{\, \prime}) ({\cal K}_2 j) +
(P \tilde \Lambda j) ({\cal K}_2 j^{\, \prime}) - 
(j \tilde \Lambda j^{\, \prime}) ({\cal K}_2 P)  
\right.
\\[3mm]
\nonumber
&&\left. + m_f 
[(j \tilde \Lambda j^{\, \prime}){\cal K}_4 +  
(j \tilde \varphi j^{\, \prime}) {\cal K}_3 ]\right ] 
{\cal I}^{\, \prime}_{n, \ell'} {\cal I}_{n, \ell} - 
(M_{\ell} + m_f) (M_{\ell'} + m_f) \left [(P \tilde \Lambda j^{\, \prime}) ({\cal K}_2 j) +
(P \tilde \Lambda j) ({\cal K}_2 j^{\, \prime})  \right.
\\[3mm]
\nonumber
&&\left. - 
(j \tilde \Lambda j^{\, \prime}) ({\cal K}_2 P) - m_f
[(j \tilde \Lambda j^{\, \prime}){\cal K}_4 +
(j \tilde \varphi j^{\, \prime}) {\cal K}_3 ] \right ] 
{\cal I}^{\, \prime}_{n-1, \ell'-1} {\cal I}_{n-1, \ell-1} 
\\[3mm]
\nonumber
&&+
2\beta \sqrt{n} [(j \tilde \Lambda j^{\, \prime}){\cal K}_4 +
(j \tilde \varphi j^{\, \prime}) {\cal K}_3 ] 
\big [\sqrt{\ell} (M_{\ell'} + m_f)
{\cal I}^{\, \prime}_{n-1, \ell'-1} {\cal I}_{n, \ell} -
\sqrt{\ell^{\, \prime}} (M_{\ell} + m_f)
{\cal I}^{\, \prime}_{n, \ell'} {\cal I}_{n-1, \ell-1} \big ] 
\\[3mm]
\nonumber
&&-
\sqrt{\frac{2 \beta \ell}{q_{\mprp}^{\, \prime 2}}} \, 
(M_{\ell'} + m_f) [(q^{\, \prime} \Lambda j^{\, \prime}) + 
\ii (q^{\, \prime} \varphi j^{\, \prime})]  
[(P \tilde \Lambda j)  {\cal K}_4 - 
 (P \tilde \varphi j) {\cal K}_3 + m_f ({\cal K}_2 j)] 
{\cal I}^{\, \prime}_{n, \ell'-1} {\cal I}_{n, \ell} 
\\[3mm]
\nonumber
&&+
\sqrt{\frac{2 \beta \ell^{\, \prime}}{q_{\mprp}^2}} \, 
(M_{\ell} + m_f) [(q \Lambda j) - 
\ii (q \varphi j)] 
[(P \tilde \Lambda j^{\, \prime})  {\cal K}_4 + 
(P \tilde \varphi j^{\, \prime})  {\cal K}_3 + 
m_f ({\cal K}_2 j^{\, \prime})] 
{\cal I}^{\, \prime}_{n, \ell'} {\cal I}_{n, \ell-1} 
\\[3mm]
\nonumber
&&+ 
\sqrt{\frac{2 \beta \ell^{\, \prime}}{q_{\mprp}^2}} \, 
(M_{\ell} + m_f) [(q \Lambda j^{\, \prime}) - 
\ii (q \varphi j^{\, \prime})] 
[(P \tilde \Lambda j)  {\cal K}_4 - (P \tilde \varphi j)  {\cal K}_3 - 
m_f ({\cal K}_2 j)] 
{\cal I}^{\, \prime}_{n-1, \ell'} {\cal I}_{n-1, \ell-1} 
\\[3mm]
\nonumber
&&-
\sqrt{\frac{2 \beta \ell}{q_{\mprp}^{\, \prime 2}}} \, 
(M_{\ell'} + m_f) [(q^{\, \prime} \Lambda j) + 
\ii (q^{\, \prime} \varphi j)] 
 [(P \tilde \Lambda j^{\, \prime})  {\cal K}_4 + 
(P \tilde \varphi j^{\, \prime})  {\cal K}_3 - 
m_f ({\cal K}_2 j^{\, \prime})] 
{\cal I}^{\, \prime}_{n-1, \ell'-1} {\cal I}_{n-1, \ell} 
\\[3mm]
\nonumber
&&-
(M_{\ell} + m_f)(M_{\ell'} + m_f)   
[(j \Lambda j^{\, \prime}) + \ii 
(j \varphi j^{\, \prime})]\;[({\cal K}_2 P) + m_f {\cal K}_4] 
\frac{(q \Lambda q^{\, \prime}) - \ii (q \varphi q^{\, \prime})}
{\sqrt{q_{\mprp}^2  q_{\mprp}^{\, \prime 2}}} 
\; {\cal I}^{\, \prime}_{n, \ell'-1} {\cal I}_{n, \ell-1} 
\\[3mm]
\nonumber
&&+
2 \beta \sqrt{\ell \ell^{\, \prime}} \; [(j \Lambda j^{\, \prime}) - \ii 
(j \varphi j^{\, \prime})]  
[({\cal K}_2 P) - m_f {\cal K}_4] 
\frac{(q \Lambda q^{\, \prime}) + \ii (q \varphi q^{\, \prime})}
{\sqrt{q_{\mprp}^2  q_{\mprp}^{\, \prime 2}}} 
\; {\cal I}^{\, \prime}_{n-1, \ell'} {\cal I}_{n-1, \ell}  
\\[3mm]
\nonumber
&&-
\sqrt{2\beta n}\; ({\cal K}_2 j) \; \big [ 2\beta
\sqrt{\ell \ell^{\, \prime}} 
\; \frac{(q^{\, \prime} \Lambda j^{\, \prime}) - 
\ii (q^{\, \prime} \varphi j^{\, \prime})}
{\sqrt{q_{\mprp}^{\, \prime 2}}}  
{\cal I}^{\, \prime}_{n-1, \ell'} {\cal I}_{n, \ell}  - 
(M_{\ell} + m_f) (M_{\ell'} + m_f) \; \frac{(q^{\, \prime} \Lambda j^{\, \prime}) + 
\ii (q^{\, \prime} \varphi j^{\, \prime})}
{\sqrt{q_{\mprp}^{\, \prime 2}}} \;
{\cal I}^{\, \prime}_{n, \ell'-1} {\cal I}_{n-1, \ell-1} \big ] 
\\[3mm]
\nonumber
&&+  
\sqrt{2\beta n}\; ({\cal K}_2 j^{\, \prime}) 
\big [(M_{\ell} + m_f) (M_{\ell'} + m_f)   
\; \frac{(q \Lambda j) - 
\ii (q \varphi j)}{\sqrt{q_{\mprp}^2}}
{\cal I}^{\, \prime}_{n-1, \ell'-1} {\cal I}_{n, \ell-1} - 
2\beta \sqrt{\ell \ell^{\, \prime}}  \frac{(q \Lambda j) + 
\ii (q \varphi j)}
{\sqrt{q_{\mprp}^2}} \;
{\cal I}^{\, \prime}_{n, \ell'} {\cal I}_{n-1, \ell} \big ] 
\\[3mm]
\nonumber
&& - 2 \beta \sqrt{n} \; {\cal K}_4 \; \big [\sqrt{\ell^{\, \prime}}\; (M_{\ell} + m_f)  
\; \frac{(q \Lambda j) - 
\ii (q \varphi j)}{\sqrt{q_{\mprp}^2}} \; \frac{(q^{\, \prime} \Lambda j^{\, \prime}) - 
\ii (q^{\, \prime} \varphi j^{\, \prime})}{\sqrt{q_{\mprp}^{\, \prime 2}}} 
{\cal I}^{\, \prime}_{n-1, \ell'} {\cal I}_{n, \ell-1}  
\\[3mm]
\nonumber
&&- \sqrt{\ell}\; (M_{\ell'} + m_f) \; \frac{(q \Lambda j) +
\ii (q \varphi j)}{\sqrt{q_{\mprp}^2}} \; \frac{(q^{\, \prime} \Lambda j^{\, \prime}) +
\ii (q^{\, \prime} \varphi j^{\, \prime})}{\sqrt{q_{\mprp}^{\, \prime 2}}} 
{\cal I}^{\, \prime}_{n, \ell'-1} {\cal I}_{n-1, \ell} \big ]
\bigg \} \, ;
\eeq

\newpage

%
\beq
\label{eq:Rvaupup2}
&&{\cal R}^{++}_{VA} =  g_v g_a^{\, \prime} \bigg \{2\beta \sqrt{\ell \ell^{\, \prime}}
\left [(P^{\, \prime} \tilde \Lambda j^{\, \prime}) ({\cal K}_2 j) +
(P^{\, \prime} \tilde \Lambda j) ({\cal K}_2 j^{\, \prime}) - 
(j \tilde \Lambda j^{\, \prime}) ({\cal K}_2 P^{\, \prime})  
\right.
\\[3mm]
\nonumber
&&\left. - m_f 
[(j \tilde \Lambda j^{\, \prime}){\cal K}_4 -  
(j \tilde \varphi j^{\, \prime}) {\cal K}_3 ]\right ] 
{\cal I}_{n, \ell'} {\cal I}^{\, \prime}_{n, \ell} - 
(M_{\ell} + m_f) (M_{\ell'} + m_f) \left [(P^{\, \prime} \tilde \Lambda j^{\, \prime}) ({\cal K}_2 j) +
(P^{\, \prime} \tilde \Lambda j) ({\cal K}_2 j^{\, \prime})  \right.
\\[3mm]
\nonumber
&&\left. - 
(j \tilde \Lambda j^{\, \prime}) ({\cal K}_2 P^{\, \prime}) + m_f
[(j \tilde \Lambda j^{\, \prime}){\cal K}_4 -
(j \tilde \varphi j^{\, \prime}) {\cal K}_3 ]\right ] 
{\cal I}_{n-1, \ell'-1} {\cal I}^{\, \prime}_{n-1, \ell-1} 
\\[3mm]
\nonumber
&&+
2\beta \sqrt{n} [(j \tilde \Lambda j^{\, \prime}){\cal K}_4 -
(j \tilde \varphi j^{\, \prime}) {\cal K}_3 ] 
\big [\sqrt{\ell} (M_{\ell'} + m_f)
{\cal I}_{n-1, \ell'-1} {\cal I}^{\, \prime}_{n, \ell} -
\sqrt{\ell^{\, \prime}} (M_{\ell} + m_f)
{\cal I}_{n, \ell'} {\cal I}^{\, \prime}_{n-1, \ell-1} \big ] 
\\[3mm]
\nonumber
&&+
\sqrt{\frac{2 \beta \ell}{q_{\mprp}^{2}}} \, 
(M_{\ell'} + m_f) [(q \Lambda j) + 
\ii (q \varphi j)]  
[(P^{\, \prime} \tilde \Lambda j^{\, \prime})  {\cal K}_4 - 
 (P^{\, \prime} \tilde \varphi j^{\, \prime}) {\cal K}_3 - m_f ({\cal K}_2 j^{\, \prime})] 
{\cal I}_{n, \ell'-1} {\cal I}^{\, \prime}_{n, \ell} 
\\[3mm]
\nonumber
&&-
\sqrt{\frac{2 \beta \ell^{\, \prime}}{q_{\mprp}^{\, \prime 2}}} \, 
(M_{\ell} + m_f) [(q^{\, \prime} \Lambda j^{\, \prime}) - 
\ii (q^{\, \prime} \varphi j^{\, \prime})] 
[(P^{\, \prime} \tilde \Lambda j)  {\cal K}_4 + 
(P^{\, \prime} \tilde \varphi j)  {\cal K}_3 - 
m_f ({\cal K}_2 j)] 
{\cal I}_{n, \ell'} {\cal I}^{\, \prime}_{n, \ell-1} 
\\[3mm]
\nonumber
&&- 
\sqrt{\frac{2 \beta \ell^{\, \prime}}{q_{\mprp}^{\, \prime 2}}} \, 
(M_{\ell} + m_f) [(q^{\, \prime} \Lambda j) - 
\ii (q^{\, \prime} \varphi j)] 
[(P^{\, \prime} \tilde \Lambda j^{\, \prime})  {\cal K}_4 - 
(P^{\, \prime} \tilde \varphi j^{\, \prime})  {\cal K}_3 + 
m_f ({\cal K}_2 j^{\, \prime})] 
{\cal I}_{n-1, \ell'} {\cal I}^{\, \prime}_{n-1, \ell-1} 
\\[3mm]
\nonumber
&&+
\sqrt{\frac{2 \beta \ell}{q_{\mprp}^{2}}} \, 
(M_{\ell'} + m_f) [(q \Lambda j^{\, \prime}) + 
\ii (q \varphi j^{\, \prime})] 
 [(P^{\, \prime} \tilde \Lambda j)  {\cal K}_4 + 
(P^{\, \prime} \tilde \varphi j)  {\cal K}_3 + 
m_f ({\cal K}_2 j)] 
{\cal I}_{n-1, \ell'-1} {\cal I}^{\, \prime}_{n-1, \ell} 
\\[3mm]
\nonumber
&&-
(M_{\ell} + m_f)(M_{\ell'} + m_f)   
[(j \Lambda j^{\, \prime}) - \ii 
(j \varphi j^{\, \prime})]\;[({\cal K}_2 P^{\, \prime}) - m_f {\cal K}_4] 
\frac{(q \Lambda q^{\, \prime}) + \ii (q \varphi q^{\, \prime})}
{\sqrt{q_{\mprp}^2  q_{\mprp}^{\, \prime 2}}} 
\; {\cal I}_{n, \ell'-1} {\cal I}^{\, \prime}_{n, \ell-1} 
\\[3mm]
\nonumber
&&+
2 \beta \sqrt{\ell \ell^{\, \prime}} \; [(j \Lambda j^{\, \prime}) + \ii 
(j \varphi j^{\, \prime})]  
[({\cal K}_2 P^{\, \prime}) + m_f {\cal K}_4] 
\frac{(q \Lambda q^{\, \prime}) - \ii (q \varphi q^{\, \prime})}
{\sqrt{q_{\mprp}^2  q_{\mprp}^{\, \prime 2}}} 
\; {\cal I}_{n-1, \ell'} {\cal I}^{\, \prime}_{n-1, \ell}  
\\[3mm]
\nonumber
&&+
\sqrt{2\beta n}\; ({\cal K}_2 j^{\, \prime}) \; \big [ 2\beta
\sqrt{\ell \ell^{\, \prime}} 
\; \frac{(q \Lambda j) - 
\ii (q \varphi j)}
{\sqrt{q_{\mprp}^{2}}}  
{\cal I}_{n-1, \ell'} {\cal I}^{\, \prime}_{n, \ell}  - 
(M_{\ell} + m_f) (M_{\ell'} + m_f) \; \frac{(q \Lambda j) + 
\ii (q \varphi j)}
{\sqrt{q_{\mprp}^{2}}} \;
{\cal I}_{n, \ell'-1} {\cal I}^{\, \prime}_{n-1, \ell-1} \big ] 
\\[3mm]
\nonumber
&&-  
\sqrt{2\beta n}\; ({\cal K}_2 j) 
\big [(M_{\ell} + m_f) (M_{\ell'} + m_f)   
\; \frac{(q^{\, \prime} \Lambda j^{\, \prime}) - 
\ii (q^{\, \prime} \varphi j^{\, \prime})}{\sqrt{q_{\mprp}^{\, \prime 2}}}
{\cal I}_{n-1, \ell'-1} {\cal I}^{\, \prime}_{n, \ell-1} - 
2\beta \sqrt{\ell \ell^{\, \prime}}  \frac{(q^{\, \prime} \Lambda j^{\, \prime}) + 
\ii (q^{\, \prime} \varphi j^{\, \prime})}
{\sqrt{q_{\mprp}^{\, \prime 2}}} \;
{\cal I}_{n, \ell'} {\cal I}^{\, \prime}_{n-1, \ell} \big ] 
\\[3mm]
\nonumber
&& - 2 \beta \sqrt{n} \; {\cal K}_4 \; \big [\sqrt{\ell^{\, \prime}}\; (M_{\ell} + m_f)  
\; \frac{(q \Lambda j) - 
\ii (q \varphi j)}{\sqrt{q_{\mprp}^2}} \; \frac{(q^{\, \prime} \Lambda j^{\, \prime}) - 
\ii (q^{\, \prime} \varphi j^{\, \prime})}{\sqrt{q_{\mprp}^{\, \prime 2}}} 
{\cal I}_{n-1, \ell'} {\cal I}^{\, \prime}_{n, \ell-1}  
\\[3mm]
\nonumber
&&- \sqrt{\ell}\; (M_{\ell'} + m_f) \; \frac{(q \Lambda j) +
\ii (q \varphi j)}{\sqrt{q_{\mprp}^2}} \; \frac{(q^{\, \prime} \Lambda j^{\, \prime}) +
\ii (q^{\, \prime} \varphi j^{\, \prime})}{\sqrt{q_{\mprp}^{\, \prime 2}}} 
{\cal I}_{n, \ell'-1} {\cal I}^{\, \prime}_{n-1, \ell} \big ]
\bigg \} \, ;
\eeq

\newpage

\beq
\label{eq:Rvaupdown}
&&{\cal R}^{+-}_{AV} =  \ii g_v g_a^{\, \prime} \bigg \{
- \sqrt{2 \beta \ell^{\, \prime}} \, (M_{\ell} + m_f)
\left [(P \tilde \Lambda j^{\, \prime}) ({\cal K}_1 j) +
(P \tilde \Lambda j) ({\cal K}_1 j^{\, \prime}) - 
(j \tilde \Lambda j^{\, \prime}) ({\cal K}_1 P) 
\right.
\\
\nonumber
&&\left. - m_f 
[(j \tilde \Lambda j^{\, \prime}){\cal K}_3 +  
(j \tilde \varphi j^{\, \prime}) {\cal K}_4 ] \right ] 
{\cal I}^{\, \prime}_{n, \ell'} {\cal I}_{n, \ell} + 
 \sqrt{2 \beta \ell} \, (M_{\ell'} + m_f) 
\left [(P \tilde \Lambda j^{\, \prime}) ({\cal K}_1 j) +
(P \tilde \Lambda j) ({\cal K}_1 j^{\, \prime})  \right.
\\
\nonumber
&&\left. -
(j \tilde \Lambda j^{\, \prime}) ({\cal K}_1 P) - m_f
[(j \tilde \Lambda j^{\, \prime}){\cal K}_3 +
(j \tilde \varphi j^{\, \prime}) {\cal K}_4 ] \right ] 
{\cal I}^{\, \prime}_{n-1, \ell'-1} {\cal I}_{n-1, \ell-1}   
\\[3mm]
\nonumber
&&+
  \sqrt{2\beta n}\, [(j \tilde \Lambda j^{\, \prime}){\cal K}_3 +
(j \tilde \varphi j^{\, \prime}) {\cal K}_4 ] 
 \big [ (M_{\ell} + m_f) (M_{\ell'} + m_f)
{\cal I}^{\, \prime}_{n-1, \ell'-1} {\cal I}_{n, \ell} +
2 \beta \sqrt{\ell \ell^{\, \prime}}
{\cal I}^{\, \prime}_{n, \ell'} {\cal I}_{n-1, \ell-1} \big ] 
\\[3mm]
\nonumber
&&-
(M_{\ell} + m_f) (M_{\ell'} + m_f) \; \frac{(q^{\, \prime} \Lambda j^{\, \prime}) + 
\ii (q^{\, \prime} \varphi j^{\, \prime})}{\sqrt{q_{\mprp}^{\, \prime 2}}} \;   
[(P \tilde \Lambda j)  {\cal K}_3 - 
 (P \tilde \varphi j) {\cal K}_4 + m_f ({\cal K}_1 j)] 
{\cal I}^{\, \prime}_{n, \ell'-1} {\cal I}_{n, \ell} 
\\[3mm]
\nonumber
&&-
\frac{ 2 \beta \sqrt{\ell \ell^{\, \prime}}}{\sqrt{q_{\mprp}^2}} \, [(q \Lambda j) - 
\ii (q \varphi j)]\; [(P \tilde \Lambda j^{\, \prime})  {\cal K}_3 + 
 (P \tilde \varphi j^{\, \prime})  {\cal K}_4 + 
m_f ({\cal K}_1 j^{\, \prime})] 
{\cal I}^{\, \prime}_{n, \ell'} {\cal I}_{n, \ell-1} 
\\[3mm]
\nonumber
&&- 
\frac{ 2 \beta \sqrt{\ell \ell^{\, \prime}}}{\sqrt{q_{\mprp}^2}} \, 
 [(q \Lambda j^{\, \prime}) - 
\ii (q \varphi j^{\, \prime})]\; [(P \tilde \Lambda j)  {\cal K}_3 - 
 (P \tilde \varphi j)  {\cal K}_4 - 
m_f ({\cal K}_1 j)] 
{\cal I}^{\, \prime}_{n-1, \ell'} {\cal I}_{n-1, \ell-1}  
\\[3mm]
\nonumber
&&-
(M_{\ell} + m_f) (M_{\ell'} + m_f) 
\frac{(q^{\, \prime} \Lambda j) + 
\ii (q^{\, \prime} \varphi j)}{\sqrt{q_{\mprp}^{\, \prime 2}}} 
 [(P \tilde \Lambda j^{\, \prime})  {\cal K}_3 + 
(P \tilde \varphi j^{\, \prime})  {\cal K}_4 - 
m_f ({\cal K}_1 j^{\, \prime})] 
{\cal I}^{\, \prime}_{n-1, \ell'-1} {\cal I}_{n-1, \ell}  
\\[3mm]
\nonumber
&&+
 \sqrt{2 \beta \ell} \; (M_{\ell'} + m_f) [(j \Lambda j^{\, \prime}) + \ii 
(j \varphi j^{\, \prime})]  
[({\cal K}_1 P) + m_f {\cal K}_3] 
\frac{(q \Lambda q^{\, \prime}) - \ii (q \varphi q^{\, \prime})}
{\sqrt{q_{\mprp}^2  q_{\mprp}^{\, \prime 2}}} 
\; {\cal I}^{\, \prime}_{n, \ell'-1} {\cal I}_{n, \ell-1}   
\\[3mm]
\nonumber
&&+
 \sqrt{2 \beta \ell^{\, \prime}} \; (M_{\ell} + m_f) [(j \Lambda j^{\, \prime}) - \ii 
(j \varphi j^{\, \prime})] \; [({\cal K}_1 P) - m_f {\cal K}_3] 
\frac{(q \Lambda q^{\, \prime}) + \ii (q \varphi q^{\, \prime})}
{\sqrt{q_{\mprp}^2  q_{\mprp}^{\, \prime 2}}} 
\; {\cal I}^{\, \prime}_{n-1, \ell'} {\cal I}_{n-1, \ell}  
\\[3mm]
\nonumber
&&-
2\beta \sqrt{n}\; ({\cal K}_1 j) \; \big [\sqrt{\ell^{\, \prime}} \, (M_{\ell} + m_f)  
\; \frac{(q^{\, \prime} \Lambda j^{\, \prime}) - 
\ii (q^{\, \prime} \varphi j^{\, \prime})}
{\sqrt{q_{\mprp}^{\, \prime 2}}}{\cal I}^{\, \prime}_{n-1, \ell'} {\cal I}_{n, \ell} +
\sqrt{\ell} \, (M_{\ell'} + m_f) \; \frac{(q^{\, \prime} \Lambda j^{\, \prime}) + 
\ii (q^{\, \prime} \varphi j^{\, \prime})}
{\sqrt{q_{\mprp}^{\, \prime 2}}} \;
{\cal I}^{\, \prime}_{n, \ell'-1} {\cal I}_{n-1, \ell-1} \big ] 
\\[3mm]
\nonumber
&&-  
2\beta \sqrt{n}\; ({\cal K}_1 j^{\, \prime}) \; 
\big [\sqrt{\ell^{\, \prime}} \, (M_{\ell} + m_f)  
\; \frac{(q \Lambda j) - 
\ii (q \varphi j)}{\sqrt{q_{\mprp}^2}} 
{\cal I}^{\, \prime}_{n-1, \ell'-1} {\cal I}_{n, \ell-1} + 
\sqrt{\ell} \, (M_{\ell'} + m_f)  \frac{(q \Lambda j) + 
\ii (q \varphi j)}
{\sqrt{q_{\mprp}^2}} \;
{\cal I}^{\, \prime}_{n, \ell'} {\cal I}_{n-1, \ell} \big ] 
\\[3mm]
\nonumber
&&+
\sqrt{2 \beta n} \; {\cal K}_3 \; \big [2 \beta \sqrt{\ell \ell^{\, \prime}}  
\frac{(q \Lambda j) - 
\ii (q \varphi j)}{\sqrt{q_{\mprp}^2}} \; \frac{(q^{\, \prime} \Lambda j^{\, \prime}) - 
\ii (q^{\, \prime} \varphi j^{\, \prime})}{\sqrt{q_{\mprp}^{\, \prime 2}}} 
{\cal I}^{\, \prime}_{n-1, \ell'} {\cal I}_{n, \ell-1} 
\\[3mm]
\nonumber
&&+ (M_{\ell} + m_f) (M_{\ell'} + m_f) \; \frac{(q \Lambda j) +
\ii (q \varphi j)}{\sqrt{q_{\mprp}^2}} \; \frac{(q^{\, \prime} \Lambda j^{\, \prime}) +
\ii (q^{\, \prime} \varphi j^{\, \prime})}{\sqrt{q_{\mprp}^{\, \prime 2}}} 
{\cal I}^{\, \prime}_{n, \ell'-1} {\cal I}_{n-1, \ell} \big ]
\bigg \} \, ;
\eeq

\newpage
%

\beq
\label{eq:Rvaupdown2}
&&{\cal R}^{+-}_{VA} =  \ii g_v g_a^{\, \prime} \bigg \{
- \sqrt{2 \beta \ell^{\, \prime}} \, (M_{\ell} + m_f)
\left [(P^{\, \prime} \tilde \Lambda j^{\, \prime}) ({\cal K}_1 j) +
(P^{\, \prime} \tilde \Lambda j) ({\cal K}_1 j^{\, \prime}) - 
(j \tilde \Lambda j^{\, \prime}) ({\cal K}_1 P^{\, \prime}) 
\right.
\\
\nonumber
&&\left. + m_f 
[(j \tilde \Lambda j^{\, \prime}){\cal K}_3 -  
(j \tilde \varphi j^{\, \prime}) {\cal K}_4 ] \right ] 
{\cal I}_{n, \ell'} {\cal I}^{\, \prime}_{n, \ell} + 
 \sqrt{2 \beta \ell} \, (M_{\ell'} + m_f) 
\left [(P^{\, \prime} \tilde \Lambda j^{\, \prime}) ({\cal K}_1 j) +
(P^{\, \prime} \tilde \Lambda j) ({\cal K}_1 j^{\, \prime})  \right.
\\
\nonumber
&&\left. -
(j \tilde \Lambda j^{\, \prime}) ({\cal K}_1 P^{\, \prime}) + m_f
[(j \tilde \Lambda j^{\, \prime}){\cal K}_3 -
(j \tilde \varphi j^{\, \prime}) {\cal K}_4 ] \right ] 
{\cal I}_{n-1, \ell'-1} {\cal I}^{\, \prime}_{n-1, \ell-1}   
\\[3mm]
\nonumber
&&+
  \sqrt{2\beta n}\, [(j \tilde \Lambda j^{\, \prime}){\cal K}_3 -
(j \tilde \varphi j^{\, \prime}) {\cal K}_4 ] 
 \big [ (M_{\ell} + m_f) (M_{\ell'} + m_f)
{\cal I}_{n-1, \ell'-1} {\cal I}^{\, \prime}_{n, \ell} +
2 \beta \sqrt{\ell \ell^{\, \prime}}
{\cal I}_{n, \ell'} {\cal I}^{\, \prime}_{n-1, \ell-1} \big ] 
\\[3mm]
\nonumber
&&+
(M_{\ell} + m_f) (M_{\ell'} + m_f) \; \frac{(q \Lambda j) + 
\ii (q \varphi j)}{\sqrt{q_{\mprp}^{2}}} \;   
[(P^{\, \prime} \tilde \Lambda j^{\, \prime})  {\cal K}_3 - 
 (P^{\, \prime} \tilde \varphi j^{\, \prime}) {\cal K}_4 - m_f ({\cal K}_1 j^{\, \prime})] 
{\cal I}_{n, \ell'-1} {\cal I}^{\, \prime}_{n, \ell} 
\\[3mm]
\nonumber
&&+
\frac{ 2 \beta \sqrt{\ell \ell^{\, \prime}}}{\sqrt{q_{\mprp}^{\, \prime 2}}} \, 
[(q^{\, \prime} \Lambda j^{\, \prime}) - 
\ii (q^{\, \prime} \varphi j^{\, \prime})]\; [(P^{\, \prime} \tilde \Lambda j)  {\cal K}_3 + 
 (P^{\, \prime} \tilde \varphi j)  {\cal K}_4 - 
m_f ({\cal K}_1 j)] 
{\cal I}_{n, \ell'} {\cal I}^{\, \prime}_{n, \ell-1} 
\\[3mm]
\nonumber
&&+ 
\frac{ 2 \beta \sqrt{\ell \ell^{\, \prime}}}{\sqrt{q_{\mprp}^{\, \prime 2}}} \, 
 [(q^{\, \prime} \Lambda j) - 
\ii (q^{\, \prime} \varphi j)]\; [(P^{\, \prime} \tilde \Lambda j^{\, \prime})  {\cal K}_3 - 
 (P^{\, \prime} \tilde \varphi j^{\, \prime})  {\cal K}_4 + 
m_f ({\cal K}_1 j^{\, \prime})] 
{\cal I}_{n-1, \ell'} {\cal I}^{\, \prime}_{n-1, \ell-1}  
\\[3mm]
\nonumber
&&+
(M_{\ell} + m_f) (M_{\ell'} + m_f) 
\frac{(q \Lambda j^{\, \prime}) + 
\ii (q \varphi j^{\, \prime})}{\sqrt{q_{\mprp}^{2}}} 
 [(P^{\, \prime} \tilde \Lambda j)  {\cal K}_3 + 
(P^{\, \prime} \tilde \varphi j)  {\cal K}_4 + 
m_f ({\cal K}_1 j)] 
{\cal I}_{n-1, \ell'-1} {\cal I}^{\, \prime}_{n-1, \ell}  
\\[3mm]
\nonumber
&&+
 \sqrt{2 \beta \ell} \; (M_{\ell'} + m_f) [(j \Lambda j^{\, \prime}) - \ii 
(j \varphi j^{\, \prime})]  
[({\cal K}_1 P^{\, \prime}) - m_f {\cal K}_3] 
\frac{(q \Lambda q^{\, \prime}) + \ii (q \varphi q^{\, \prime})}
{\sqrt{q_{\mprp}^2  q_{\mprp}^{\, \prime 2}}} 
\; {\cal I}_{n, \ell'-1} {\cal I}^{\, \prime}_{n, \ell-1}   
\\[3mm]
\nonumber
&&+
 \sqrt{2 \beta \ell^{\, \prime}} \; (M_{\ell} + m_f) [(j \Lambda j^{\, \prime}) + \ii 
(j \varphi j^{\, \prime})] \; [({\cal K}_1 P^{\, \prime}) + m_f {\cal K}_3] 
\frac{(q \Lambda q^{\, \prime}) - \ii (q \varphi q^{\, \prime})}
{\sqrt{q_{\mprp}^2  q_{\mprp}^{\, \prime 2}}} 
\; {\cal I}_{n-1, \ell'} {\cal I}^{\, \prime}_{n-1, \ell}  
\\[3mm]
\nonumber
&&+
2\beta \sqrt{n}\; ({\cal K}_1 j^{\, \prime}) \; \big [\sqrt{\ell^{\, \prime}} \, (M_{\ell} + m_f)  
\; \frac{(q \Lambda j) - 
\ii (q \varphi j)}
{\sqrt{q_{\mprp}^{2}}}{\cal I}_{n-1, \ell'} {\cal I}^{\, \prime}_{n, \ell} +
\sqrt{\ell} \, (M_{\ell'} + m_f) \; \frac{(q \Lambda j) + 
\ii (q \varphi j)}
{\sqrt{q_{\mprp}^{2}}} \;
{\cal I}_{n, \ell'-1} {\cal I}^{\, \prime}_{n-1, \ell-1} \big ] 
\\[3mm]
\nonumber
&&+  
2\beta \sqrt{n}\; ({\cal K}_1 j) \; 
\big [\sqrt{\ell^{\, \prime}} \, (M_{\ell} + m_f)  
\; \frac{(q^{\, \prime} \Lambda j^{\, \prime}) - 
\ii (q^{\, \prime} \varphi j^{\, \prime})}{\sqrt{q_{\mprp}^{\, \prime 2}}} 
{\cal I}_{n-1, \ell'-1} {\cal I}^{\, \prime}_{n, \ell-1} + 
\sqrt{\ell} \, (M_{\ell'} + m_f)  \frac{(q^{\, \prime} \Lambda j^{\, \prime}) + 
\ii (q^{\, \prime} \varphi j^{\, \prime})}
{\sqrt{q_{\mprp}^{\, \prime 2}}} \;
{\cal I}_{n, \ell'} {\cal I}^{\, \prime}_{n-1, \ell} \big ] 
\\[3mm]
\nonumber
&&+
\sqrt{2 \beta n} \; {\cal K}_3 \; \big [2 \beta \sqrt{\ell \ell^{\, \prime}}  
\frac{(q \Lambda j) - 
\ii (q \varphi j)}{\sqrt{q_{\mprp}^2}} \; \frac{(q^{\, \prime} \Lambda j^{\, \prime}) - 
\ii (q^{\, \prime} \varphi j^{\, \prime})}{\sqrt{q_{\mprp}^{\, \prime 2}}} 
{\cal I}_{n-1, \ell'} {\cal I}^{\, \prime}_{n, \ell-1} 
\\[3mm]
\nonumber
&&+ (M_{\ell} + m_f) (M_{\ell'} + m_f) \; \frac{(q \Lambda j) +
\ii (q \varphi j)}{\sqrt{q_{\mprp}^2}} \; \frac{(q^{\, \prime} \Lambda j^{\, \prime}) +
\ii (q^{\, \prime} \varphi j^{\, \prime})}{\sqrt{q_{\mprp}^{\, \prime 2}}} 
{\cal I}_{n, \ell'-1} {\cal I}^{\, \prime}_{n-1, \ell} \big ]
\bigg \} \, ;
\eeq

\newpage
%

\beq
\label{eq:Rvadownup}
&&{\cal R}^{-+}_{AV} = \ii g_v g_a^{\, \prime} \bigg \{
-  \sqrt{2 \beta \ell} \, (M_{\ell} + m_f^{\, \prime})
\left [(P \tilde \Lambda j^{\, \prime}) ({\cal K}_1 j) +
(P \tilde \Lambda j) ({\cal K}_1 j^{\, \prime}) - 
(j \tilde \Lambda j^{\, \prime}) ({\cal K}_1 P)  
\right.
\\
\nonumber
&&\left.- m_f 
[(j \tilde \Lambda j^{\, \prime}){\cal K}_3 +  
(j \tilde \varphi j^{\, \prime}) {\cal K}_4 ]\right ] 
{\cal I}^{\, \prime}_{n, \ell'} {\cal I}_{n, \ell} - 
 \sqrt{2 \beta \ell^{\, \prime}} \, (M_{\ell} + m_f) 
\left [(P \tilde \Lambda j^{\, \prime}) ({\cal K}_1 j) +
(P \tilde \Lambda j) ({\cal K}_1 j^{\, \prime})  \right.
\\
\nonumber
&&\left.
- (j \tilde \Lambda j^{\, \prime}) ({\cal K}_1 P) + m_f
[(j \tilde \Lambda j^{\, \prime}){\cal K}_3 +
(j \tilde \varphi j^{\, \prime}) {\cal K}_4 ] \right ] 
{\cal I}^{\, \prime}_{n-1, \ell'-1} {\cal I}_{n-1, \ell-1} 
\\[3mm]
\nonumber
&&-
\sqrt{2\beta n}\, [(j \tilde \Lambda j^{\, \prime}){\cal K}_3 +
(j \tilde \varphi j^{\, \prime}) {\cal K}_4 ] 
\big [2 \beta \sqrt{\ell \ell^{\, \prime}} 
{\cal I}^{\, \prime}_{n-1, \ell'-1} {\cal I}_{n, \ell} +
(M_{\ell} + m_f) (M_{\ell'} + m_f)
{\cal I}^{\, \prime}_{n, \ell'} {\cal I}_{n-1, \ell-1} \big ] 
\\[3mm]
\nonumber
&&+
\frac{ 2 \beta \sqrt{\ell \ell^{\, \prime}}}{\sqrt{q_{\mprp}^{\, \prime 2}}} \, 
 [(q^{\, \prime} \Lambda j^{\, \prime}) + 
\ii (q^{\, \prime} \varphi j^{\, \prime})]  
[(P \tilde \Lambda j) {\cal K}_3 - 
(P \tilde \varphi j)  {\cal K}_4 - m_f ({\cal K}_1 j)] 
{\cal I}^{\, \prime}_{n, \ell'-1} {\cal I}_{n, \ell} 
\\[3mm]
\nonumber
&&+
(M_{\ell} + m_f) (M_{\ell'} + m_f) \; \frac{(q \Lambda j) - 
\ii (q \varphi j)}{\sqrt{q_{\mprp}^2}} \;  
[(P \tilde \Lambda j^{\, \prime})  {\cal K}_3 + 
(P \tilde \varphi j^{\, \prime})  {\cal K}_4 - 
m_f ({\cal K}_1 j^{\, \prime})] 
{\cal I}^{\, \prime}_{n, \ell'} {\cal I}_{n, \ell-1}  
\\[3mm]
\nonumber
&&+ 
 (M_{\ell} + m_f) (M_{\ell'} + m_f) 
\frac{(q \Lambda j^{\, \prime}) - \ii (q \varphi j^{\, \prime})}{\sqrt{q_{\mprp}^2}}\; 
[(P \tilde \Lambda j)  {\cal K}_3 - 
(P \tilde \varphi j)  {\cal K}_4 + 
m_f ({\cal K}_1 j)] 
{\cal I}^{\, \prime}_{n-1, \ell'} {\cal I}_{n-1, \ell-1}  
\\ [3mm]
\nonumber
&& +
\frac{ 2 \beta \sqrt{\ell \ell^{\, \prime}}}{\sqrt{q_{\mprp}^{\, \prime 2}}} 
[(q^{\, \prime} \Lambda j) + 
\ii (q^{\, \prime} \varphi j)]\; 
[(P \tilde \Lambda j^{\, \prime}) {\cal K}_3 + 
(P \tilde \varphi j^{\, \prime})  {\cal K}_4 + 
m_f ({\cal K}_1 j^{\, \prime})] 
{\cal I}^{\, \prime}_{n-1, \ell'-1} {\cal I}_{n-1, \ell}  
\\[3mm]
\nonumber
&& -
\sqrt{2 \beta \ell^{\, \prime}} 
(M_{\ell} + m_f) [(j \Lambda j^{\, \prime}) + \ii 
(j \varphi j^{\, \prime})] 
[({\cal K}_1 P) - m_f {\cal K}_3] 
\frac{(q \Lambda q^{\, \prime}) - \ii (q \varphi q^{\, \prime})}
{\sqrt{q_{\mprp}^2  q_{\mprp}^{\, \prime 2}}} 
\; {\cal I}^{\, \prime}_{n, \ell'-1} {\cal I}_{n, \ell-1} 
\\[3mm]
\nonumber
&&-
 \sqrt{2 \beta \ell} \; (M_{\ell'} + m_f) 
[(j \Lambda j^{\, \prime}) - \ii 
(j \varphi j^{\, \prime})] \; [({\cal K}_1 P) + m_f {\cal K}_3] 
\frac{(q \Lambda q^{\, \prime}) + \ii (q \varphi q^{\, \prime})}
{\sqrt{q_{\mprp}^2  q_{\mprp}^{\, \prime 2}}} 
\; {\cal I}^{\, \prime}_{n-1, \ell'} {\cal I}_{n-1, \ell}  
\\[3mm]
\nonumber
&&+
 2\beta \sqrt{n}\; ({\cal K}_1 j) \; \big [\sqrt{\ell} \, (M_{\ell'} + m_f) 
\frac{(q^{\, \prime} \Lambda j^{\, \prime}) - 
\ii (q^{\, \prime} \varphi j^{\, \prime})}
{\sqrt{q_{\mprp}^{\, \prime 2}}}
{\cal I}^{\, \prime}_{n-1, \ell'} {\cal I}_{n, \ell} +
\sqrt{\ell^{\, \prime}} \, (M_{\ell} + m_f) \; \frac{(q^{\, \prime} \Lambda j^{\, \prime}) + 
\ii (q^{\, \prime} \varphi j^{\, \prime})}
{\sqrt{q_{\mprp}^{\, \prime 2}}} \;
{\cal I}^{\, \prime}_{n, \ell'-1} {\cal I}_{n-1, \ell-1} \big ] 
\\[3mm]
\nonumber
&&+ 
 2\beta \sqrt{n}\; ({\cal K}_1 j^{\, \prime})  
\big [\sqrt{\ell} \, (M_{\ell'} + m_f)  
\; \frac{(q \Lambda j) - 
\ii (q \varphi j)}{\sqrt{q_{\mprp}^2}}
{\cal I}^{\, \prime}_{n-1, \ell'-1} {\cal I}_{n, \ell-1} + 
\sqrt{\ell^{\, \prime}} \, (M_{\ell} + m_f)  \frac{(q \Lambda j) + 
\ii (q \varphi j)}
{\sqrt{q_{\mprp}^2}} \;
{\cal I}^{\, \prime}_{n, \ell'} {\cal I}_{n-1, \ell} \big ] 
\\[3mm]
\nonumber
&&- \sqrt{2 \beta n} \; {\cal K}_3 \; \big [(M_{\ell} + m_f) (M_{\ell'} + m_f)  
\; \frac{(q \Lambda j) - 
\ii (q \varphi j)}{\sqrt{q_{\mprp}^2}} \; \frac{(q^{\, \prime} \Lambda j^{\, \prime}) - 
\ii (q^{\, \prime} \varphi j^{\, \prime})}{\sqrt{q_{\mprp}^{\, \prime 2}}} 
{\cal I}^{\, \prime}_{n-1, \ell'} {\cal I}_{n, \ell-1}  
\\[3mm]
\nonumber
&&+2 \beta \sqrt{\ell \ell^{\, \prime}} \; \frac{(q \Lambda j) +
\ii (q \varphi j)}{\sqrt{q_{\mprp}^2}} \; \frac{(q^{\, \prime} \Lambda j^{\, \prime}) +
\ii (q^{\, \prime} \varphi j^{\, \prime})}{\sqrt{q_{\mprp}^{\, \prime 2}}} 
{\cal I}^{\, \prime}_{n, \ell'-1} {\cal I}_{n-1, \ell} \big ]
\bigg \} \, ;
\eeq

\newpage

\beq
\label{eq:Rvadownu2}
&&{\cal R}^{-+}_{VA} = \ii g_v g_a^{\, \prime} \bigg \{
-  \sqrt{2 \beta \ell} \, (M_{\ell} + m_f^{\, \prime})
\left [(P^{\, \prime} \tilde \Lambda j^{\, \prime}) ({\cal K}_1 j) +
(P^{\, \prime} \tilde \Lambda j) ({\cal K}_1 j^{\, \prime}) - 
(j \tilde \Lambda j^{\, \prime}) ({\cal K}_1 P^{\, \prime})  
\right.
\\
\nonumber
&&\left.+ m_f 
[(j \tilde \Lambda j^{\, \prime}){\cal K}_3 -  
(j \tilde \varphi j^{\, \prime}) {\cal K}_4 ]\right ] 
{\cal I}_{n, \ell'} {\cal I}^{\, \prime}_{n, \ell} - 
 \sqrt{2 \beta \ell^{\, \prime}} \, (M_{\ell} + m_f) 
\left [(P^{\, \prime} \tilde \Lambda j^{\, \prime}) ({\cal K}_1 j) +
(P^{\, \prime} \tilde \Lambda j) ({\cal K}_1 j^{\, \prime})  \right.
\\
\nonumber
&&\left.
- (j \tilde \Lambda j^{\, \prime}) ({\cal K}_1 P^{\, \prime}) - m_f
[(j \tilde \Lambda j^{\, \prime}){\cal K}_3 -
(j \tilde \varphi j^{\, \prime}) {\cal K}_4 ] \right ] 
{\cal I}_{n-1, \ell'-1} {\cal I}^{\, \prime}_{n-1, \ell-1} 
\\[3mm]
\nonumber
&&-
\sqrt{2\beta n}\, [(j \tilde \Lambda j^{\, \prime}){\cal K}_3 -
(j \tilde \varphi j^{\, \prime}) {\cal K}_4 ] 
\big [2 \beta \sqrt{\ell \ell^{\, \prime}} 
{\cal I}_{n-1, \ell'-1} {\cal I}^{\, \prime}_{n, \ell} +
(M_{\ell} + m_f) (M_{\ell'} + m_f)
{\cal I}_{n, \ell'} {\cal I}^{\, \prime}_{n-1, \ell-1} \big ] 
\\[3mm]
\nonumber
&&-
\frac{ 2 \beta \sqrt{\ell \ell^{\, \prime}}}{\sqrt{q_{\mprp}^{2}}} \, 
 [(q \Lambda j) + 
\ii (q \varphi j)]  
[(P^{\, \prime} \tilde \Lambda j^{\, \prime}) {\cal K}_3 - 
(P^{\, \prime} \tilde \varphi j^{\, \prime})  {\cal K}_4 + m_f ({\cal K}_1 j^{\, \prime})] 
{\cal I}_{n, \ell'-1} {\cal I}^{\, \prime}_{n, \ell} 
\\[3mm]
\nonumber
&&-
(M_{\ell} + m_f) (M_{\ell'} + m_f) \; \frac{(q^{\, \prime} \Lambda j^{\, \prime}) - 
\ii (q^{\, \prime} \varphi j^{\, \prime})}{\sqrt{q_{\mprp}^{\, \prime 2}}} \;  
[(P^{\, \prime} \tilde \Lambda j)  {\cal K}_3 + 
(P^{\, \prime} \tilde \varphi j)  {\cal K}_4 + 
m_f ({\cal K}_1 j)] 
{\cal I}_{n, \ell'} {\cal I}^{\, \prime}_{n, \ell-1}  
\\[3mm]
\nonumber
&&- 
 (M_{\ell} + m_f) (M_{\ell'} + m_f) 
\frac{(q^{\, \prime} \Lambda j) - \ii (q^{\, \prime} \varphi j)}{\sqrt{q_{\mprp}^{\, \prime 2}}}\; 
[(P^{\, \prime} \tilde \Lambda j^{\, \prime})  {\cal K}_3 - 
(P^{\, \prime} \tilde \varphi j^{\, \prime})  {\cal K}_4 - 
m_f ({\cal K}_1 j^{\, \prime})] 
{\cal I}_{n-1, \ell'} {\cal I}^{\, \prime}_{n-1, \ell-1}  
\\ [3mm]
\nonumber
&& -
\frac{ 2 \beta \sqrt{\ell \ell^{\, \prime}}}{\sqrt{q_{\mprp}^{2}}} 
[(q \Lambda j^{\, \prime}) + 
\ii (q \varphi j^{\, \prime})]\; 
[(P^{\, \prime} \tilde \Lambda j) {\cal K}_3 + 
(P^{\, \prime} \tilde \varphi j)  {\cal K}_4 - 
m_f ({\cal K}_1 j)] 
{\cal I}_{n-1, \ell'-1} {\cal I}^{\, \prime}_{n-1, \ell}  
\\[3mm]
\nonumber
&& -
\sqrt{2 \beta \ell^{\, \prime}} 
(M_{\ell} + m_f) [(j \Lambda j^{\, \prime}) - \ii 
(j \varphi j^{\, \prime})] 
[({\cal K}_1 P^{\, \prime}) + m_f {\cal K}_3] 
\frac{(q \Lambda q^{\, \prime}) + \ii (q \varphi q^{\, \prime})}
{\sqrt{q_{\mprp}^2  q_{\mprp}^{\, \prime 2}}} 
\; {\cal I}_{n, \ell'-1} {\cal I}^{\, \prime}_{n, \ell-1} 
\\[3mm]
\nonumber
&&-
 \sqrt{2 \beta \ell} \; (M_{\ell'} + m_f) 
[(j \Lambda j^{\, \prime}) + \ii 
(j \varphi j^{\, \prime})] \; [({\cal K}_1 P^{\, \prime}) - m_f {\cal K}_3] 
\frac{(q \Lambda q^{\, \prime}) - \ii (q \varphi q^{\, \prime})}
{\sqrt{q_{\mprp}^2  q_{\mprp}^{\, \prime 2}}} 
\; {\cal I}_{n-1, \ell'} {\cal I}^{\, \prime}_{n-1, \ell}  
\\[3mm]
\nonumber
&&-
 2\beta \sqrt{n}\; ({\cal K}_1 j^{\, \prime}) \; \big [\sqrt{\ell} \, (M_{\ell'} + m_f) 
\frac{(q \Lambda j) - 
\ii (q \varphi j)}
{\sqrt{q_{\mprp}^{2}}}
{\cal I}_{n-1, \ell'} {\cal I}^{\, \prime}_{n, \ell} +
\sqrt{\ell^{\, \prime}} \, (M_{\ell} + m_f) \; \frac{(q \Lambda j) + 
\ii (q \varphi j)}
{\sqrt{q_{\mprp}^{2}}} \;
{\cal I}_{n, \ell'-1} {\cal I}^{\, \prime}_{n-1, \ell-1} \big ] 
\\[3mm]
\nonumber
&&- 
 2\beta \sqrt{n}\; ({\cal K}_1 j)  
\big [\sqrt{\ell} \, (M_{\ell'} + m_f)  
\; \frac{(q^{\, \prime} \Lambda j^{\, \prime}) - 
\ii (q^{\, \prime} \varphi j^{\, \prime})}{\sqrt{q_{\mprp}^{\, \prime 2}}}
{\cal I}_{n-1, \ell'-1} {\cal I}^{\, \prime}_{n, \ell-1} + 
\sqrt{\ell^{\, \prime}} \, (M_{\ell} + m_f)  \frac{(q^{\, \prime} \Lambda j^{\, \prime}) + 
\ii (q^{\, \prime} \varphi j^{\, \prime})}
{\sqrt{q_{\mprp}^{\, \prime 2}}} \;
{\cal I}_{n, \ell'} {\cal I}^{\, \prime}_{n-1, \ell} \big ] 
\\[3mm]
\nonumber
&&- \sqrt{2 \beta n} \; {\cal K}_3 \; \big [(M_{\ell} + m_f) (M_{\ell'} + m_f)  
\; \frac{(q \Lambda j) - 
\ii (q \varphi j)}{\sqrt{q_{\mprp}^2}} \; \frac{(q^{\, \prime} \Lambda j^{\, \prime}) - 
\ii (q^{\, \prime} \varphi j^{\, \prime})}{\sqrt{q_{\mprp}^{\, \prime 2}}} 
{\cal I}_{n-1, \ell'} {\cal I}^{\, \prime}_{n, \ell-1}  
\\[3mm]
\nonumber
&&+2 \beta \sqrt{\ell \ell^{\, \prime}} \; \frac{(q \Lambda j) +
\ii (q \varphi j)}{\sqrt{q_{\mprp}^2}} \; \frac{(q^{\, \prime} \Lambda j^{\, \prime}) +
\ii (q^{\, \prime} \varphi j^{\, \prime})}{\sqrt{q_{\mprp}^{\, \prime 2}}} 
{\cal I}_{n, \ell'-1} {\cal I}^{\, \prime}_{n-1, \ell} \big ]
\bigg \} \, ;
\eeq

\newpage

%
\beq
\label{eq:Rvadowndown}
&&{\cal R}^{--}_{AV} =  g_v g_a^{\, \prime} \bigg \{(M_{\ell} + m_f) (M_{\ell'} + m_f)
\left [(P \tilde \Lambda j^{\, \prime}) ({\cal K}_2 j) +
(P \tilde \Lambda j) ({\cal K}_2 j^{\, \prime}) - 
(j \tilde \Lambda j^{\, \prime}) ({\cal K}_2 P)  
\right.
\\
\nonumber
&&\left. - m_f 
[(j \tilde \Lambda j^{\, \prime}){\cal K}_4 +  
(j \tilde \varphi j^{\, \prime}) {\cal K}_3 ]\right ] 
{\cal I}^{\, \prime}_{n, \ell'} {\cal I}_{n, \ell} - 
2\beta \sqrt{\ell \ell^{\, \prime}} \left [(P \tilde \Lambda j^{\, \prime}) ({\cal K}_2 j) +
(P \tilde \Lambda j) ({\cal K}_2 j^{\, \prime})  \right.
\\
\nonumber
&&\left. - 
(j \tilde \Lambda j^{\, \prime}) ({\cal K}_2 P) + m_f
[(j \tilde \Lambda j^{\, \prime}){\cal K}_4 +
(j \tilde \varphi j^{\, \prime}) {\cal K}_3 ] \right ] 
{\cal I}^{\, \prime}_{n-1, \ell'-1} {\cal I}_{n-1, \ell-1} +
2\beta \sqrt{n} [(j \tilde \Lambda j^{\, \prime}){\cal K}_4 +
(j \tilde \varphi j^{\, \prime}) {\cal K}_3 ]  
\\[3mm]
\nonumber
&&\times \big [\sqrt{\ell^{\, \prime}} (M_{\ell} + m_f)
{\cal I}^{\, \prime}_{n-1, \ell'-1} {\cal I}_{n, \ell} -
\sqrt{\ell} (M_{\ell'} + m_f)
{\cal I}^{\, \prime}_{n, \ell'} {\cal I}_{n-1, \ell-1} \big ] 
\\[3mm]
\nonumber
&& 
-
\sqrt{\frac{2 \beta \ell^{\, \prime}}{q_{\mprp}^{\, \prime 2}}} \, 
(M_{\ell} + m_f) [(q^{\, \prime} \Lambda j^{\, \prime}) + 
\ii (q^{\, \prime} \varphi j^{\, \prime})]
[(P \tilde \Lambda j)  {\cal K}_4 - 
(P \tilde \varphi j)  {\cal K}_3 - m_f ({\cal K}_2 j)] 
{\cal I}^{\, \prime}_{n, \ell'-1} {\cal I}_{n, \ell} 
\\[3mm]
\nonumber
&&+
\sqrt{\frac{2 \beta \ell}{q_{\mprp}^2}} \, 
(M_{\ell'} + m_f) [(q \Lambda j) - 
\ii (q \varphi j)]  
[(P \tilde \Lambda j^{\, \prime})  {\cal K}_4 + 
(P \tilde \varphi j^{\, \prime})  {\cal K}_3 - 
m_f ({\cal K}_2 j^{\, \prime})] 
{\cal I}^{\, \prime}_{n, \ell'} {\cal I}_{n, \ell-1} 
\\[3mm]
\nonumber
&&+ 
\sqrt{\frac{2 \beta \ell}{q_{\mprp}^2}} \, 
(M_{\ell'} + m_f) 
[(q \Lambda j^{\, \prime}) - 
\ii (q \varphi j^{\, \prime})]\; [(P \tilde \Lambda j)  {\cal K}_4 - 
(P \tilde \varphi j)  {\cal K}_3 + 
m_f ({\cal K}_2 j)] 
{\cal I}^{\, \prime}_{n-1, \ell'} {\cal I}_{n-1, \ell-1} 
\\[3mm]
\nonumber
&& -
\sqrt{\frac{2 \beta \ell^{\, \prime}}{q_{\mprp}^{\, \prime 2}}} \, 
(M_{\ell} + m_f)  
[(q^{\, \prime} \Lambda j) + 
\ii (q^{\, \prime} \varphi j)]\; 
[(P \tilde \Lambda j^{\, \prime})  {\cal K}_4 + 
(P \tilde \varphi j^{\, \prime})  {\cal K}_3 + 
m_f ({\cal K}_2 j^{\, \prime})] 
{\cal I}^{\, \prime}_{n-1, \ell'-1} {\cal I}_{n-1, \ell} 
\\[3mm]
\nonumber
&&-
2 \beta \sqrt{\ell \ell^{\, \prime}}  
[(j \Lambda j^{\, \prime}) + \ii 
(j \varphi j^{\, \prime})] \; [({\cal K}_2 P) - m_f {\cal K}_4] 
\frac{(q \Lambda q^{\, \prime}) - \ii (q \varphi q^{\, \prime})}
{\sqrt{q_{\mprp}^2  q_{\mprp}^{\, \prime 2}}} 
\; {\cal I}^{\, \prime}_{n, \ell'-1} {\cal I}_{n, \ell-1} \\[3mm]
\nonumber
&&+
(M_{\ell} + m_f)(M_{\ell'} + m_f) \; [(j \Lambda j^{\, \prime}) - \ii 
(j \varphi j^{\, \prime})]  
[({\cal K}_2 P) + m_f {\cal K}_4] 
\frac{(q \Lambda q^{\, \prime}) + \ii (q \varphi q^{\, \prime})}
{\sqrt{q_{\mprp}^2  q_{\mprp}^{\, \prime 2}}} 
\; {\cal I}^{\, \prime}_{n-1, \ell'} {\cal I}_{n-1, \ell} 
\\[3mm]
\nonumber
&&-
\sqrt{2\beta n}\; ({\cal K}_2 j) \; \big [(M_{\ell} + m_f) (M_{\ell'} + m_f)  
\; \frac{(q^{\, \prime} \Lambda j^{\, \prime}) - 
\ii (q^{\, \prime} \varphi j^{\, \prime})}
{\sqrt{q_{\mprp}^{\, \prime 2}}}  
{\cal I}^{\, \prime}_{n-1, \ell'} {\cal I}_{n, \ell} - 
2\beta
\sqrt{\ell \ell^{\, \prime}} \; \frac{(q^{\, \prime} \Lambda j^{\, \prime}) + 
\ii (q^{\, \prime} \varphi j^{\, \prime})}
{\sqrt{q_{\mprp}^{\, \prime 2}}} \;
{\cal I}^{\, \prime}_{n, \ell'-1} {\cal I}_{n-1, \ell-1} \big ] 
\\[3mm]
\nonumber
&& +  
\sqrt{2\beta n}\; ({\cal K}_2 j^{\, \prime}) 
\big [ 2\beta \sqrt{\ell \ell^{\, \prime}} \; 
 \frac{(q \Lambda j) - 
\ii (q \varphi j)}{\sqrt{q_{\mprp}^2}}
{\cal I}^{\, \prime}_{n-1, \ell'-1} {\cal I}_{n, \ell-1} - 
(M_{\ell} + m_f) (M_{\ell'} + m_f)  \frac{(q \Lambda j) + 
\ii (q \varphi j)}
{\sqrt{q_{\mprp}^2}} \;
{\cal I}^{\, \prime}_{n, \ell'} {\cal I}_{n-1, \ell} \big ] 
\\[3mm]
\nonumber
&&-
2 \beta \sqrt{n} \; {\cal K}_4  
  \big [\sqrt{\ell}\; (M_{\ell'} + m_f)  
\; \frac{(q \Lambda j) - 
\ii (q \varphi j)}{\sqrt{q_{\mprp}^2}} \; \frac{(q^{\, \prime} \Lambda j^{\, \prime}) - 
\ii (q^{\, \prime} \varphi j^{\, \prime})}{\sqrt{q_{\mprp}^{\, \prime 2}}}  
{\cal I}^{\, \prime}_{n-1, \ell'} {\cal I}_{n, \ell-1} 
\\[3mm]
\nonumber
&&- 
\sqrt{\ell^{\, \prime}}\; (M_{\ell} + m_f) \; \frac{(q \Lambda j) +
\ii (q \varphi j)}{\sqrt{q_{\mprp}^2}} \;  
\frac{(q^{\, \prime} \Lambda j^{\, \prime}) +
\ii (q^{\, \prime} \varphi j^{\, \prime})}{\sqrt{q_{\mprp}^{\, \prime 2}}} 
{\cal I}^{\, \prime}_{n, \ell'-1} {\cal I}_{n-1, \ell} \big ]
\bigg \} \, ;
\eeq

\newpage

%
\beq
\label{eq:Rvadowndown2}
&&{\cal R}^{--}_{VA} =  g_v g_a^{\, \prime} \bigg \{(M_{\ell} + m_f) (M_{\ell'} + m_f)
\left [(P^{\, \prime} \tilde \Lambda j^{\, \prime}) ({\cal K}_2 j) +
(P^{\, \prime} \tilde \Lambda j) ({\cal K}_2 j^{\, \prime}) - 
(j \tilde \Lambda j^{\, \prime}) ({\cal K}_2 P^{\, \prime})  
\right.
\\
\nonumber
&&\left. + m_f 
[(j \tilde \Lambda j^{\, \prime}){\cal K}_4 -  
(j \tilde \varphi j^{\, \prime}) {\cal K}_3 ]\right ] 
{\cal I}_{n, \ell'} {\cal I}^{\, \prime}_{n, \ell} - 
2\beta \sqrt{\ell \ell^{\, \prime}} \left [(P^{\, \prime} \tilde \Lambda j^{\, \prime}) ({\cal K}_2 j) +
(P^{\, \prime} \tilde \Lambda j) ({\cal K}_2 j^{\, \prime})  \right.
\\
\nonumber
&&\left. - 
(j \tilde \Lambda j^{\, \prime}) ({\cal K}_2 P^{\, \prime}) - m_f
[(j \tilde \Lambda j^{\, \prime}){\cal K}_4 -
(j \tilde \varphi j^{\, \prime}) {\cal K}_3 ] \right ] 
{\cal I}_{n-1, \ell'-1} {\cal I}^{\, \prime}_{n-1, \ell-1} +
2\beta \sqrt{n} [(j \tilde \Lambda j^{\, \prime}){\cal K}_4 -
(j \tilde \varphi j^{\, \prime}) {\cal K}_3 ]  
\\[3mm]
\nonumber
&&\times \big [\sqrt{\ell^{\, \prime}} (M_{\ell} + m_f)
{\cal I}_{n-1, \ell'-1} {\cal I}^{\, \prime}_{n, \ell} -
\sqrt{\ell} (M_{\ell'} + m_f)
{\cal I}_{n, \ell'} {\cal I}^{\, \prime}_{n-1, \ell-1} \big ] 
\\[3mm]
\nonumber
&& 
+
\sqrt{\frac{2 \beta \ell^{\, \prime}}{q_{\mprp}^{2}}} \, 
(M_{\ell} + m_f) [(q \Lambda j) + 
\ii (q \varphi j)]
[(P^{\, \prime} \tilde \Lambda j^{\, \prime})  {\cal K}_4 - 
(P^{\, \prime} \tilde \varphi j^{\, \prime})  {\cal K}_3 + m_f ({\cal K}_2 j^{\, \prime})] 
{\cal I}_{n, \ell'-1} {\cal I}^{\, \prime}_{n, \ell} 
\\[3mm]
\nonumber
&&-
\sqrt{\frac{2 \beta \ell}{q_{\mprp}^{\, \prime 2}}} \, 
(M_{\ell'} + m_f) [(q^{\, \prime} \Lambda j^{\, \prime}) - 
\ii (q^{\, \prime} \varphi j^{\, \prime})]  
[(P^{\, \prime} \tilde \Lambda j)  {\cal K}_4 + 
(P^{\, \prime} \tilde \varphi j)  {\cal K}_3 + 
m_f ({\cal K}_2 j)] 
{\cal I}_{n, \ell'} {\cal I}^{\, \prime}_{n, \ell-1} 
\\[3mm]
\nonumber
&&- 
\sqrt{\frac{2 \beta \ell}{q_{\mprp}^{\, \prime 2}}} \, 
(M_{\ell'} + m_f) 
[(q^{\, \prime} \Lambda j) - 
\ii (q^{\, \prime} \varphi j)]\; [(P^{\, \prime} \tilde \Lambda j^{\, \prime})  {\cal K}_4 - 
(P^{\, \prime} \tilde \varphi j^{\, \prime})  {\cal K}_3 - 
m_f ({\cal K}_2 j^{\, \prime})] 
{\cal I}_{n-1, \ell'} {\cal I}^{\, \prime}_{n-1, \ell-1} 
\\[3mm]
\nonumber
&& +
\sqrt{\frac{2 \beta \ell^{\, \prime}}{q_{\mprp}^{2}}} \, 
(M_{\ell} + m_f)  
[(q \Lambda j^{\, \prime}) + 
\ii (q \varphi j^{\, \prime})]\; 
[(P^{\, \prime} \tilde \Lambda j)  {\cal K}_4 + 
(P^{\, \prime} \tilde \varphi j)  {\cal K}_3 - 
m_f ({\cal K}_2 j)] 
{\cal I}_{n-1, \ell'-1} {\cal I}^{\, \prime}_{n-1, \ell} 
\\[3mm]
\nonumber
&&-
2 \beta \sqrt{\ell \ell^{\, \prime}}  
[(j \Lambda j^{\, \prime}) - \ii 
(j \varphi j^{\, \prime})] \; [({\cal K}_2 P^{\, \prime}) + m_f {\cal K}_4] 
\frac{(q \Lambda q^{\, \prime}) + \ii (q \varphi q^{\, \prime})}
{\sqrt{q_{\mprp}^2  q_{\mprp}^{\, \prime 2}}} 
\; {\cal I}_{n, \ell'-1} {\cal I}^{\, \prime}_{n, \ell-1} \\[3mm]
\nonumber
&&+
(M_{\ell} + m_f)(M_{\ell'} + m_f) \; [(j \Lambda j^{\, \prime}) + \ii 
(j \varphi j^{\, \prime})]  
[({\cal K}_2 P^{\, \prime}) - m_f {\cal K}_4] 
\frac{(q \Lambda q^{\, \prime}) - \ii (q \varphi q^{\, \prime})}
{\sqrt{q_{\mprp}^2  q_{\mprp}^{\, \prime 2}}} 
\; {\cal I}_{n-1, \ell'} {\cal I}^{\, \prime}_{n-1, \ell} 
\\[3mm]
\nonumber
&&+
\sqrt{2\beta n}\; ({\cal K}_2 j^{\, \prime}) \; \big [(M_{\ell} + m_f) (M_{\ell'} + m_f)  
\; \frac{(q \Lambda j) - 
\ii (q \varphi j)}
{\sqrt{q_{\mprp}^{2}}}  
{\cal I}_{n-1, \ell'} {\cal I}^{\, \prime}_{n, \ell} - 
2\beta
\sqrt{\ell \ell^{\, \prime}} \; \frac{(q \Lambda j) + 
\ii (q \varphi j)}
{\sqrt{q_{\mprp}^{ 2}}} \;
{\cal I}_{n, \ell'-1} {\cal I}^{\, \prime}_{n-1, \ell-1} \big ] 
\\[3mm]
\nonumber
&& -  
\sqrt{2\beta n}\; ({\cal K}_2 j) 
\big [ 2\beta \sqrt{\ell \ell^{\, \prime}} \; 
 \frac{(q^{\, \prime} \Lambda j^{\, \prime}) - 
\ii (q^{\, \prime} \varphi j^{\, \prime})}{\sqrt{q^{\, \prime 2}_{\mprp}}}
{\cal I}_{n-1, \ell'-1} {\cal I}^{\, \prime}_{n, \ell-1} - 
(M_{\ell} + m_f) (M_{\ell'} + m_f)  \frac{(q^{\, \prime} \Lambda j^{\, \prime}) + 
\ii (q^{\, \prime} \varphi j^{\, \prime})}
{\sqrt{q^{\, \prime 2}_{\mprp}}} \;
{\cal I}_{n, \ell'} {\cal I}^{\, \prime}_{n-1, \ell} \big ] 
\\[3mm]
\nonumber
&&-
2 \beta \sqrt{n} \; {\cal K}_4  
  \big [\sqrt{\ell}\; (M_{\ell'} + m_f)  
\; \frac{(q \Lambda j) - 
\ii (q \varphi j)}{\sqrt{q_{\mprp}^2}} \; \frac{(q^{\, \prime} \Lambda j^{\, \prime}) - 
\ii (q^{\, \prime} \varphi j^{\, \prime})}{\sqrt{q_{\mprp}^{\, \prime 2}}}  
{\cal I}_{n-1, \ell'} {\cal I}^{\, \prime}_{n, \ell-1} 
\\[3mm]
\nonumber
&&- 
\sqrt{\ell^{\, \prime}}\; (M_{\ell} + m_f) \; \frac{(q \Lambda j) +
\ii (q \varphi j)}{\sqrt{q_{\mprp}^2}} \;  
\frac{(q^{\, \prime} \Lambda j^{\, \prime}) +
\ii (q^{\, \prime} \varphi j^{\, \prime})}{\sqrt{q_{\mprp}^{\, \prime 2}}} 
{\cal I}_{n, \ell'-1} {\cal I}^{\, \prime}_{n-1, \ell} \big ]
\bigg \} \, .
\eeq

\newpage

\item
In the case where $j$ and $j^{\, \prime}$ are pseudovector currents 
($k =  k^{\, \prime} = A$) we obtain

%
%
\beq
\nonumber
&&{\cal R}^{++}_{AA} =  g_a g_a^{\, \prime} \bigg \{2\beta \sqrt{\ell \ell^{\, \prime}}
\left [(P \tilde \Lambda j^{\, \prime}) ({\cal K}_1 j) +
(P \tilde \Lambda j) ({\cal K}_1 j^{\, \prime}) - 
(j \tilde \Lambda j^{\, \prime}) ({\cal K}_1 P) +  m_f 
[(j \tilde \Lambda j^{\, \prime}){\cal K}_3 +  
(j \tilde \varphi j^{\, \prime}) {\cal K}_4 ] \right ] 
{\cal I}^{\, \prime}_{n, \ell'} {\cal I}_{n, \ell} 
\\
\label{eq:Raaupup}
&& + 
(M_{\ell} + m_f) (M_{\ell'} + m_f) \left [(P \tilde \Lambda j^{\, \prime}) ({\cal K}_1 j) +
(P \tilde \Lambda j) ({\cal K}_1 j^{\, \prime}) - (j \tilde \Lambda j^{\, \prime}) ({\cal K}_1 P)
- m_f [(j \tilde \Lambda j^{\, \prime}){\cal K}_3 +
(j \tilde \varphi j^{\, \prime}) {\cal K}_4 ] \right ]
\\[3mm]
\nonumber
&&\times  {\cal I}^{\, \prime}_{n-1, \ell'-1} {\cal I}_{n-1, \ell-1} +
2\beta \sqrt{n} [(j \tilde \Lambda j^{\, \prime}){\cal K}_3 +
(j \tilde \varphi j^{\, \prime}) {\cal K}_4 ]  \big [\sqrt{\ell} (M_{\ell'} + m_f)
{\cal I}^{\, \prime}_{n-1, \ell'-1} {\cal I}_{n, \ell}  +
\sqrt{\ell^{\, \prime}} (M_{\ell} + m_f)
{\cal I}^{\, \prime}_{n, \ell'} {\cal I}_{n-1, \ell-1} \big ] 
\\[3mm]
\nonumber
&& -
\sqrt{\frac{2 \beta \ell}{q_{\mprp}^{\, \prime 2}}} \, 
(M_{\ell'} + m_f) [(q^{\, \prime} \Lambda j^{\, \prime}) + 
\ii (q^{\, \prime} \varphi j^{\, \prime})]
[(P \tilde \Lambda j) {\cal K}_3 - 
(P \tilde \varphi j) {\cal K}_4 + m_f ({\cal K}_1 j)] 
{\cal I}^{\, \prime}_{n, \ell'-1} {\cal I}_{n, \ell}
\\[3mm]
\nonumber
&& -\sqrt{\frac{2 \beta \ell^{\, \prime}}{q_{\mprp}^2}} \, 
(M_{\ell} + m_f) [(q \Lambda j) - 
\ii (q \varphi j)] [(P \tilde \Lambda j^{\, \prime}) {\cal K}_3 + 
(P \tilde \varphi j^{\, \prime}) {\cal K}_4 + 
m_f ({\cal K}_1 j^{\, \prime})] 
{\cal I}^{\, \prime}_{n, \ell'} {\cal I}_{n, \ell-1} 
\\[3mm]
\nonumber
&& - \sqrt{\frac{2 \beta \ell^{\, \prime}}{q_{\mprp}^2}} \, 
(M_{\ell} + m_f) [(q \Lambda j^{\, \prime}) - 
\ii (q \varphi j^{\, \prime})] [(P \tilde \Lambda j) {\cal K}_3 - 
(P \tilde \varphi j)  {\cal K}_4 - m_f ({\cal K}_1 j)] 
{\cal I}^{\, \prime}_{n-1, \ell'} {\cal I}_{n-1, \ell-1} 
\\[3mm]
\nonumber
&& - \sqrt{\frac{2 \beta \ell}{q_{\mprp}^{\, \prime 2}}} \, 
(M_{\ell'} + m_f) [(q^{\, \prime} \Lambda j) + 
\ii (q^{\, \prime} \varphi j)]  
[(P \tilde \Lambda j^{\, \prime}) {\cal K}_3 + 
(P \tilde \varphi j^{\, \prime})  {\cal K}_4 - m_f ({\cal K}_1 j^{\, \prime})] 
{\cal I}^{\, \prime}_{n-1, \ell'-1} {\cal I}_{n-1, \ell} 
\\[3mm]
\nonumber
&&  +
(M_{\ell} + m_f)(M_{\ell'} + m_f) [(j \Lambda j^{\, \prime}) + \ii 
(j \varphi j^{\, \prime})] \; [({\cal K}_1 P) + m_f {\cal K}_3] 
\frac{(q \Lambda q^{\, \prime}) - \ii (q \varphi q^{\, \prime})}
{\sqrt{q_{\mprp}^2  q_{\mprp}^{\, \prime 2}}} 
\; {\cal I}^{\, \prime}_{n, \ell'-1} {\cal I}_{n, \ell-1}
\\[3mm]
\nonumber
&&  +
2 \beta \sqrt{\ell \ell^{\, \prime}} \; [(j \Lambda j^{\, \prime}) - \ii 
(j \varphi j^{\, \prime})] [({\cal K}_1 P) - m_f {\cal K}_3] 
\frac{(q \Lambda q^{\, \prime}) + \ii (q \varphi q^{\, \prime})}
{\sqrt{q_{\mprp}^2  q_{\mprp}^{\, \prime 2}}} 
\; {\cal I}^{\, \prime}_{n-1, \ell'} {\cal I}_{n-1, \ell} 
\\[3mm]
\nonumber
&& -
\sqrt{2\beta n}\; ({\cal K}_1 j) \; \big [ 2\beta
\sqrt{\ell \ell^{\, \prime}} 
\; \frac{(q^{\, \prime} \Lambda j^{\, \prime}) - 
\ii (q^{\, \prime} \varphi j^{\, \prime})}
{\sqrt{q_{\mprp}^{\, \prime 2}}} 
{\cal I}^{\, \prime}_{n-1, \ell'} {\cal I}_{n, \ell} + 
(M_{\ell} + m_f) (M_{\ell'} + m_f) \; \frac{(q^{\, \prime} \Lambda j^{\, \prime}) + 
\ii (q^{\, \prime} \varphi j^{\, \prime})}
{\sqrt{q_{\mprp}^{\, \prime 2}}} \;
{\cal I}^{\, \prime}_{n, \ell'-1} {\cal I}_{n-1, \ell-1} \big ]  
\\[3mm]
\nonumber
&& -  
\sqrt{2\beta n}\; ({\cal K}_1 j^{\, \prime}) \big [(M_{\ell} + m_f) (M_{\ell'} + m_f)   
\; \frac{(q \Lambda j) - 
\ii (q \varphi j)}{\sqrt{q_{\mprp}^2}}
{\cal I}^{\, \prime}_{n-1, \ell'-1} {\cal I}_{n, \ell-1} +
2\beta \sqrt{\ell \ell^{\, \prime}}  \frac{(q \Lambda j) + 
\ii (q \varphi j)}
{\sqrt{q_{\mprp}^2}} \;
{\cal I}^{\, \prime}_{n, \ell'} {\cal I}_{n-1, \ell} \big ] 
\\[3mm]
\nonumber
&& + 2 \beta \sqrt{n} \; {\cal K}_3 \; \big [\sqrt{\ell^{\, \prime}}\; (M_{\ell} + m_f)  
\; \frac{(q \Lambda j) - 
\ii (q \varphi j)}{\sqrt{q_{\mprp}^2}} \; \frac{(q^{\, \prime} \Lambda j^{\, \prime}) - 
\ii (q^{\, \prime} \varphi j^{\, \prime})}{\sqrt{q_{\mprp}^{\, \prime 2}}} 
{\cal I}^{\, \prime}_{n-1, \ell'} {\cal I}_{n, \ell - 1} 
\\[3mm]
\nonumber
&&+\sqrt{\ell}\; (M_{\ell'} + m_f) \; \frac{(q \Lambda j) +
\ii (q \varphi j)}{\sqrt{q_{\mprp}^2}} \; \frac{(q^{\, \prime} \Lambda j^{\, \prime}) +
\ii (q^{\, \prime} \varphi j^{\, \prime})}{\sqrt{q_{\mprp}^{\, \prime 2}}} 
{\cal I}^{\, \prime}_{n, \ell'-1} {\cal I}_{n-1, \ell} \big ]
\bigg \} \, ;
\eeq

\newpage
\beq
\label{eq:Raaupdown}
&&{\cal R}^{+-}_{AA} = \ii g_a g_a^{\, \prime} \bigg \{
 \sqrt{2 \beta \ell^{\, \prime}} \, (M_{\ell} + m_f)
\left [(P \tilde \Lambda j^{\, \prime}) ({\cal K}_2 j) +
(P \tilde \Lambda j) ({\cal K}_2 j^{\, \prime}) - 
(j \tilde \Lambda j^{\, \prime}) ({\cal K}_2 P) \right.
\\
\nonumber
&&\left. +  m_f 
[(j \tilde \Lambda j^{\, \prime}){\cal K}_4 +  
(j \tilde \varphi j^{\, \prime}) {\cal K}_3 ]\right ] 
{\cal I}^{\, \prime}_{n, \ell'} {\cal I}_{n, \ell} - 
 \sqrt{2 \beta \ell} \, (M_{\ell'} + m_f) \left [(P \tilde \Lambda j^{\, \prime}) ({\cal K}_2 j) +
(P \tilde \Lambda j) ({\cal K}_2 j^{\, \prime})  \right.
\\
\nonumber
&&\left. -
(j \tilde \Lambda j^{\, \prime}) ({\cal K}_2 P) - m_f
[(j \tilde \Lambda j^{\, \prime}){\cal K}_4 +
(j \tilde \varphi j^{\, \prime}) {\cal K}_3 ]\right ] 
{\cal I}^{\, \prime}_{n-1, \ell'-1} {\cal I}_{n-1, \ell-1} +
  \sqrt{2\beta n}\, [(j \tilde \Lambda j^{\, \prime}){\cal K}_4 +
(j \tilde \varphi j^{\, \prime}) {\cal K}_3 ] 
\\[3mm]
\nonumber
&&\times  \big [ (M_{\ell} + m_f) (M_{\ell'} + m_f)
{\cal I}^{\, \prime}_{n-1, \ell'-1} {\cal I}_{n, \ell} -
2 \beta \sqrt{\ell \ell^{\, \prime}}
{\cal I}^{\, \prime}_{n, \ell'} {\cal I}_{n-1, \ell-1} \big ]  
\\[3mm]
\nonumber
&&-
(M_{\ell} + m_f) (M_{\ell'} + m_f) \frac{(q^{\, \prime} \Lambda j^{\, \prime}) + 
\ii (q^{\, \prime} \varphi j^{\, \prime})}{\sqrt{q_{\mprp}^{\, \prime 2}}}\; 
[(P \tilde \Lambda j)  {\cal K}_4 - 
 (P \tilde \varphi j)  {\cal K}_3 + m_f ({\cal K}_2 j)] 
{\cal I}^{\, \prime}_{n, \ell'-1} {\cal I}_{n, \ell} 
\\[3mm]
\nonumber
&&+
\frac{2 \beta \sqrt{\ell \ell^{\, \prime}}}{\sqrt{q_{\mprp}^2}} \, [(q \Lambda j) - 
\ii (q \varphi j)] [(P \tilde \Lambda j^{\, \prime})  {\cal K}_4 + 
 (P \tilde \varphi j^{\, \prime})  {\cal K}_3 + 
m_f ({\cal K}_2 j^{\, \prime})] 
{\cal I}^{\, \prime}_{n, \ell'} {\cal I}_{n, \ell-1} 
\\[3mm]
\nonumber
&&+ 
\frac{2 \beta \sqrt{\ell \ell^{\, \prime}}}{\sqrt{q_{\mprp}^2}} \, 
 [(q \Lambda j^{\, \prime}) - 
\ii (q \varphi j^{\, \prime})] [(P \tilde \Lambda j)  {\cal K}_4 - 
 (P \tilde \varphi j) {\cal K}_3 - 
m_f ({\cal K}_2 j)] 
{\cal I}^{\, \prime}_{n-1, \ell'} {\cal I}_{n-1, \ell-1} 
\\[3mm]
\nonumber
&& -
(M_{\ell} + m_f) (M_{\ell'} + m_f) \frac{(q^{\, \prime} \Lambda j) + 
\ii (q^{\, \prime} \varphi j)}{\sqrt{q_{\mprp}^{\, \prime 2}}}\; 
[(P \tilde \Lambda j^{\, \prime}) {\cal K}_4 + 
 (P \tilde \varphi j^{\, \prime})  {\cal K}_3 - 
m_f ({\cal K}_2 j^{\, \prime})] 
{\cal I}^{\, \prime}_{n-1, \ell'-1} {\cal I}_{n-1, \ell} 
\\[3mm]
\nonumber
&&-
 \sqrt{2 \beta \ell}   
(M_{\ell'} + m_f) [(j \Lambda j^{\, \prime}) + \ii 
(j \varphi j^{\, \prime})] \; [({\cal K}_2 P) + m_f {\cal K}_4] 
\frac{(q \Lambda q^{\, \prime}) - \ii (q \varphi q^{\, \prime})}
{\sqrt{q_{\mprp}^2  q_{\mprp}^{\, \prime 2}}} 
\; {\cal I}^{\, \prime}_{n, \ell'-1} {\cal I}_{n, \ell-1} 
\\[3mm]
\nonumber
&& +
\sqrt{2 \beta \ell^{\, \prime}} \; (M_{\ell} + m_f) [(j \Lambda j^{\, \prime}) - \ii 
(j \varphi j^{\, \prime})] \; [({\cal K}_2 P) - m_f {\cal K}_4] 
\frac{(q \Lambda q^{\, \prime}) + \ii (q \varphi q^{\, \prime})}
{\sqrt{q_{\mprp}^2  q_{\mprp}^{\, \prime 2}}} 
\; {\cal I}^{\, \prime}_{n-1, \ell'} {\cal I}_{n-1, \ell} 
\\[3mm]
\nonumber
&&- 2\beta \sqrt{n}\; ({\cal K}_2 j) \; \big [\sqrt{\ell^{\, \prime}} \, (M_{\ell} + m_f)  
\frac{(q^{\, \prime} \Lambda j^{\, \prime}) -
\ii (q^{\, \prime} \varphi j^{\, \prime})}
{\sqrt{q_{\mprp}^{\, \prime 2}}} {\cal I}^{\, \prime}_{n-1, \ell'} {\cal I}_{n, \ell} - 
\sqrt{\ell} \, (M_{\ell'} + m_f) \; \frac{(q^{\, \prime} \Lambda j^{\, \prime}) + 
\ii (q^{\, \prime} \varphi j^{\, \prime})}
{\sqrt{q_{\mprp}^{\, \prime 2}}} \;
{\cal I}^{\, \prime}_{n, \ell'-1} {\cal I}_{n-1, \ell-1} \big ]   
\\[3mm]
\nonumber
&&+2\beta \sqrt{n}\; ({\cal K}_2 j^{\, \prime}) 
\big [\sqrt{\ell^{\, \prime}} \, (M_{\ell} + m_f)  
\; \frac{(q \Lambda j) - 
\ii (q \varphi j)}{\sqrt{q_{\mprp}^2}}
{\cal I}^{\, \prime}_{n-1, \ell'-1} {\cal I}_{n, \ell-1} - 
\sqrt{\ell} \, (M_{\ell'} + m_f)  \frac{(q \Lambda j) + 
\ii (q \varphi j)}
{\sqrt{q_{\mprp}^2}} \;
{\cal I}^{\, \prime}_{n, \ell'} {\cal I}_{n-1, \ell} \big ] 
\\[3mm]
\nonumber
&&- \sqrt{2 \beta n} \; {\cal K}_4 \big [2 \beta \sqrt{\ell \ell^{\, \prime}}  
\; \frac{(q \Lambda j) - 
\ii (q \varphi j)}{\sqrt{q_{\mprp}^2}} \; \frac{(q^{\, \prime} \Lambda j^{\, \prime}) - 
\ii (q^{\, \prime} \varphi j^{\, \prime})}{\sqrt{q_{\mprp}^{\, \prime 2}}} 
{\cal I}^{\, \prime}_{n-1, \ell'} {\cal I}_{n, \ell-1}  
\\[3mm]
\nonumber
&&- (M_{\ell} + m_f) (M_{\ell'} + m_f) \; \frac{(q \Lambda j) +
\ii (q \varphi j)}{\sqrt{q_{\mprp}^2}} 
\frac{(q^{\, \prime} \Lambda j^{\, \prime}) +
\ii (q^{\, \prime} \varphi j^{\, \prime})}{\sqrt{q_{\mprp}^{\, \prime 2}}} 
{\cal I}^{\, \prime}_{n, \ell'-1} {\cal I}_{n-1, \ell} \big ]
\bigg \} \, ;
\eeq
%
\beq
\label{eq:Raadownup}
&&{\cal R}^{-+}_{AA} =  \ii g_a g_a^{\, \prime} \bigg \{
-  \sqrt{2 \beta \ell} \, (M_{\ell} + m_f^{\, \prime})
\left [(P \tilde \Lambda j^{\, \prime}) ({\cal K}_2 j) +
(P \tilde \Lambda j) ({\cal K}_2 j^{\, \prime}) - 
(j \tilde \Lambda j^{\, \prime}) ({\cal K}_2 P)  
\right.
\\
\nonumber
&&\left. - m_f 
[(j \tilde \Lambda j^{\, \prime}){\cal K}_4 +  
(j \tilde \varphi j^{\, \prime}) {\cal K}_3 ]\right ] 
{\cal I}^{\, \prime}_{n, \ell'} {\cal I}_{n, \ell} + 
 \sqrt{2 \beta \ell^{\, \prime}} \, (M_{\ell} + m_f) 
\left [(P \tilde \Lambda j^{\, \prime}) ({\cal K}_2 j) +
(P \tilde \Lambda j) ({\cal K}_2 j^{\, \prime})  \right.
\\
\nonumber
&&\left.
-(j \tilde \Lambda j^{\, \prime}) ({\cal K}_2 P) + m_f
[(j \tilde \Lambda j^{\, \prime}){\cal K}_4 +
(j \tilde \varphi j^{\, \prime}) {\cal K}_3 ]\right ] 
{\cal I}^{\, \prime}_{n-1, \ell'-1} {\cal I}_{n-1, \ell-1} 
\\[3mm]
\nonumber
&&-
 \sqrt{2\beta n}\, [(j \tilde \Lambda j^{\, \prime}){\cal K}_4 +
(j \tilde \varphi j^{\, \prime}) {\cal K}_3 ] 
\big [2 \beta \sqrt{\ell \ell^{\, \prime}} 
{\cal I}^{\, \prime}_{n-1, \ell'-1} {\cal I}_{n, \ell} -
(M_{\ell} + m_f) (M_{\ell'} + m_f)
{\cal I}^{\, \prime}_{n, \ell'} {\cal I}_{n-1, \ell-1} \big ] 
\\[3mm]
\nonumber
&&+\frac{2 \beta \sqrt{\ell \ell^{\, \prime}}}{\sqrt{q_{\mprp}^{\, \prime 2}}} \, 
 [(q^{\, \prime} \Lambda j^{\, \prime}) + 
\ii (q^{\, \prime} \varphi j^{\, \prime})]  
[(P \tilde \Lambda j)  {\cal K}_4 - 
(P \tilde \varphi j)  {\cal K}_3 - m_f ({\cal K}_2 j)] 
{\cal I}^{\, \prime}_{n, \ell'-1} {\cal I}_{n, \ell} 
\\[3mm]
\nonumber
&&-
(M_{\ell} + m_f) (M_{\ell'} + m_f) \frac{(q \Lambda j) - 
\ii (q \varphi j)}{\sqrt{q_{\mprp}^2}}\; [(P \tilde \Lambda j^{\, \prime})  {\cal K}_4 + 
(P \tilde \varphi j^{\, \prime}) {\cal K}_3 - 
m_f ({\cal K}_2 j^{\, \prime})] 
{\cal I}^{\, \prime}_{n, \ell'} {\cal I}_{n, \ell-1} 
\\[3mm]
\nonumber
&& - 
 (M_{\ell} + m_f) (M_{\ell'} + m_f)  
\frac{(q \Lambda j^{\, \prime}) - 
\ii (q \varphi j^{\, \prime})}{\sqrt{q_{\mprp}^2}} \; [(P \tilde \Lambda j)  {\cal K}_4 - 
(P \tilde \varphi j) {\cal K}_3 + m_f ({\cal K}_2 j)] 
{\cal I}^{\, \prime}_{n-1, \ell'} {\cal I}_{n-1, \ell-1} 
\\[3mm]
\nonumber
&& +
\frac{2 \beta \sqrt{\ell \ell^{\, \prime}}}{\sqrt{q_{\mprp}^{\, \prime 2}}} \; 
[(q^{\, \prime} \Lambda j) + 
\ii (q^{\, \prime} \varphi j)]\; 
[(P \tilde \Lambda j^{\, \prime}) {\cal K}_4 + 
(P \tilde \varphi j^{\, \prime})  {\cal K}_3 + 
m_f ({\cal K}_2 j^{\, \prime})] 
{\cal I}^{\, \prime}_{n-1, \ell'-1} {\cal I}_{n-1, \ell}   
\\[3mm]
\nonumber
&&+
 \sqrt{2 \beta \ell^{\, \prime}} \; (M_{\ell} + m_f)  [(j \Lambda j^{\, \prime}) + \ii 
(j \varphi j^{\, \prime})] \; [({\cal K}_2 P) - m_f {\cal K}_4] 
\frac{(q \Lambda q^{\, \prime}) - \ii (q \varphi q^{\, \prime})}
{\sqrt{q_{\mprp}^2  q_{\mprp}^{\, \prime 2}}} 
\; {\cal I}^{\, \prime}_{n, \ell'-1} {\cal I}_{n, \ell-1} 
\\[3mm]
\nonumber
&&-
 \sqrt{2 \beta \ell} \; (M_{\ell'} + m_f) [(j \Lambda j^{\, \prime}) - \ii 
(j \varphi j^{\, \prime})]  
[({\cal K}_2 P) + m_f {\cal K}_4] 
\frac{(q \Lambda q^{\, \prime}) + \ii (q \varphi q^{\, \prime})}
{\sqrt{q_{\mprp}^2  q_{\mprp}^{\, \prime 2}}} 
\; {\cal I}^{\, \prime}_{n-1, \ell'} {\cal I}_{n-1, \ell}  
\\[3mm]
\nonumber
&&+
 2\beta \sqrt{n}\; ({\cal K}_2 j) \; \big [\sqrt{\ell} \, (M_{\ell'} + m_f)  
\; \frac{(q^{\, \prime} \Lambda j^{\, \prime}) - 
\ii (q^{\, \prime} \varphi j^{\, \prime})}
{\sqrt{q_{\mprp}^{\, \prime 2}}} {\cal I}^{\, \prime}_{n-1, \ell'} {\cal I}_{n, \ell} - 
\sqrt{\ell^{\, \prime}} \, (M_{\ell} + m_f) \; \frac{(q^{\, \prime} \Lambda j^{\, \prime}) + 
\ii (q^{\, \prime} \varphi j^{\, \prime})}
{\sqrt{q_{\mprp}^{\, \prime 2}}} \;
{\cal I}^{\, \prime}_{n, \ell'-1} {\cal I}_{n-1, \ell-1} \big ] 
\\[3mm]
\nonumber
&&-  
 2\beta \sqrt{n}\; ({\cal K}_2 j^{\, \prime}) \; 
\big [\sqrt{\ell} \, (M_{\ell'} + m_f)   
\frac{(q \Lambda j) - 
\ii (q \varphi j)}{\sqrt{q_{\mprp}^2}}
{\cal I}^{\, \prime}_{n-1, \ell'-1} {\cal I}_{n, \ell-1} - 
\sqrt{\ell^{\, \prime}} \, (M_{\ell} + m_f)  \frac{(q \Lambda j) + 
\ii (q \varphi j)}
{\sqrt{q_{\mprp}^2}} \;
{\cal I}^{\, \prime}_{n, \ell'} {\cal I}_{n-1, \ell} \big ] 
\\[3mm]
\nonumber
&&+
\sqrt{2 \beta n} \; {\cal K}_4  \big [(M_{\ell} + m_f) (M_{\ell'} + m_f)  
\; \frac{(q \Lambda j) - 
\ii (q \varphi j)}{\sqrt{q_{\mprp}^2}} \; \frac{(q^{\, \prime} \Lambda j^{\, \prime}) - 
\ii (q^{\, \prime} \varphi j^{\, \prime})}{\sqrt{q_{\mprp}^{\, \prime 2}}} 
{\cal I}^{\, \prime}_{n-1, \ell'} {\cal I}_{n, \ell-1}  
\\[3mm]
\nonumber
&&- 2 \beta \sqrt{\ell \ell^{\, \prime}} \; \frac{(q \Lambda j) +
\ii (q \varphi j)}{\sqrt{q_{\mprp}^2}}  
\frac{(q^{\, \prime} \Lambda j^{\, \prime}) +
\ii (q^{\, \prime} \varphi j^{\, \prime})}{\sqrt{q_{\mprp}^{\, \prime 2}}} 
{\cal I}^{\, \prime}_{n, \ell'-1} {\cal I}_{n-1, \ell} \big ]
\bigg \} \, ;
\eeq

%
\beq
\label{eq:Raadowndown}
&&{\cal R}^{--}_{AA} =  g_a g_a^{\, \prime} \bigg \{(M_{\ell} + m_f) (M_{\ell'} + m_f)
\left [(P \tilde \Lambda j^{\, \prime}) ({\cal K}_1 j) +
(P \tilde \Lambda j) ({\cal K}_1 j^{\, \prime}) - 
(j \tilde \Lambda j^{\, \prime}) ({\cal K}_1 P)  
\right.
\\
\nonumber
&&\left. -m_f 
[(j \tilde \Lambda j^{\, \prime}){\cal K}_3 +  
(j \tilde \varphi j^{\, \prime}) {\cal K}_4 ]\right ] 
{\cal I}^{\, \prime}_{n, \ell'} {\cal I}_{n, \ell} + 
2\beta \sqrt{\ell \ell^{\, \prime}} 
\left [(P \tilde \Lambda j^{\, \prime}) ({\cal K}_1 j) +
(P \tilde \Lambda j) ({\cal K}_1 j^{\, \prime})  \right.
\\
\nonumber
&&\left. -
(j \tilde \Lambda j^{\, \prime}) ({\cal K}_1 P) + m_f
[(j \tilde \Lambda j^{\, \prime}){\cal K}_3 +
(j \tilde \varphi j^{\, \prime}) {\cal K}_4 ]\right ] 
{\cal I}^{\, \prime}_{n-1, \ell'-1} {\cal I}_{n-1, \ell-1}  
\\[3mm]
\nonumber
&&+
2\beta \sqrt{n} [(j \tilde \Lambda j^{\, \prime}){\cal K}_3 +
(j \tilde \varphi j^{\, \prime}) {\cal K}_4 ]  
\big [\sqrt{\ell^{\, \prime}} (M_{\ell} + m_f)
{\cal I}^{\, \prime}_{n-1, \ell'-1} {\cal I}_{n, \ell} +
\sqrt{\ell} (M_{\ell'} + m_f)
{\cal I}^{\, \prime}_{n, \ell'} {\cal I}_{n-1, \ell-1} \big ] 
\\[3mm]
\nonumber
&&-
\sqrt{\frac{2 \beta \ell^{\, \prime}}{q_{\mprp}^{\, \prime 2}}} \, 
(M_{\ell} + m_f) [(q^{\, \prime} \Lambda j^{\, \prime}) + 
\ii (q^{\, \prime} \varphi j^{\, \prime})]  
[(P \tilde \Lambda j) {\cal K}_3 - 
(P \tilde \varphi j)  {\cal K}_4 - m_f ({\cal K}_1 j)] 
{\cal I}^{\, \prime}_{n, \ell'-1} {\cal I}_{n, \ell} 
\\[3mm]
\nonumber
&&-
\sqrt{\frac{2 \beta \ell}{q_{\mprp}^2}} \, 
(M_{\ell'} + m_f) [(q \Lambda j) - 
\ii (q \varphi j)] 
 [(P \tilde \Lambda j^{\, \prime})  {\cal K}_3 + 
(P \tilde \varphi j^{\, \prime}) {\cal K}_4 - 
m_f ({\cal K}_1 j^{\, \prime})] 
{\cal I}^{\, \prime}_{n, \ell'} {\cal I}_{n, \ell-1} 
\\[3mm]
\nonumber
&&- 
\sqrt{\frac{2 \beta \ell}{q_{\mprp}^2}} \, 
(M_{\ell'} + m_f) 
[(q \Lambda j^{\, \prime}) - 
\ii (q \varphi j^{\, \prime})]\; [(P \tilde \Lambda j) {\cal K}_3 - 
 (P \tilde \varphi j) {\cal K}_4 + 
m_f ({\cal K}_1 j)] 
{\cal I}^{\, \prime}_{n-1, \ell'} {\cal I}_{n-1, \ell-1} 
\\[3mm]
\nonumber
&&-
\sqrt{\frac{2 \beta \ell^{\, \prime}}{q_{\mprp}^{\, \prime 2}}} \, 
(M_{\ell} + m_f) 
 [(q^{\, \prime} \Lambda j) + 
\ii (q^{\, \prime} \varphi j)]\; 
[(P \tilde \Lambda j^{\, \prime}) {\cal K}_3 + 
(P \tilde \varphi j^{\, \prime})  {\cal K}_4 + 
m_f ({\cal K}_1 j^{\, \prime})] 
{\cal I}^{\, \prime}_{n-1, \ell'-1} {\cal I}_{n-1, \ell} 
\\[3mm]
\nonumber
&&+
2 \beta \sqrt{\ell \ell^{\, \prime}}   
[(j \Lambda j^{\, \prime}) + \ii 
(j \varphi j^{\, \prime})] \; [({\cal K}_1 P) - m_f {\cal K}_3] 
\frac{(q \Lambda q^{\, \prime}) - \ii (q \varphi q^{\, \prime})}
{\sqrt{q_{\mprp}^2  q_{\mprp}^{\, \prime 2}}} 
\; {\cal I}^{\, \prime}_{n, \ell'-1} {\cal I}_{n, \ell-1} 
\\[3mm]
\nonumber
&&+
(M_{\ell} + m_f)(M_{\ell'} + m_f) \; [(j \Lambda j^{\, \prime}) - \ii 
(j \varphi j^{\, \prime})]  
[({\cal K}_1 P) + m_f {\cal K}_3] 
\frac{(q \Lambda q^{\, \prime}) + \ii (q \varphi q^{\, \prime})}
{\sqrt{q_{\mprp}^2  q_{\mprp}^{\, \prime 2}}} 
\; {\cal I}^{\, \prime}_{n-1, \ell'} {\cal I}_{n-1, \ell}  
\\[3mm]
\nonumber
&&-
\sqrt{2\beta n}\; ({\cal K}_1 j) \; 
\big [(M_{\ell} + m_f) (M_{\ell'} + m_f)  
\frac{(q^{\, \prime} \Lambda j^{\, \prime}) - 
\ii (q^{\, \prime} \varphi j^{\, \prime})}
{\sqrt{q_{\mprp}^{\, \prime 2}}}
{\cal I}^{\, \prime}_{n-1, \ell'} {\cal I}_{n, \ell}  + 
2\beta
\sqrt{\ell \ell^{\, \prime}} \; \frac{(q^{\, \prime} \Lambda j^{\, \prime}) + 
\ii (q^{\, \prime} \varphi j^{\, \prime})}{\sqrt{q_{\mprp}^{\, \prime 2}}} \;
{\cal I}^{\, \prime}_{n, \ell'-1} {\cal I}_{n-1, \ell-1} \big ] 
\\[3mm]
\nonumber
&&+  
\sqrt{2\beta n}\; ({\cal K}_1 j^{\, \prime}) 
\big [ 2\beta \sqrt{\ell \ell^{\, \prime}}  
\; \frac{-(q \Lambda j) + 
\ii (q \varphi j)}{\sqrt{q_{\mprp}^2}}
{\cal I}^{\, \prime}_{n-1, \ell'-1} {\cal I}_{n, \ell-1} - 
(M_{\ell} + m_f) (M_{\ell'} + m_f)  \frac{(q \Lambda j) + 
\ii (q \varphi j)}
{\sqrt{q_{\mprp}^2}} \;
{\cal I}^{\, \prime}_{n, \ell'} {\cal I}_{n-1, \ell} \big ] +
\\[3mm]
\nonumber
&& 2 \beta \sqrt{n} \; {\cal K}_3 \; \big [\sqrt{\ell}\; (M_{\ell'} + m_f)  
\; \frac{(q \Lambda j) - 
\ii (q \varphi j)}{\sqrt{q_{\mprp}^2}} \; \frac{(q^{\, \prime} \Lambda j^{\, \prime}) - 
\ii (q^{\, \prime} \varphi j^{\, \prime})}{\sqrt{q_{\mprp}^{\, \prime 2}}} 
{\cal I}^{\, \prime}_{n-1, \ell'}  {\cal I}_{n, \ell-1}  + 
\\[3mm]
\nonumber
&&\sqrt{\ell^{\, \prime}}\; (M_{\ell} + m_f) \; \frac{(q \Lambda j) +
\ii (q \varphi j)}{\sqrt{q_{\mprp}^2}} \; \frac{(q^{\, \prime} \Lambda j^{\, \prime}) +
\ii (q^{\, \prime} \varphi j^{\, \prime})}{\sqrt{q_{\mprp}^{\, \prime 2}}} 
{\cal I}^{\, \prime}_{n, \ell'-1} {\cal I}_{n-1, \ell} \big ]
\bigg \} \, .
\eeq

For second diagram we have the following replacement $P_{\alpha} \to P_{\alpha}^{\, \prime}$, 
$q_{\alpha} \leftrightarrow -q_{\alpha}^{\, \prime}$, 
$j_{\alpha} \leftrightarrow j_{\alpha}^{\, \prime}$ 
${\cal I}_{m,n} \leftrightarrow {\cal I}_{m,n}^{\, \prime}$.

\end{enumerate}

\end{widetext}



\section{Ground Landau Level}\label{Sec:3}


The obtained results can be essentially simplified in several special cases.  
In the present section we consider the strong field limit, where the
magnetic field strength $B$ is the maximal physical parameter,
namely, $\sqrt{eB} \gg  \omega, \, E$, etc.  In this case $n,\; \ell,\; \ell^{\, \prime} =0$,
$M_{\ell} = M_{\ell'} = m_f$ and we obtain the following expressions for the amplitudes, 
for different spin states  
of the initial and final fermions and for generalized vertices of the scalar, pseudoscalar, vector or  
axial vector types
\beq
\label{eq:M00}
&&{\cal M}^{--}_{k^{\, \prime} k} = - \exp{\left [-\ii\theta \right ]}
\exp{\left [-\frac{q_{\mprp}^2+q^{\, \prime 2}_{\mprp}}
{4 \beta}\right ]}
\\
\nonumber
&&\times \left \{ 
 \frac{\eee^{\ii (q \varphi q^{\, \prime})/(2\beta)} \; 
{\cal R}^{(1)}_{0 \; k^{\, \prime} k}}{P^2_{\mprl} - m_f^2} + 
 \frac{\eee^{-\ii (q \varphi q^{\, \prime})/(2\beta)} \; 
{\cal R}^{(2)}_{0 \; k k^{\, \prime}}}{P^{\, \prime 2}_{\mprl} - m_f^2} \right \} \;  ,
\eeq
\noindent   where
\beq
\label{eq:Rss01}
{\cal R}^{(1)}_{0 SS} = g_s g_s^{\, \prime} j_s j_s^{\, \prime}  
\left [({\cal K}_1 P) + m_f {\cal K}_3 \right ] \, ; 
\eeq
\beq
\label{eq:Rss02}
{\cal R}^{(2)}_{0 SS} = g_s g_s^{\, \prime} j_s j_s^{\, \prime}  
\left [({\cal K}_1 P^{\, \prime}) + m_f {\cal K}_3 \right ] \, ; 
\eeq
%
\beq
\label{eq:Rsp01}
{\cal R}^{(1)}_{0 PS} =  g_s g_p^{\, \prime} j_s j_p^{\, \prime}         
\left [({\cal K}_2 P) - m_f {\cal K}_4 \right ] \, ; 
\eeq
\beq
\label{eq:Rsp02}
{\cal R}^{(2)}_{0 SP} = - g_s g_p^{\, \prime} j_s j_p^{\, \prime}        
\left [({\cal K}_2 P^{\, \prime}) + m_f {\cal K}_4 \right ] \, ; 
\eeq
\beq
\label{eq:Rsv01}
&&
{\cal R}^{(1)}_{0 VS} = g_s g_v^{\, \prime} j_s  
\big [(P \widetilde \Lambda j^{\, \prime}){\cal K}_3 + 
(P \widetilde \varphi j^{\, \prime}) {\cal K}_4 
\\
\nonumber
&&
+ m_f ({\cal K}_1 j^{\, \prime}) \big ] \, ; 
\eeq
\beq
\label{eq:Rsv02}
&&
{\cal R}^{(2)}_{0 SV} = g_s g_v^{\, \prime} j_s  
\big [(P^{\, \prime} \widetilde \Lambda j^{\, \prime}){\cal K}_3 - 
(P^{\, \prime} \widetilde \varphi j^{\, \prime}) {\cal K}_4 
\\
\nonumber
&&+ m_f ({\cal K}_1 j^{\, \prime}) \big ] \, ; 
\eeq
%
\beq
\label{eq:Rsa01}
&&{\cal R}^{(1)}_{0 AS} =   g_s g_a^{\, \prime} j_s                
\big [(P \widetilde \Lambda j^{\, \prime}) {\cal K}_4 + 
(P \widetilde \varphi j^{\, \prime})  {\cal K}_3 
\\
\nonumber
&& - m_f ({\cal K}_2 j^{\, \prime}) \big ]\, ; 
\eeq
\beq
\label{eq:Rsa02}
&&{\cal R}^{(2)}_{0 SA} =   g_s g_a^{\, \prime} j_s 
\big [(P^{\, \prime} \widetilde \Lambda j^{\, \prime}) {\cal K}_4 - 
(P^{\, \prime} \widetilde \varphi j^{\, \prime})  {\cal K}_3 
\\
\nonumber
&& + m_f ({\cal K}_2 j^{\, \prime}) \big ]\, ; 
\eeq
%
\beq
\label{eq:Rpp01}
{\cal R}^{(1)}_{0 PP} = - g_p g_p^{\, \prime} j_p j_p^{\, \prime}  
\left [({\cal K}_1 P) - m_f {\cal K}_3 \right ] \, ; 
\eeq
\beq
\label{eq:Rpp02}
{\cal R}^{(2)}_{0 PP} = - g_p g_p^{\, \prime} j_p j_p^{\, \prime}  
\left [({\cal K}_1 P^{\, \prime}) - m_f {\cal K}_3 \right ] \, ; 
\eeq
%
\beq
\label{eq:Rpv01}
&&{\cal R}^{(1)}_{0 VP} = -  g_p g_v^{\, \prime} j_p                 
\big [(P \widetilde \Lambda j^{\, \prime}) {\cal K}_4 + 
(P \widetilde \varphi j^{\, \prime})  {\cal K}_3  
\\
\nonumber
&&+ m_f ({\cal K}_2 j^{\, \prime}) \big ]\, ; 
\eeq
\beq
\label{eq:Rpv02}
&&{\cal R}^{(2)}_{0 PV} =   g_p g_v^{\, \prime} j_p                  
\big [(P^{\, \prime} \widetilde \Lambda j^{\, \prime}) {\cal K}_4 - 
(P^{\, \prime} \widetilde \varphi j^{\, \prime})  {\cal K}_3  
\\
\nonumber
&&- m_f ({\cal K}_2 j^{\, \prime}) \big ]\, ; 
\eeq
%
\beq
\label{eq:Rpa01}
&&{\cal R}^{(1)}_{0 AP} = -g_p g_a^{\, \prime} j_p  
\big [(P \widetilde \Lambda j^{\, \prime}) {\cal K}_3 + 
(P \widetilde \varphi j^{\, \prime})  {\cal K}_4  
\\
\nonumber
&&- m_f ({\cal K}_1 j^{\, \prime}) \big ] \, ; 
\eeq
\beq
\label{eq:Rpa02} 
&&{\cal R}^{(2)}_{0 PA} =  g_p g_a^{\, \prime} j_p                    
\big [(P^{\, \prime} \widetilde \Lambda j^{\, \prime}) {\cal K}_3 - 
(P^{\, \prime} \widetilde \varphi j^{\, \prime})  {\cal K}_4  
\\
\nonumber
&&- m_f ({\cal K}_1 j^{\, \prime}) \big ] \, ; 
\eeq
%
\beq
\label{eq:Rvv01}
&&{\cal R}^{(1)}_{0 VV} =  g_v g_v^{\, \prime}  
\left \{(P \widetilde \Lambda j^{\, \prime}) ({\cal K}_1 j) +
(P \widetilde \Lambda j) ({\cal K}_1 j^{\, \prime})  
\right.
\\
\nonumber
&&\left. - (j \widetilde \Lambda j^{\, \prime}) ({\cal K}_1 P) +  m_f 
\big [(j \widetilde \Lambda j^{\, \prime}){\cal K}_3 +  
(j \widetilde \varphi j^{\, \prime}) {\cal K}_4 \big ]\right \}\, ; 
\eeq
\beq
\label{eq:Rvv02}
&&{\cal R}^{(2)}_{0 VV} =  g_v g_v^{\, \prime}  
\left \{(P^{\, \prime} \widetilde \Lambda j^{\, \prime}) ({\cal K}_1 j) +
(P^{\, \prime} \widetilde \Lambda j) ({\cal K}_1 j^{\, \prime})  
\right.
\\
\nonumber
&&\left. - (j \widetilde \Lambda j^{\, \prime}) ({\cal K}_1 P^{\, \prime}) +  m_f 
\big [(j \widetilde \Lambda j^{\, \prime}){\cal K}_3 -  
(j \widetilde \varphi j^{\, \prime}) {\cal K}_4 \big ]\right \}\, ; 
\eeq
%
\beq
\label{eq:Rav01} 
&&{\cal R}^{(1)}_{0 AV} =  g_v g_a^{\, \prime}  
\left \{(P \widetilde \Lambda j^{\, \prime}) ({\cal K}_2 j) +
(P \widetilde \Lambda j) ({\cal K}_2 j^{\, \prime})  
\right.
\\
\nonumber
&&\left. - (j \widetilde \Lambda j^{\, \prime}) ({\cal K}_2 P) -  m_f 
\big [(j \widetilde \Lambda j^{\, \prime}){\cal K}_4 +  
(j \widetilde \varphi j^{\, \prime}) {\cal K}_3 \big ]\right \}\, ; 
\eeq
\beq
\label{eq:Rva01}
&&{\cal R}^{(2)}_{0 VA} =  g_v g_a^{\, \prime}  
\left \{(P^{\, \prime} \widetilde \Lambda j^{\, \prime}) ({\cal K}_2 j) +
(P^{\, \prime} \widetilde \Lambda j) ({\cal K}_2 j^{\, \prime})  
\right.
\\
\nonumber
&&\left. - (j \widetilde \Lambda j^{\, \prime}) ({\cal K}_2 P^{\, \prime}) +  m_f 
\big [(j \widetilde \Lambda j^{\, \prime}){\cal K}_4 -  
(j \widetilde \varphi j^{\, \prime}) {\cal K}_3 \big ]\right \}\, ; 
\eeq
%
\beq
\label{eq:Raa01}
&&{\cal R}^{(1)}_{0 AA} =  g_a g_a^{\, \prime}  
\left \{(P \widetilde \Lambda j^{\, \prime}) ({\cal K}_1 j) +
(P \widetilde \Lambda j) ({\cal K}_1 j^{\, \prime})  
\right.
\\
\nonumber
&&\left. - (j \widetilde \Lambda j^{\, \prime}) ({\cal K}_1 P) - 
m_f 
\big [(j \widetilde \Lambda j^{\, \prime}){\cal K}_3 +  
(j \widetilde \varphi j^{\, \prime}) {\cal K}_4 \big ]\right \}\, . 
\eeq
\beq
\label{eq:Raa02}
&&{\cal R}^{(2)}_{0 AA} =  g_a g_a^{\, \prime}  
\left \{(P^{\, \prime} \widetilde \Lambda j^{\, \prime}) ({\cal K}_1 j) +
(P^{\, \prime} \widetilde \Lambda j) ({\cal K}_1 j^{\, \prime})  
\right.
\\
\nonumber
&&\left. - (j \widetilde \Lambda j^{\, \prime}) ({\cal K}_1 P^{\, \prime}) - 
m_f 
\big [(j \widetilde \Lambda j^{\, \prime}){\cal K}_3 -  
(j \widetilde \varphi j^{\, \prime}) {\cal K}_4 \big ]\right \}\, . 
\eeq

We note that the obtained results for ${\cal M}^{--}_{VV}$ and ${\cal M}^{--}_{VA}$ 
exactly coincide with the amplitude of the photo-neutrino process 
from Ref.~\cite{RCh08} (see also~\cite{RCh09}) after taking account of the second diagram and 
of the momentum conservation law.

\section{Forward scattering} \label{Sec:4}


For generalization of the results obtained in Ref.~\cite{Borovkov:1999}, to the case 
of magnetized plasma we consider the process of a coherent scattering of the generalized current $j$ 
off the real fermions without change of their states (the ``forward'' scattering). 
In this case, under the generalized current $j$ in the initial state we mean only the field operator 
of a single particle, while the generalized current $j^{\, \prime}$ in the final state could be both 
the field operator of a single particle and e.g. the neutrino current.  
In this case:  
$s=s^{\, \prime}$, $q^\mu = q^{\, \prime \mu}$, 
$p^\mu = p^{\, \prime \mu}$, ${\cal K}_{1 \alpha} = 2 (p \tilde \Lambda)_{\alpha}$,   
${\cal K}_{2\alpha} = 2 (\tilde \varphi p)_\alpha$, ${\cal K}_3 = 2 M_{\ell}$, ${\cal K}_4 = 0$.  
We obtain the follwing results for the amplitudes: 
\beq
\label{eq:FS1}
&&{\cal M}_{k^{\, \prime} k} = - \frac{\beta}{2 \pi^2} \; \sum_{\ell,n = 0}^{\infty} 
\int \frac{\dd p_z}{E_{\ell}} \; 
f_{f}(E_{\ell})\;   
\\
\nonumber
&&\times \left \{ \frac{{\cal D}^{(1)}_{k^{\, \prime} k}}
{(p+q)^2_{\mprl} - m_f^2 - 2\beta n} +  \frac{{\cal D}^{(2)}_{kk^{\, \prime}}}
{(p-q)^2_{\mprl} - m_f^2 - 2\beta n}\right \} \, ,
\eeq
\noindent where $f_{f}(E_{\ell}) = [1+\exp{(E_{\ell} - \mu_f)/T}]^{-1}$ is the fermion 
distribution function,  $T$ and $\mu_f$ are the temperature and the chemical potential of plasma 
correspondingly, 
%
\beq
\label{eq:FSS}
&&{\cal D}^{(1)}_{SS} = g_s g_s^{\, \prime} j_s j_s^{\, \prime} \left \{ [(q \widetilde \Lambda p) + 
2 \beta \ell + 2 m_f^2]
\right.
\\[2mm]
\nonumber
&&\times \left.({\cal I}_{n,\ell}^2 + {\cal I}_{n-1,\ell-1}^2)  - 
 4 \beta \sqrt{n \ell}\, {\cal I}_{n,\ell} {\cal I}_{n-1,\ell-1} \right \} ;
\\[2mm]
\nonumber
&&{\cal D}^{(2)}_{SS} = {\cal D}^{(1)}_{SS} (q\to -q) \, ;
\eeq
%
%
\beq
\label{eq:FSP}
{\cal D}^{(1)}_{SP} = {\cal D}^{(2)}_{PS} = 
 g_s g_p^{\, \prime} j_s j_p^{\, \prime} (q \widetilde \varphi p)  
\left [{\cal I}_{n,\ell}^2 - {\cal I}_{n-1,\ell-1}^2 \right ] ;
\eeq
%
%
\beq
\nonumber
&&{\cal D}^{(1)}_{VS} =  g_s g_v^{\, \prime} j_s m_f   
\left \{ [2 (p \widetilde \Lambda j^{\, \prime}) + (q \widetilde \Lambda j^{\, \prime})] 
\left [{\cal I}_{n,\ell}^2 + {\cal I}_{n-1,\ell-1}^2 \right ]  \right.
\\[3mm]
\label{eq:FSV}
&&\left. - \sqrt{\frac{2 \beta \ell}{q_{\mprp}^2}} \, \left [[(q \Lambda j^{\, \prime}) + 
\ii (q \varphi j^{\, \prime})] {\cal I}_{n,\ell} {\cal I}_{n,\ell-1}  
\right.\right.
\\[3mm]
\nonumber
&&\left.\left. +[(q \Lambda j^{\, \prime}) - 
\ii (q \varphi j^{\, \prime})] {\cal I}_{n-1,\ell} {\cal I}_{n-1,\ell-1}  \right]   \right. 
\\[3mm]
\nonumber
&&\left. - 
 \sqrt{\frac{2 \beta n}{q_{\mprp}^2}} \left  [[(q \Lambda j^{\, \prime}) +
\ii (q \varphi j^{\, \prime})] {\cal I}_{n,\ell-1} {\cal I}_{n-1,\ell-1} 
\right. \right. 
\\[3mm]
\nonumber
&&\left. \left. + [(q \Lambda j^{\, \prime}) -
\ii (q \varphi j^{\, \prime})] {\cal I}_{n,\ell} {\cal I}_{n-1,\ell}  \right] 
\right \};
\eeq

%
\beq
\nonumber
&&{\cal D}^{(2)}_{SV} =  g_s g_v^{\, \prime} j_s m_f   
\left \{ [2 (p \widetilde \Lambda j^{\, \prime}) - (q \widetilde \Lambda j^{\, \prime})] 
\left [{\cal I}_{n,\ell}^2 + {\cal I}_{n-1,\ell-1}^2 \right ]  \right.
\\[3mm]
\label{eq:FSV2}
&&\left. + \sqrt{\frac{2 \beta \ell}{q_{\mprp}^2}} \, \left [[(q \Lambda j^{\, \prime}) - 
\ii (q \varphi j^{\, \prime})] {\cal I}_{n,\ell} {\cal I}_{n,\ell-1}  
\right.\right.
\\[3mm]
\nonumber
&&\left.\left. +[(q \Lambda j^{\, \prime}) + 
\ii (q \varphi j^{\, \prime})] {\cal I}_{n-1,\ell} {\cal I}_{n-1,\ell-1}  \right]   \right. 
\\[3mm]
\nonumber
&&\left. + 
 \sqrt{\frac{2 \beta n}{q_{\mprp}^2}} \left  [[(q \Lambda j^{\, \prime}) -
\ii (q \varphi j^{\, \prime})] {\cal I}_{n,\ell-1} {\cal I}_{n-1,\ell-1} 
\right. \right. 
\\[3mm]
\nonumber
&&\left. \left. + [(q \Lambda j^{\, \prime}) +
\ii (q \varphi j^{\, \prime})] {\cal I}_{n,\ell} {\cal I}_{n-1,\ell}  \right] 
\right \};
\eeq

%
%
\beq
\label{eq:FSA}
&&{\cal D}^{(1)}_{AS} =  g_s g_a^{\, \prime} j_s m_f 
[2 (p \widetilde \varphi j^{\, \prime}) + (q \widetilde \varphi j^{\, \prime})] 
\\[2mm]
\nonumber 
&&\times \left [{\cal I}_{n,\ell}^2 - {\cal I}_{n-1,\ell-1}^2 \right ] ;
\\[2mm]
\nonumber
&&{\cal D}^{(2)}_{SA} = {\cal D}^{(1)}_{AS} (q\to -q) \, ;
\eeq
%
%

\beq
\nonumber
&&{\cal D}^{(1)}_{PP} = - g_p g_p^{\, \prime} j_p j_p^{\, \prime} 
\left \{ [(q \widetilde \Lambda p) + 
2 \beta \ell ] 
\left [{\cal I}_{n,\ell}^2 + {\cal I}_{n-1,\ell-1}^2 \right ]   
\right.
\\[2mm]
\label{eq:FPP}
&&\left. - 4 \beta \sqrt{n \ell}\, {\cal I}_{n,\ell} {\cal I}_{n-1,\ell-1} \right \} ;
\\[2mm]
\nonumber
&&{\cal D}^{(2)}_{PP} = {\cal D}^{(1)}_{PP} (q\to -q) \, ;
\eeq
%
%
\beq
\label{eq:FPV}
{\cal D}^{(1)}_{VP} &=& {\cal D}^{(2)}_{PV} 
\\
\nonumber
&=& - g_p g_v^{\, \prime} j_p m_f (q \widetilde \varphi j^{\, \prime})  
\left [{\cal I}_{n,\ell}^2 - {\cal I}_{n-1,\ell-1}^2 \right ] ;
\eeq

%
%
\beq
\label{eq:FPA1}
&&{\cal D}^{(1)}_{AP} = - g_p g_a^{\, \prime} j_p m_f   
\left \{ (q \widetilde \Lambda j^{\, \prime}) 
\left [{\cal I}_{n,\ell}^2 + {\cal I}_{n-1,\ell-1}^2 \right ]   \right.
\\[3mm]
\nonumber
&&\left. +\sqrt{\frac{2 \beta \ell}{q_{\mprp}^2}} \, \left [[(q \Lambda j^{\, \prime}) + 
\ii (q \varphi j^{\, \prime})] {\cal I}_{n,\ell} {\cal I}_{n,\ell-1}  \right. \right.
\\[3mm]
\nonumber
&&\left. \left. +
[(q \Lambda j^{\, \prime}) - 
\ii (q \varphi j^{\, \prime})] {\cal I}_{n-1,\ell} {\cal I}_{n-1,\ell-1}  \right]  \right. 
\\[3mm]
\nonumber
&&\left. - \sqrt{\frac{2 \beta n}{q_{\mprp}^2}} \, \left [[(q \Lambda j^{\, \prime}) +
\ii (q \varphi j^{\, \prime})] {\cal I}_{n,\ell-1} {\cal I}_{n-1,\ell-1}  \right. \right.
\\[3mm]
\nonumber
&&\left. \left. + 
[(q \Lambda j^{\, \prime}) -
\ii (q \varphi j^{\, \prime})] {\cal I}_{n,\ell} {\cal I}_{n-1,\ell}  \right] 
\right \};
\eeq

\beq
\label{eq:FPA2}
&&{\cal D}^{(2)}_{PA} = - g_p g_a^{\, \prime} j_p m_f   
\left \{  (q \widetilde \Lambda j^{\, \prime}) 
\left [{\cal I}_{n,\ell}^2 + {\cal I}_{n-1,\ell-1}^2 \right ]   \right.
\\[3mm]
\nonumber
&&\left. -\sqrt{\frac{2 \beta \ell}{q_{\mprp}^2}} \, \left [[(q \Lambda j^{\, \prime}) - 
\ii (q \varphi j^{\, \prime})] {\cal I}_{n,\ell} {\cal I}_{n,\ell-1}  \right. \right.
\\[3mm]
\nonumber
&&\left. \left. -
[(q \Lambda j^{\, \prime}) + 
\ii (q \varphi j^{\, \prime})] {\cal I}_{n-1,\ell} {\cal I}_{n-1,\ell-1}  \right]  \right. 
\\[3mm]
\nonumber
&&\left. - \sqrt{\frac{2 \beta n}{q_{\mprp}^2}} \, \left [[(q \Lambda j^{\, \prime}) -
\ii (q \varphi j^{\, \prime})] {\cal I}_{n,\ell-1} {\cal I}_{n-1,\ell-1}  \right. \right.
\\[3mm]
\nonumber
&&\left. \left. + 
[(q \Lambda j^{\, \prime}) +
\ii (q \varphi j^{\, \prime})] {\cal I}_{n,\ell} {\cal I}_{n-1,\ell}  \right] 
\right \};
\eeq

%
\begin{widetext}

\beq
\label{eq:FVV}
&&{\cal D}^{(1)}_{VV} =  g_v g_v^{\, \prime}    
\left \{ \left [ (p \widetilde \Lambda j) (P \widetilde \Lambda j^{\, \prime}) + 
(P \widetilde \Lambda j) (p \widetilde \Lambda j^{\, \prime})  - 
(j \widetilde \Lambda j^{\, \prime})[2 \beta \ell + (p \widetilde \Lambda q)] \right ] 
\left [{\cal I}_{n,\ell}^2 + {\cal I}_{n-1,\ell-1}^2 \right ]   
\right.
\\[3mm]
\nonumber
&&\left. + 4 \beta \sqrt{n\ell} \, (j \widetilde \Lambda j^{\, \prime}) 
{\cal I}_{n,\ell} {\cal I}_{n-1,\ell-1} - 
 \sqrt{\frac{2 \beta \ell}{q_{\mprp}^2}} \, \left [
(P \widetilde \Lambda j) [(q \Lambda j^{\, \prime}) + 
\ii (q \varphi j^{\, \prime})]  + (P \widetilde \Lambda j^{\, \prime}) [(q \Lambda j) - 
\ii (q \varphi j)] \right] {\cal I}_{n,\ell} {\cal I}_{n,\ell-1} \right. 
\\[3mm]
\nonumber
&&\left.  - 
 \sqrt{\frac{2 \beta \ell}{q_{\mprp}^2}} \, \left [
(P \widetilde \Lambda j) [(q \Lambda j^{\, \prime}) - 
\ii (q \varphi j^{\, \prime})]  + 
 (P \widetilde \Lambda j^{\, \prime}) [(q \Lambda j) + 
\ii (q \varphi j)] \right] {\cal I}_{n-1,\ell-1} {\cal I}_{n-1,\ell} \right. 
\\[3mm]
\nonumber
&&\left. - 
\sqrt{\frac{2 \beta n}{q_{\mprp}^2}} \, \left [
(p \widetilde \Lambda j) [(q \Lambda j^{\, \prime}) - 
\ii (q \varphi j^{\, \prime})]  +  (p \widetilde \Lambda j^{\, \prime}) 
[(q \Lambda j) + 
\ii (q \varphi j)] \right] {\cal I}_{n,\ell} {\cal I}_{n-1,\ell} 
\right. 
\\[3mm]
\nonumber
&&\left. - \sqrt{\frac{2 \beta n}{q_{\mprp}^2}}  \left [
(p \widetilde \Lambda j) [(q \Lambda j^{\, \prime}) + 
\ii (q \varphi j^{\, \prime})]  +  (p \widetilde \Lambda j^{\, \prime}) 
[(q \Lambda j) - 
\ii (q \varphi j)] \right] {\cal I}_{n-1,\ell-1} {\cal I}_{n,\ell-1} + 
 [2 \beta \ell + (p \widetilde \Lambda q)] 
\right. 
\\[3mm]
\nonumber
&&\left. \times \left [[(j \Lambda j^{\, \prime}) + 
\ii (j \varphi j^{\, \prime})] {\cal I}_{n,\ell-1}^2 + [(j \Lambda j^{\, \prime}) - 
\ii (j \varphi j^{\, \prime})] {\cal I}_{n-1,\ell}^2 \right ] + 
 \frac{4 \beta \sqrt{n\ell}}{q_{\mprp}^2} \, 
[(q \Lambda j) (q \Lambda j^{\, \prime}) - 
(q \varphi j) (q \varphi j^{\, \prime})] {\cal I}_{n,\ell-1} {\cal I}_{n-1,\ell}
\right \};
\\[2mm]
\nonumber
&&{\cal D}^{(2)}_{VV} = {\cal D}^{(1)}_{VV} (q\to -q ,\, \, j \leftrightarrow j^{\, \prime}) \, ;
\eeq

%
%
\beq
\nonumber
&&{\cal D}^{(1)}_{AV} =  g_v g_a^{\, \prime}    
\left \{ [(P \widetilde \Lambda j)  (j^{\, \prime} \widetilde \varphi p) + 
(P \widetilde \Lambda j^{\, \prime})  (j \widetilde \varphi p) - 
(j \widetilde \Lambda j^{\, \prime}) (q \widetilde \varphi p) - 
m_f^2 (j \widetilde \varphi j^{\, \prime})] 
\left [{\cal I}_{n,\ell}^2 - {\cal I}_{n-1,\ell-1}^2 \right ]  \right.
\\[3mm]
\label{eq:FVA}
&&\left. + \sqrt{\frac{2 \beta \ell}{q_{\mprp}^2}} \, \left [
(P \widetilde \varphi j) [(q \Lambda j^{\, \prime}) + 
\ii (q \varphi j^{\, \prime})]  + (P \widetilde \varphi j^{\, \prime}) 
[(q \Lambda j) - 
\ii (q \varphi j)] \right] {\cal I}_{n,\ell} {\cal I}_{n,\ell-1}  \right. 
\\[3mm]
\nonumber
&&\left. - \sqrt{\frac{2 \beta \ell}{q_{\mprp}^2}} \, \left [
(P \widetilde \varphi j) [(q \Lambda j^{\, \prime}) - 
\ii (q \varphi j^{\, \prime})]  + (P \widetilde \varphi j^{\, \prime}) 
[(q \Lambda j) + 
\ii (q \varphi j)] \right] {\cal I}_{n-1,\ell-1} {\cal I}_{n-1,\ell}  \right. 
\\[3mm]
\nonumber
&&\left. + \sqrt{\frac{2 \beta n}{q_{\mprp}^2}} \, \left [
(p \widetilde \varphi j) [(q \Lambda j^{\, \prime}) - 
\ii (q \varphi j^{\, \prime})]  +  (p \widetilde \varphi j^{\, \prime}) 
[(q \Lambda j) + 
\ii (q \varphi j)] \right] {\cal I}_{n,\ell} {\cal I}_{n-1,\ell}  \right. 
\\[3mm]
\nonumber
&&\left. - \sqrt{\frac{2 \beta n}{q_{\mprp}^2}} \, \left [
(p \widetilde \varphi j) [(q \Lambda j^{\, \prime}) + 
\ii (q \varphi j^{\, \prime})]  +  (p \widetilde \varphi j^{\, \prime}) 
[(q \Lambda j) - 
\ii (q \varphi j)] \right] {\cal I}_{n-1,\ell-1} {\cal I}_{n,\ell-1}  \right. 
\\[3mm]
\nonumber
&&\left. + (p \widetilde \varphi q) \left [[(j \Lambda j^{\, \prime}) + 
\ii (j \varphi j^{\, \prime})]{\cal I}_{n,\ell-1}^2 - [(j \Lambda j^{\, \prime}) - 
\ii (j \varphi j^{\, \prime})]{\cal I}_{n-1,\ell}^2 \right ] 
\right \};
\eeq
\beq
\nonumber
&&{\cal D}^{(2)}_{VA} =  g_v g_a^{\, \prime}    
\left \{ [(P^{\, \prime} \widetilde \Lambda j)  (j^{\, \prime} \widetilde \varphi p) + 
(P^{\, \prime} \widetilde \Lambda j^{\, \prime})  (j \widetilde \varphi p) + 
(j \widetilde \Lambda j^{\, \prime}) (q \widetilde \varphi p) - 
m_f^2 (j \widetilde \varphi j^{\, \prime})] 
\left [{\cal I}_{n,\ell}^2 - {\cal I}_{n-1,\ell-1}^2 \right ]  \right.
\\[3mm]
\label{eq:FVA2}
&&\left. - \sqrt{\frac{2 \beta \ell}{q_{\mprp}^2}} \, \left [
(P^{\, \prime} \widetilde \varphi j^{\, \prime}) [(q \Lambda j) + 
\ii (q \varphi j)]  + (P^{\, \prime} \widetilde \varphi j) 
[(q \Lambda j^{\, \prime}) - 
\ii (q \varphi j^{\, \prime})] \right] {\cal I}_{n,\ell} {\cal I}_{n,\ell-1}  \right. 
\\[3mm]
\nonumber
&&\left. + \sqrt{\frac{2 \beta \ell}{q_{\mprp}^2}} \, \left [
(P^{\, \prime} \widetilde \varphi j^{\, \prime}) [(q \Lambda j) - 
\ii (q \varphi j)]  + (P^{\, \prime} \widetilde \varphi j) 
[(q \Lambda j^{\, \prime}) + 
\ii (q \varphi j^{\, \prime})] \right] {\cal I}_{n-1,\ell-1} {\cal I}_{n-1,\ell}  \right. 
\\[3mm]
\nonumber
&&\left. - \sqrt{\frac{2 \beta n}{q_{\mprp}^2}} \, \left [
(p \widetilde \varphi j^{\, \prime}) [(q \Lambda j) - 
\ii (q \varphi j)]  +  (p \widetilde \varphi j) 
[(q \Lambda j^{\, \prime}) + 
\ii (q \varphi j^{\, \prime})] \right] {\cal I}_{n,\ell} {\cal I}_{n-1,\ell}  \right. 
\\[3mm]
\nonumber
&&\left. + \sqrt{\frac{2 \beta n}{q_{\mprp}^2}} \, \left [
(p \widetilde \varphi j^{\, \prime}) [(q \Lambda j) + 
\ii (q \varphi j)]  +  (p \widetilde \varphi j) 
[(q \Lambda j^{\, \prime}) - 
\ii (q \varphi j^{\, \prime})] \right] {\cal I}_{n-1,\ell-1} {\cal I}_{n,\ell-1}  \right. 
\\[3mm]
\nonumber
&&\left. - (p \widetilde \varphi q) \left [[(j \Lambda j^{\, \prime}) - 
\ii (j \varphi j^{\, \prime})]{\cal I}_{n,\ell-1}^2 - [(j \Lambda j^{\, \prime}) +
\ii (j \varphi j^{\, \prime})]{\cal I}_{n-1,\ell}^2 \right ] 
\right \};
\eeq

%
%
\beq
\label{eq:FAA}
&&{\cal D}^{(1)}_{AA} =  g_a g_a^{\, \prime}    
\bigg \{ [(P \widetilde \Lambda j) (p \widetilde \Lambda j^{\, \prime}) + 
(p \widetilde \Lambda j) (P \widetilde \Lambda j^{\, \prime})  - 
(j \widetilde \Lambda j^{\, \prime})(M_{\ell}^2 + m^2 + (p \widetilde \Lambda q))] 
[{\cal I}_{n,\ell}^2 + {\cal I}_{n-1,\ell-1}^2]  
\\[3mm]
\nonumber
&& + 4 \beta \sqrt{n\ell} \, (j \widetilde \Lambda j^{\, \prime}) 
{\cal I}_{n,\ell} {\cal I}_{n-1,\ell-1} - 
\sqrt{\frac{2 \beta \ell}{q_{\mprp}^2}} \, \left [
(P \widetilde \Lambda j) [(q \Lambda j^{\, \prime}) + 
\ii (q \varphi j^{\, \prime})]  + (P \widetilde \Lambda j^{\, \prime}) 
[(q \Lambda j) - \ii (q \varphi j)] \right] {\cal I}_{n,\ell} {\cal I}_{n,\ell-1} 
\\[3mm]
\nonumber
&& - 
 \sqrt{\frac{2 \beta \ell}{q_{\mprp}^2}} \, \left [
(P \widetilde \Lambda j)[(q \Lambda j^{\, \prime}) - 
\ii (q \varphi j^{\, \prime})]  + (P \widetilde \Lambda j^{\, \prime})  
[(q \Lambda j) + 
\ii (q \varphi j)] \right] {\cal I}_{n-1,\ell-1} {\cal I}_{n-1,\ell} 
\eeq
\beq
\nonumber
&& - 
\sqrt{\frac{2 \beta n}{q_{\mprp}^2}} \, \left [
(p \widetilde \Lambda j)[(q \Lambda j^{\, \prime}) - 
\ii (q \varphi j^{\, \prime})]  +  (p \widetilde \Lambda j^{\, \prime}) 
[(q \Lambda j) + 
\ii (q \varphi j)] \right] {\cal I}_{n,\ell} {\cal I}_{n-1,\ell}   
\\[3mm]
\nonumber
&&- \sqrt{\frac{2 \beta n}{q_{\mprp}^2}} \, \left [
(p \widetilde \Lambda j) [(q \Lambda j^{\, \prime}) + 
\ii (q \varphi j^{\, \prime})]  +  (p \widetilde \Lambda j^{\, \prime}) 
[(q \Lambda j) - 
\ii (q \varphi j)] \right] {\cal I}_{n-1,\ell-1} {\cal I}_{n,\ell-1} + 
 (M_{\ell}^2 + m^2 + (p \widetilde \Lambda q))   
\\[3mm]
\nonumber
&&\times \left [[(j \Lambda j^{\, \prime}) + 
\ii (j \varphi j^{\, \prime})] {\cal I}_{n,\ell-1}^2 + [(j \Lambda j^{\, \prime}) - 
\ii (j \varphi j^{\, \prime})] {\cal I}_{n-1,\ell}^2 \right ] + 
\frac{4 \beta \sqrt{n\ell}}{q_{\mprp}^2} \, 
[(q \Lambda j) (q \Lambda j^{\, \prime}) - 
(q \varphi j) (q \varphi j^{\, \prime})] {\cal I}_{n,\ell-1} {\cal I}_{n-1,\ell}
\bigg \};
\\[2mm]
\nonumber
&&{\cal D}^{(2)}_{AA} = {\cal D}^{(1)}_{AA} (q\to -q ,\, \, j \leftrightarrow j^{\, \prime}) \, .
\eeq

We notice, that the expressions  for amplitudes ${\cal M}_{VS}$, ${\cal M}_{VP}$, ${\cal M}_{VV}$ 
and ${\cal M}_{AV}$ are manifestly gauge invariant.

\end{widetext}


\section{Discussion}\label{Sec:5}


In this paper, we have calculated the tree-level two-point amplitudes for the transitions 
$jf \to j^{\, \prime} f^{\, \prime}$ in a constant uniform magnetic field of an 
arbitrary strength, and in
charged fermion plasma, for generalized vertices of the scalar, pseudoscalar, vector or  
axial types.
It is remarkable, that all the amplitudes obtained are manifestly Lorentz invariant, due to the choice 
of the Dirac equation solutions as the eigenfunctions of the covariant operator $\hat{\mu}_z$. 
In this case, partial contributions to the amplitude from the channels with different fermion 
polarization states 
are calculated separately, by direct multiplication of the bispinors and the Dirac matrices. 
This approach is an alternative to the method where the amplitudes squared are calculated, 
summed over the fermion polarization states, with using the fermion density matrices, 
see, e.g.~\cite{Andreev:2010,Gvozdev:2012}.

{\bf Acknowledgements}                   

The study was supported in part by the Russian Foundation for Basic Research (Project No. \mbox{11-02-00394-a}).


\end{document}